\documentclass[12pt]{article}
\pdfoutput=1

\usepackage{draft} 
\usepackage{hyperref}
\usepackage{graphicx,color,subfig}
\usepackage{cite}
\usepackage{mciteplus}
\usepackage{skak}
\usepackage{empheq}
\DeclareFontFamily{OT1}{pzc}{}
\DeclareFontShape{OT1}{pzc}{m}{it}{<-> s * [1.10] pzcmi7t}{}
\DeclareMathAlphabet{\mathpzc}{OT1}{pzc}{m}{it}

\def\be#1\ee{\begin{align}#1\end{align}}

\begin{document}

\unitlength = .8mm

\begin{titlepage}

\begin{center}

\hfill \\
\hfill \\
\vskip 1cm

\title{Modular Bootstrap Revisited}

\author{Scott Collier, Ying-Hsuan Lin,  Xi Yin}

\address{
Jefferson Physical Laboratory, Harvard University, \\
Cambridge, MA 02138 USA
}

\email{scollier@physics.harvard.edu, yinhslin@gmail.com,
xiyin@fas.harvard.edu}

\end{center}

\abstract{We constrain the spectrum of two-dimensional unitary, compact conformal field theories with central charge $c>1$ using modular bootstrap. Upper bounds on the gap in the dimension of primary operators of any spin, as well as in the dimension of scalar primaries, are computed numerically as functions of the central charge using semi-definite programming. Our bounds refine those of Hellerman and Friedan-Keller, and are in some cases saturated by known CFTs. In particular, we show that unitary CFTs with $c<8$ must admit relevant deformations, and that a nontrivial bound on the gap of scalar primaries exists for $c<25$. We also study bounds on the dimension gap in the presence of twist gaps, bounds on the degeneracy of operators, and demonstrate how  ``extremal spectra" which maximize the degeneracy at the gap can be determined numerically.}

\vfill

\end{titlepage}

\eject

\tableofcontents

\section{Introduction} 

It is commonly believed that unitary, compact conformal field theories in two dimensions with central charge $c>1$ and no conserved Virasoro primary currents (of any spin) are ubiquitous.\footnote{After all, spin-1 currents are always governed by current algebra in a unitary theory, and can be gauged away, while higher spin currents hint at some sort of integrability \cite{Maldacena:2011jn, Perlmutter:2016pkf}.} In particular, the holographic duals of {\it generic} quantum theories of gravity in $AdS_3$ are expected to have such properties \cite{Aharony:1999ti, Witten:2007kt, Hartman:2014oaa, Keller:2014xba, Fitzpatrick:2014vua, Fitzpatrick:2015zha, Chang:2015qfa}. Such theories are hard to construct using standard CFT techniques, however. For instance, rational CFTs always contain an extended chiral algebra and therefore extra conserved currents \cite{Witten:1983ar, Moore:1988qv, Verlinde:1988sn}.
Product orbifolds and (GSO projected) superconformal theories also contain higher spin conserved currents \cite{Eguchi:1988af}. Evidently there is a tremendous gap in our knowledge of two-dimensional CFTs: the vast majority of CFTs that are believed to exist are inaccessible with available analytic methods.\footnote{See however \cite{Halpern:1995js, Dotsenko:aa} for families of candidate irrational CFTs that potentially admit no extra conserved currents.} In this paper, we undertake initial steps towards closing this gap, by refining the modular bootstrap approach of \cite{Hellerman:2009bu,Hellerman:2010qd,Keller:2012mr, Friedan:2013cba,Qualls:2013eha,Kim:2015oca}. A surprisingly rich set of new constraints on the CFT spectrum will be uncovered from unitarity and the modular invariance of the torus partition function alone.

To begin with, we consider the Virasoro character decomposition of the torus partition function of a CFT with no conserved current primaries (the latter assumption may be relaxed, as we will consider later), of the form
\ie\label{decomp}
Z(\tau, \bar\tau) &= \chi_0(\tau) \bar \chi_0(\bar\tau) + \sum_{h, \bar h>0} d(h,\bar h) \chi_{h}(\tau) \bar\chi_{\bar h}(\bar\tau),
\fe
and impose positivity on the coefficients $d(h,\bar h)$. The characters $\chi_h$ with $h>0$ are non-degenerate, and the spins $h-\bar h$ are assumed to be integers. Modular invariance can be formulated as the ``modular crossing equation"
\ie\label{mc}
E_{0,0}(\tau,\bar\tau) + \sum_{h, \bar h>0} d(h,\bar h) E_{h,\bar h}(\tau,\bar\tau)  = 0,
\fe 
where
\ie
E_{h,\bar h}(\tau,\bar\tau) \equiv \chi_h(\tau) \bar \chi_{\bar h}(\bar\tau) - \chi_h(-1/\tau) \bar \chi_{\bar h}(-1/\bar\tau) .
\fe
Constraints on the spectrum of Virasoro primaries, namely the set of $(h,\bar h)$ in the sum as well as the coefficients $d(h,\bar h)$, due to the positivity condition on $d(h,\bar h)$, can be extracted using the powerful numerical method of semi-definite programming. One proceeds by assuming certain properties of the spectrum, say the presence of a gap in the dimensions, and seeks a linear functional $\A$ that acts on functions of $(\tau,\bar\tau)$, such that $\A[E_{0,0}]$ takes the same sign as $\A[E_{h,\bar h}]$ for all $(h,\bar h)$ in the hypothetical spectrum.\footnote{Typically, there are infinitely many primaries and we would only make assumptions on the properties of a finite subset of operators. We need $\A[E_{h,\bar h}]$ to have the same sign for sufficiently large $h$ or $\bar h$ in order to derive a useful constraint.} If such a linear functional is found, the positivity condition on $d(h,\bar h)$ cannot be satisfied and the proposed spectrum would be ruled out.

In \cite{Hellerman:2009bu,Friedan:2013cba,Qualls:2013eha}, the linear functional $\A$ was taken to be a linear combination of derivatives in $\tau_2$ up to a certain order, evaluated at the self-dual point $\tau=i$ (i.e., modular invariance was imposed only on the restriction of the partition function to the imaginary $\tau$ axis). Using such linear functionals, one finds an upper bound on the scaling dimension of the primaries that is insensitive to the spin. We denote this bound by $\Delta_{\rm HFK}^{(N)}(c)$, where $N$ is the maximal derivative order of the linear functional.\footnote{$N$ was taken to be 3 in \cite{Hellerman:2009bu} and up to 23 in \cite{Friedan:2013cba}.} It was found in \cite{Friedan:2013cba} that, for {\it fixed} $N$,  $\Delta_{\rm HFK}^{(N)}(c)={c\over 6} + {\cal O}(1)$ in the $c\to \infty$ limit.

In this paper, we make two important refinements of the analysis of \cite{Friedan:2013cba}. Firstly, we find substantial numerical evidence that the $N\to \infty$ limit and $c\to \infty$ limit of $\Delta_{\rm HFK}^{(N)}(c)$ do not commute. The optimal gap $\Delta_{\rm HFK}(c)\equiv \Delta_{\rm HFK}^{(\infty)}(c)$ appears to have the property that its slope $d\Delta_{\rm HFK}/dc$ decreases monotonically with $c$, and asymptotes to a value $b_{\rm HFK}$ that lies between ${1\over 9}$ and ${1\over 12}$. Secondly, we find a stronger bound on the gap, $\Delta_{\rm mod}^{(N)}(c)$, using linear functionals built out of derivatives in both $\tau$ and $\bar\tau$ up to total derivative order $N$. The optimal bound on the gap, $\Delta_{\rm mod}(c)\equiv \Delta_{\rm mod}^{(\infty)}(c)$, appears to have the same property that  $d\Delta_{\rm HFK}/dc$ decreases monotonically with $c$, and asymptotes to a value $b_{\rm mod}\leq b_{\rm HFK}$.\footnote{Our numerical precision at large values of $c$ is insufficient in resolving the difference between $b_{\rm mod}$ and $b_{\rm HFK}$, if there is any.}

We can also obtain spin-dependent bounds, by applying the most general linear functional to the modular crossing equation. For instance, we have computed an upper bound $\Delta_{\rm mod}^{s=0}(c)$ on the gap in the dimension of scalar primaries, regardless of the gaps in the spectra of higher-spin primaries. While initially we assume the absence of extra conserved currents, this assumption will later be relaxed, with very little difference in the resulting bounds.\footnote{More precisely, when conserved primary currents are allowed, we find no difference in the bound $\Delta_{\rm mod}^{s=0}$ within the numerical error of the binary search for the optimal bound based on semi-definite programming for $1\le c \le 8$ and a slightly weaker bound for $c>8$ obtained using linear functionals up to a {\it fixed derivative order}. Our numerical extrapolation to infinite derivative order is not accurate enough to resolve the difference between the two bounds.} It is observed that $\Delta_{\rm mod}^{s=0}(c)<2$ for $c<8$, that is, unitary CFTs with $c<8$ must admit a relevant deformation. There is a kink on the bounding curve at $c=8$, with scalar gap $\Delta_{\rm mod}^{s=0}=2$, which we believe is exact. The gap at this kink is saturated by the $E_8$ WZW model at level $1$.\footnote{Note that while this CFT does contain extra conserved currents, its partition function happens to admit a formal decomposition in terms of {\it non-degenerate} Virasoro characters with non-negative coefficients, due to the contributions from twist-2 primaries. We refer to such a partition function as that of the {\bf generic type}, which may be viewed as a limiting case of partition functions with no conserved current primaries.} Based on this, we conclude that ``perfect metals" \cite{Plamadeala:2014roa} do not exist for $c\leq 8$. Interestingly, $\Delta_{\rm mod}^{s=0}(c)$ diverges at $c=25$, and in fact modular invariance is compatible with a spectrum that contains {\it no} scalar primaries for $c>25$. We will give an explicit example of a spectrum that has such a property.

It is also possible to place upper bounds on the degeneracies of the lightest primaries, provided that their dimension lies in between ${c-1\over 12}$ and $\Delta_{\rm mod}$ (the former is the upper bound on the twist gap). When this upper bound on the degeneracy is saturated, the entire spectrum of primaries is determined as zeroes of the optimal linear functional acting on the characters as a function of the weights.\footnote{A priori, the degeneracies of higher-dimension primaries in the extremal spectrum are not fixed by this procedure. However, in several examples of CFTs that realize the extremal spectrum, we find that the degeneracies of higher dimension operators agree with the respective upper bounds subject to the assumed gap.}
We refer to this spectrum as the {\bf extremal spectrum}.\footnote{Note that in our definition of the extremal spectrum, the degeneracies are only required to be positive, and not necessarily integers. Demanding the latter would slightly refine our bounds.}  We will demonstrate the extraction of the extremal spectrum in a number of examples. It is observed that when the upper bound on the (scalar) dimension gap is saturated for $1\le c\le 4$, the extremal spectrum always contains conserved spin-1 currents and marginal scalar primaries. Rather strikingly, we will uncover precisely the spectra of $SU(2),\, SU(3),\, G_2$ and $SO(8)$ WZW models at level 1 from the extremal spectra with maximal gap at the respective values of central charge. 

\section{Constraining Gaps in the Operator Spectrum}

\subsection{Basic setup}

The vacuum Virasoro character $\chi_0$ and the non-degenerate Virasoro character $\chi_h$ are given by
\ie
\chi_0(\tau) = { q^{-{c-1\over24}} \over \eta(\tau) } (1-q), \quad \chi_h(\tau) = { q^{h-{c-1\over24}} \over \eta(\tau) },
\fe
where $q = e^{2\pi i\tau}$, $\eta(\tau)$ is the Dedekind eta function, and $c$ is the central charge. In general, the partition function of a compact, unitary CFT admits the character decomposition
\ie\label{genchar}
Z(\tau, \bar\tau) &= \chi_0(\tau)\bar \chi_0(\bar\tau) + \sum_{j=1}^\infty \left[ d_j \chi_j(\tau) \bar\chi_0(\bar\tau) + \widetilde d_j \chi_0(\tau) \bar\chi_j(\bar\tau) \right] + \sum_{h,\bar h} d(h,\bar h) \chi_{h}(\tau)\bar\chi_{\bar h}(\bar\tau) ,
\fe
where $d_j$ and $\widetilde d_j$ are the degeneracies of holomorphic and anti-holomorphic currents of spin $j$, and $d(h,\bar h)$ is the degeneracy of primary operators of weight $(h,\bar h)$. For the rest of this paper, we will assume that the spectrum is parity-invariant. The constraints we derive on the parity invariant-spectrum can be applied to parity non-invariant theories as well, if we simply consider the projection of the partition function $Z(\tau,\bar\tau)$ onto its parity invariant part, ${1\over 2} \left[Z(\tau,\bar\tau) + Z^*(\bar \tau,\tau)\right]$. For the parity invariant spectrum, we will label the primaries by their dimension $\Delta=h+\bar h$ and spin $s=|h-\bar h|$ ($\in \mathbb{Z}_{\geq 0}$), and write the degeneracies as
\ie
d_{\Delta,s} \equiv d({\Delta+s\over 2},{\Delta-s\over 2}) = d({\Delta-s\over 2},{\Delta+s\over 2}) .
\fe
As explained in the introduction, we will be primarily interested in CFTs with no conserved currents, i.e., $d_j=\widetilde d_j=0$, and all nontrivial primaries obey $\Delta>s$. In deriving various numerical bounds, it will be convenient to allow for the limiting case $\Delta\to s$, which corresponds to a contribution to the partition function of the form 
\ie
\lim_{\Delta\to s} \left(\chi_{\Delta+s\over 2} \bar\chi_{\Delta-s\over 2} + \chi_{\Delta-s\over 2} \bar\chi_{\Delta+s\over 2}\right) = \chi_s (\bar\chi_0+\bar\chi_1) + (\chi_0+\chi_1)\bar\chi_s .
\fe
That is, the spectrum may contain conserved currents of spin $s$, along with twist-2 primaries of dimension $s+1$ and spin $s-1$. We refer to such a partition function as that of the {\bf generic type}. We will see in several instances that certain rational CFTs with partition functions of the generic type appear at kinks on the boundary of the domain of allowed spectra.

As in \cite{Friedan:2013cba}, it is convenient to work with the reduced partition function
\begin{equation}\label{ReducedPartition}
	\hat{Z}(\tau,\bar\tau) \equiv |\tau|^{1\over 2}|\eta(\tau)|^{2}Z(\tau,\bar\tau),
\end{equation}
decomposed into the reduced characters
\begin{align}\label{ReducedCharacters}
\hat{\chi}_0(\tau)\hat{\bar\chi}_{0}(\bar\tau) &= |\tau|^{1\over 2}|q^{-{c-1\over 24}}(1-q)|^2,~~~~	\hat{\chi}_{h}(\tau)\hat{\bar\chi}_{\bar h}(\bar\tau) = |\tau|^{1\over 2}q^{h-{c-1\over 24}}\bar q^{\bar h-{c-1\over 24}}.
\end{align}
Since $|\tau|^{1\over 2}|\eta(\tau)|^2$ is invariant\footnote{It is not invariant under the $T$ transform $\tau\to \tau+1$, but we have already taken into account $T$-invariance of $Z(\tau,\bar\tau)$ by demanding that the spins are integers, so it suffices to examine the $S$ transform.} under $\tau\to -1/\tau$, it suffices to consider the modular crossing equation for $\hat{Z}$. 

It is convenient to introduce the variable $z$ defined by $\tau = i\exp(z)$, so that the modular $S$ transformation $\tau\rightarrow-{1\over \tau}$ takes $z\rightarrow -z$.  Modular invariance amounts to the statement that the partition function is an even function in $(z,\bar z)$ (and that all spins are integers). To implement semi-definite programming on the modular crossing equation, we will apply to it the basis of linear functionals
\ie\label{basis}
\partial_z^m \partial_{\bar z}^n|_{z = \bar z = 0}, \quad m+n \text{ odd}.
\fe
In other words, we consider linear functionals of the form
\ie\label{LinearFunctionalZ}
\A \equiv \sum_{m+n \text{ odd}} \A_{m, n} \partial_z^m \partial_{\bar z}^n|_{z = \bar z = 0},
\fe
and turn the modular crossing equation (\ref{mc}) into
\ie
0 = \A[\hat Z(\tau, \bar\tau)] = \A[\hat \chi_0(\tau) \hat{\bar\chi}_0(\bar\tau)] + \sum_{s=0}^\infty \sum_{\Delta \in {\cal I}_s} d_{\Delta, s} \A\left[ \hat \chi_{\Delta-s \over 2}(\tau)\hat{\bar \chi}_{\Delta+s \over 2}(\bar\tau) + \hat \chi_{\Delta+s \over 2}(\tau) \hat {\bar\chi}_{\Delta-s \over 2}(\bar\tau) \right],
\fe
where ${\cal I}_s$ is a (typically infinite) discrete set that consists of the dimensions of primaries of spin $s$.

One may proceed by hypothesizing, for instance, that there is a gap $\Delta_s^*$ ($\geq s$) in the spectrum of spin-$s$ primaries, i.e., ${\cal I}_s$ consists of dimensions $\Delta\geq\Delta_s^*$ only. If we can find a functional $\A$, which amounts to a set of $\A_{m, n} \in \bR$ in (\ref{LinearFunctionalZ}), such that
\ie\label{ModularHypothesis}
 \A[\hat \chi_0(\tau) \hat {\bar\chi}_0(\bar\tau)] >& 0,
\\
 \A\left[ \hat \chi_{\Delta-s \over 2}(\tau) \hat {\bar\chi}_{\Delta+s \over 2}(\bar\tau) + \hat \chi_{\Delta+s \over 2}(\tau) \hat {\bar\chi}_{\Delta-s \over 2}(\bar\tau) \right] \geq & 0, \quad \Delta \geq \Delta_s^*,~~~\forall s\in \mathbb{Z}_{\geq 0},
\fe
we would then arrive at a contradiction with the non-negativity of the degeneracies $d_{\Delta,s}$ in the modular crossing equation, thereby ruling out the putative spectrum. In other words, we would have proven that the gap in the spin-$s$ spectrum cannot exceed $\Delta_s^*$ simultaneously for all $s$. 

As examples, we may take

\noindent $\bullet$ $\Delta_s^*={\rm max}(\Delta_{\rm mod},s)$, where $\Delta_{\rm mod}$ is the maximal gap in the scaling dimension spectrum of all primaries.

\noindent $\bullet$ $\Delta_0^*= \Delta_{\rm mod}^{s=0}$, $\Delta_s^*=s$ ($s\geq 1$), where $\Delta_{\rm mod}^{s=0}$ is the maximal gap in the dimension of scalar primaries.

\noindent $\bullet$ $\Delta_s^*=s+t_{\rm mod}$, where $t_{\rm mod}$ is the maximal twist gap in the spectrum of all primaries.

To implement the above procedure numerically, we must restrict to a finite subset of the basis of linear functionals, say $\partial_z^m \partial_{\bar z}^n|_{z=\bar z=0}$ with odd $m+n \leq N$. We will refer to $N$ as the ``derivative order" of the linear functional. The upper bound on the gap $\Delta_{\rm mod}$ derived by exclusion using linear functionals up to derivative order $N$, for instance, will be denoted $\Delta_{\rm mod}^{(N)}$. In other words, $\Delta_{\rm mod}^{(N)}$ is the smallest $\Delta^*$ such that \eqref{ModularHypothesis} can be satisfied with $\Delta_s^*={\rm max}(\Delta_{\rm mod},s)$, by a functional $\A$ restricted to derivative order $N$.  While for each positive integer $N$, $\Delta_{\rm mod}^{(N)}$ is a rigorous bound on the gap, the optimal bound would be obtained by taking the $N\to \infty$ limit.

Our numerical analysis is performed using the SDPB package \cite{Simmons-Duffin:2015qma}. In practice, we also need to truncate the spectrum: while SDPB allows us to consider a spectrum that consists of operators of all dimensions (say, above a hypothetical gap), we will need to restrict the spins of operators to a (sufficiently large) finite range, $s\leq s_{max}$. In seeking linear functionals for the purpose of excluding a given hypothetic spectrum, increasing $s_{max}$ puts more constraints on the linear functional, and in principle we need to take the limit $s_{max} \to \infty$.  In practice, however, the bound $\Delta_{\rm mod}^{(N)}$ derived at a given derivative order $N$ stabilizes to within numerical precision once $s_{max}$ exceeds a certain value (typically of order $N$). 

In optimizing the bounds with increasing derivative order $N$, we find that at larger values of the central charge $c$, one must also work to higher values of $N$ for the bound $\Delta_{\rm mod}^{(N)}$ to stabilize. When such a stabilization is unattainable due to the computational complexity, we will need to numerically extrapolate $\Delta_{\rm mod}^{(N)}$ to $N=\infty$, by fitting with a polynomial in $1/N$ (say, of linear or quadratic order).

\subsection{The twist gap}

In a unitary, compact two-dimensional CFT with central charge $c>1$, the twist gap\footnote{In two-dimensional CFTs, the twist $t$ (here we do not use $\tau$ to avoid confusion with the modular parameter) is defined by $t = \Delta-s = 2~\text{min}(h,\bar{h})$.}  $t_{gap}$ can be no larger than $c-1\over 12$. Furthermore, in the absence of conserved currents, there must be infinitely many high spin primaries whose twists accumulate to $c-1\over 12$. This can be seen as follows.\footnote{This argument is due to Tom Hartman.}

\begin{figure}[h!]{
\centering
\subfloat{
\includegraphics[width=.6\textwidth]{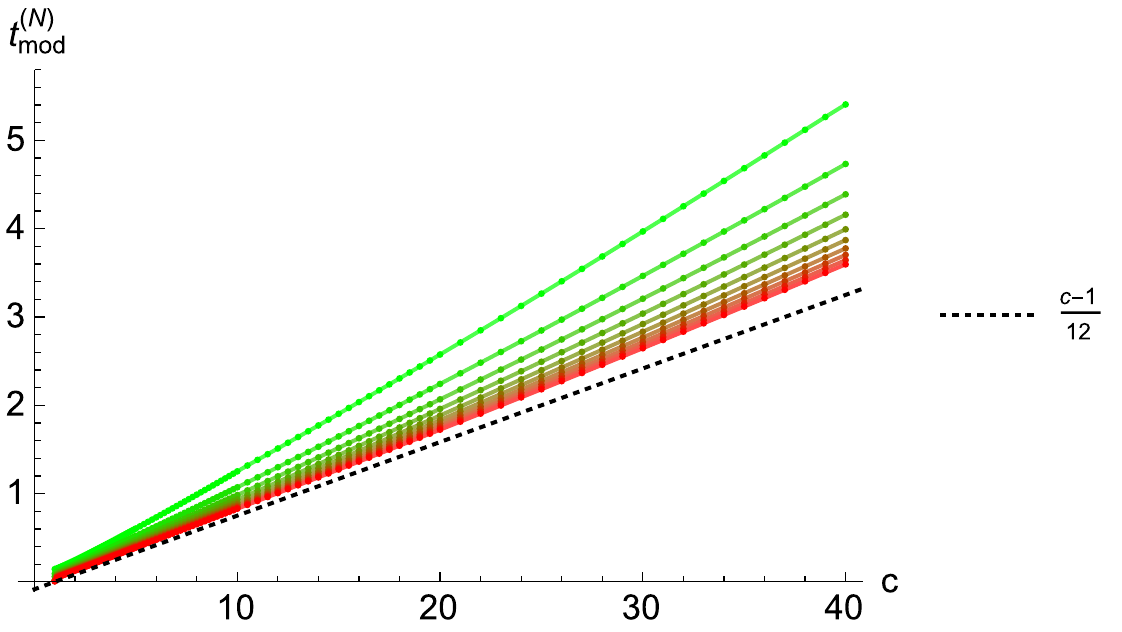}
}
\\
\subfloat{
\includegraphics[width=.45\textwidth]{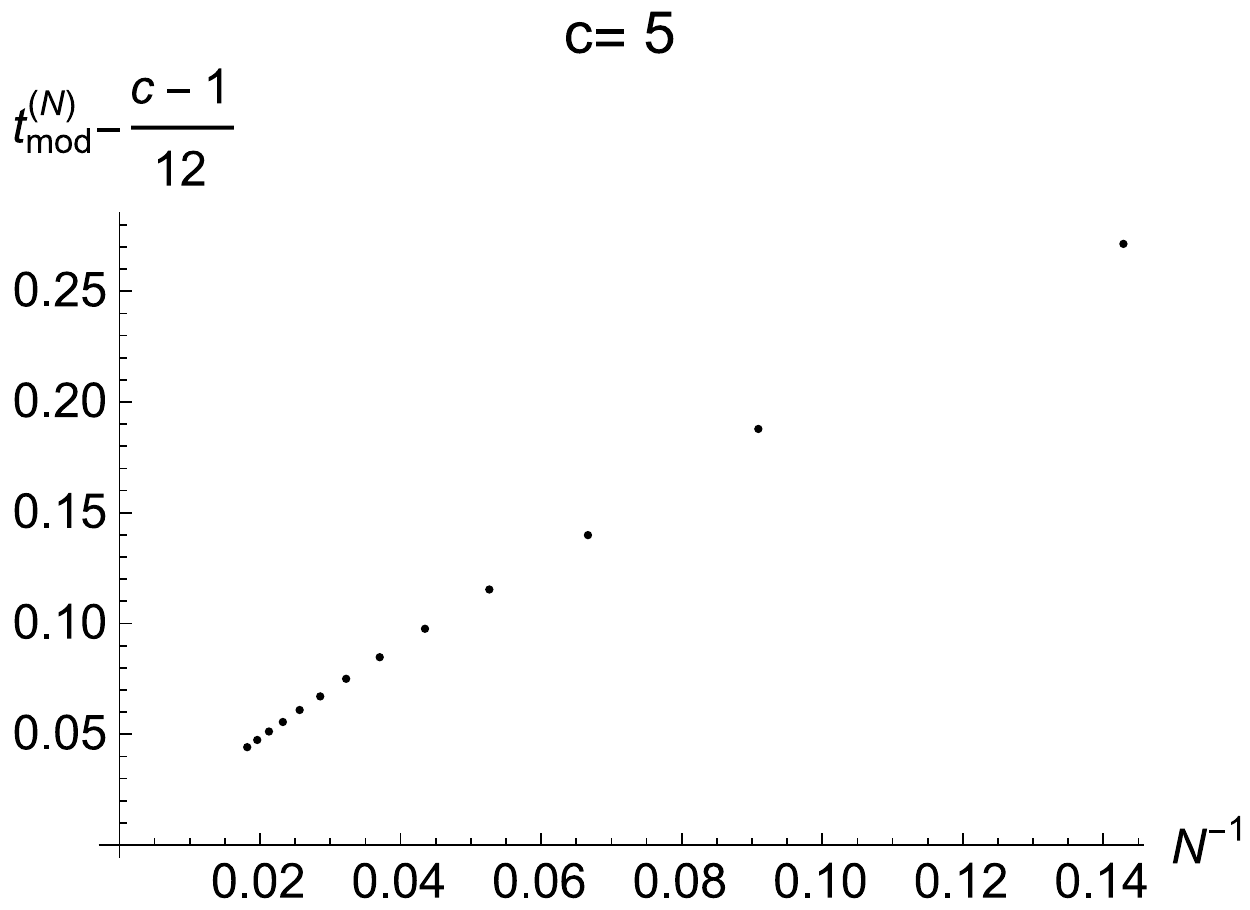}
}
\subfloat{
\includegraphics[width=.45\textwidth]{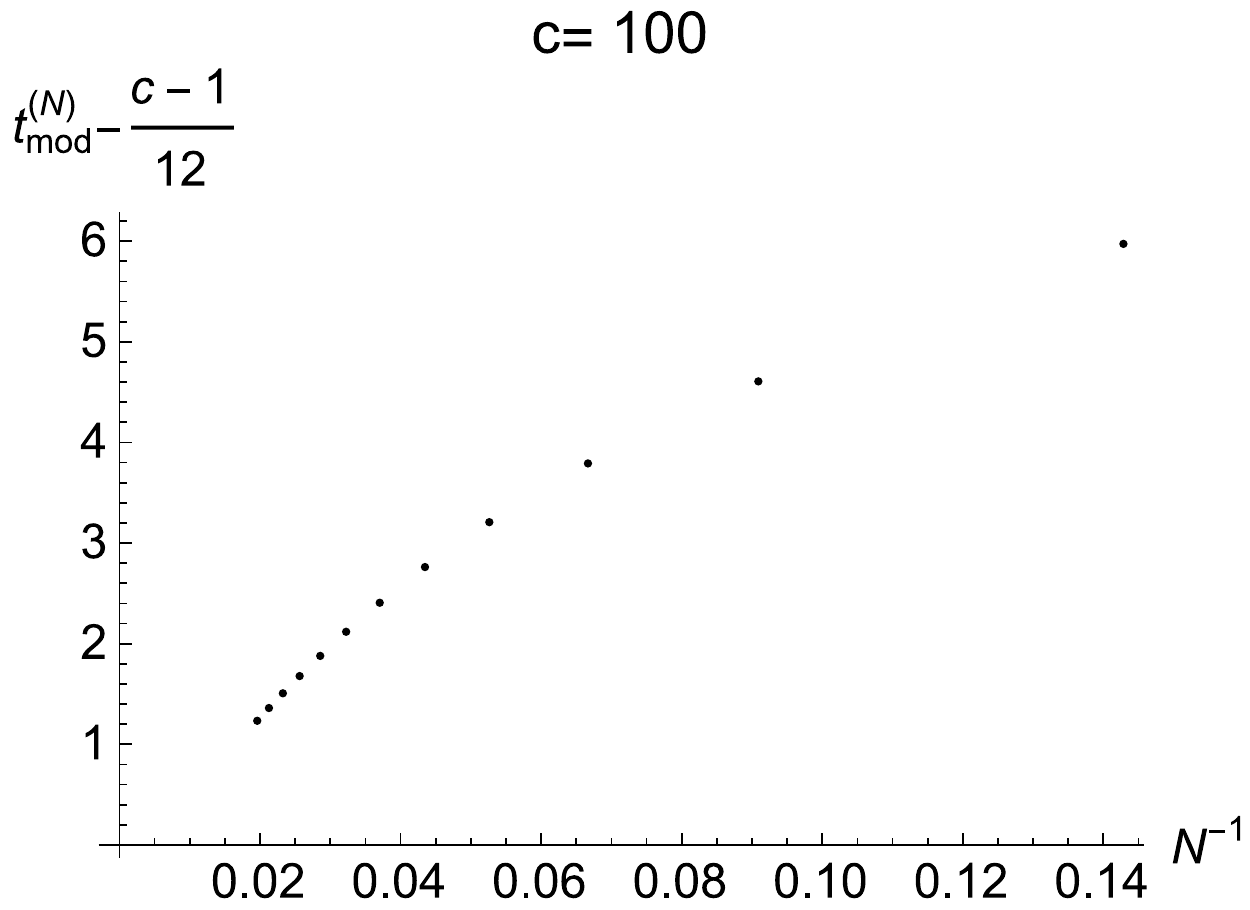}
}
\caption{{\bf Top:} The colored curves are the upper bound on the twist gap as a function of the central charge for increasing derivative order (up to $43$) from green to red. The results are consistent with the convergence to the bound on the twist gap $t_{\rm mod} = {c-1\over 12}$ predicted by the analytic argument in the infinite-$N$ limit. {\bf Bottom:} The bound on the twist gap $t^{(N)}_{\rm mod}$ as a function of inverse derivative order for $c=5$ and $c=100$.}\label{fig:TwistGap}}
\end{figure}

Consider the character decomposition of the partition function (\ref{decomp}).
In the limit $\bar\tau \to -i0^+$, $Z(\tau,\bar\tau)=Z(-1/\tau,-1/\bar\tau)$ is dominated by the modular transformed vacuum character, which may be expressed as
\ie\label{twista}
\lim_{\bar\tau\to -i0^+}{Z(-1/\tau,-1/\bar\tau)\over \bar\chi_0(-1/\bar \tau)}=\chi_0(-1/\tau)  = \int_{c-1\over24}^\infty dh' K(0, h')\chi_{h'}(\tau),
\fe
where $K(h, h')$ is the modular kernel \cite{Keller:2014xba, Hartman:2014oaa}. On the other hand, we have
\ie\label{twistb}
\lim_{\bar\tau\to -i0^+}{Z(\tau,\bar\tau)\over \bar\chi_0(-1/\bar \tau)}=\lim_{\bar\tau\rightarrow-i0^+} \Bigg[\chi_0(\tau) {\bar\chi_0(\bar\tau)\over\bar\chi_0(-{1/\bar\tau})} + \sum_{h,\bar h>0} d(h,\bar h) \chi_{h}(\tau) {\bar\chi_{\bar h}(\bar\tau)\over \bar\chi_0(-{1/\bar\tau})} \Bigg].
\fe
In the limit $\bar\tau\to -i0^+$, 
$\bar\chi_0(-{1/\bar\tau})\sim [\bar q(-{1/\bar\tau})]^{-{c\over 24}}$
 while $\bar\chi_{\bar h}(\bar\tau)\sim (i\bar\tau)^{1\over2} [\bar q(-{1/\bar\tau})]^{-{1\over 24}}$, and thus $\lim_{\bar\tau\to -i0^+} {\bar\chi_{\bar h}(\bar\tau)\over \bar\chi_0(-1/\bar\tau)}=0$ for $c>1$. If we could exchange the limit and summation over primaries in (\ref{twistb}), we would have concluded that the right-hand side of (\ref{twistb}) vanished, which would contradict (\ref{twista}). In particular, in the $\tau\to i\infty$ limit (after taking $\bar\tau\to -i0^+$ first), the right-hand side of (\ref{twista}) is dominated by the character of weight $h'={c-1\over 24}$. For (\ref{twistb}) to be consistent, there must be infinitely many primaries with left conformal weight $h$ accumulating to ${c-1\over 24}$, or equivalently, their twists accumulating to $c-1\over 12$.

As a test of our numerical approach to modular bootstrap, we can indeed reproduce this twist gap bound, by seeking linear functionals with the following positivity properties,
\begin{align}
	\alpha\bigg[\hat{\chi}_0(\tau)\hat{\bar\chi}_0(\bar\tau)\bigg]>&0\nonumber\\
	\alpha\bigg[\hat{\chi}_{{t\over 2}+s}(\tau)\hat{\bar\chi}_{t\over 2}(\bar\tau) + \hat{\chi}_{t\over 2}(\tau)\hat{\bar\chi}_{{t\over 2}+s}(\bar\tau)\bigg]\ge& 0,\quad t\ge t^*_{\rm mod},~~~\forall s\in \mathbb{Z}_{\geq 0}.
\end{align}
The smallest $t^*_{\rm mod}$ such that there exists a linear functional $\alpha$ satisfying the above equation then yields the strongest upper bound on the twist gap, which we denote $t_{\rm mod}$. Figure~\ref{fig:TwistGap} shows $t^{(N)}_{\rm mod}(c)$ as a function of the central charge. Indeed it appears that the bounds are converging to the twist gap predicted by the above argument as the derivative order is taken to infinity. However, we observe that the convergence of the bound with $N^{-1}$ is slower as the central charge is increased. This will turn out to be a generic feature of bounds we obtain from semi-definite programming, and will prevent accurate determinations of asymptotic bounds as $c\rightarrow \infty$.

\subsection{Refinement of Hellerman-Friedan-Keller bounds}

Bounds on the gap in the dimension of all primaries, regardless of spin, were derived in \cite{Hellerman:2009bu, Friedan:2013cba} by applying linear functionals of the form
\begin{equation}\label{LinearFunctionalTau}
	\alpha = \sum_{{\rm odd~} n\le N} \left.\alpha_n (\tau\partial_\tau + \bar\tau\partial_{\bar\tau})^n\right|_{\tau=i}
\end{equation}
to the modular crossing equation. We refer to (\ref{LinearFunctionalTau}) as HFK functionals, and the resulting bound on the dimension gap $\Delta_{\rm HFK}^{(N)}$.

The HFK functional amounts to restricting the partition function to the imaginary $\tau$-axis. Writing $\tau = {i\beta}$, for real $\beta$, the character decomposition is blind to the spins of operators in the spectrum. In particular, the reduced partition function takes the form
\begin{equation}
	\hat{Z}(\beta) = \hat{Z}_0(\beta) + \sum_{\Delta}d_\Delta\hat{Z}_\Delta(\beta),
\end{equation}
where $d_\Delta$ is the degeneracy including \emph{all} primary operators with dimension $\Delta = h + \bar{h}$, and
\ie
\hat{Z}_0(\beta) = \beta^{1\over 2}e^{2\pi\beta{c-1\over 12}}(1-e^{-2\pi\beta})^2, ~~~~	\hat{Z}_\Delta(\beta) = \beta^{1\over 2}e^{-2\pi\beta(\Delta-{c-1\over 12})}.
\fe
The HFK functional can be written as
\begin{equation}\label{LinearFunctionalBeta}
	\alpha = \sum_{{\rm odd~}n\le N}\alpha_n\left.(\beta\partial_\beta)^n\right|_{\beta=1}.
\end{equation}
To place upper bounds on the dimension of the lightest operator in the spectrum, we search for linear functionals that satisfy the following positivity properties
\begin{align}\label{PositivityOfAlpha}
	\alpha[\hat{Z}_0(\beta)]>0&\nonumber\\
	\alpha[\hat{Z}_{\Delta}(\beta)]\ge 0&,\quad\Delta\ge\Delta_{\rm{\rm HFK}}^*.
\end{align}
If such a functional can be found, then $\Delta_{\rm HFK}^*$ is a rigorous upper bound on the gap. The bound $\Delta_{\rm HFK}^{(N)}$ is obtained as the smallest $\Delta_{\rm{\rm HFK}}^*$ such that (\ref{PositivityOfAlpha}) holds.

In \cite{Friedan:2013cba}, the bound on the gap in the large $c$ limit was of primary interest. It was found that, for {\it fixed} $N$ (taken to be 3 in \cite{Hellerman:2009bu} and 23 in \cite{ Friedan:2013cba}), in the large $c$ limit,
\begin{align}
	\Delta_{\rm HFK}^{(N)} = {c\over 6} + \mathcal{O}(1).
\end{align}

\begin{figure}[h!]
\subfloat{\includegraphics[width=.49\textwidth]{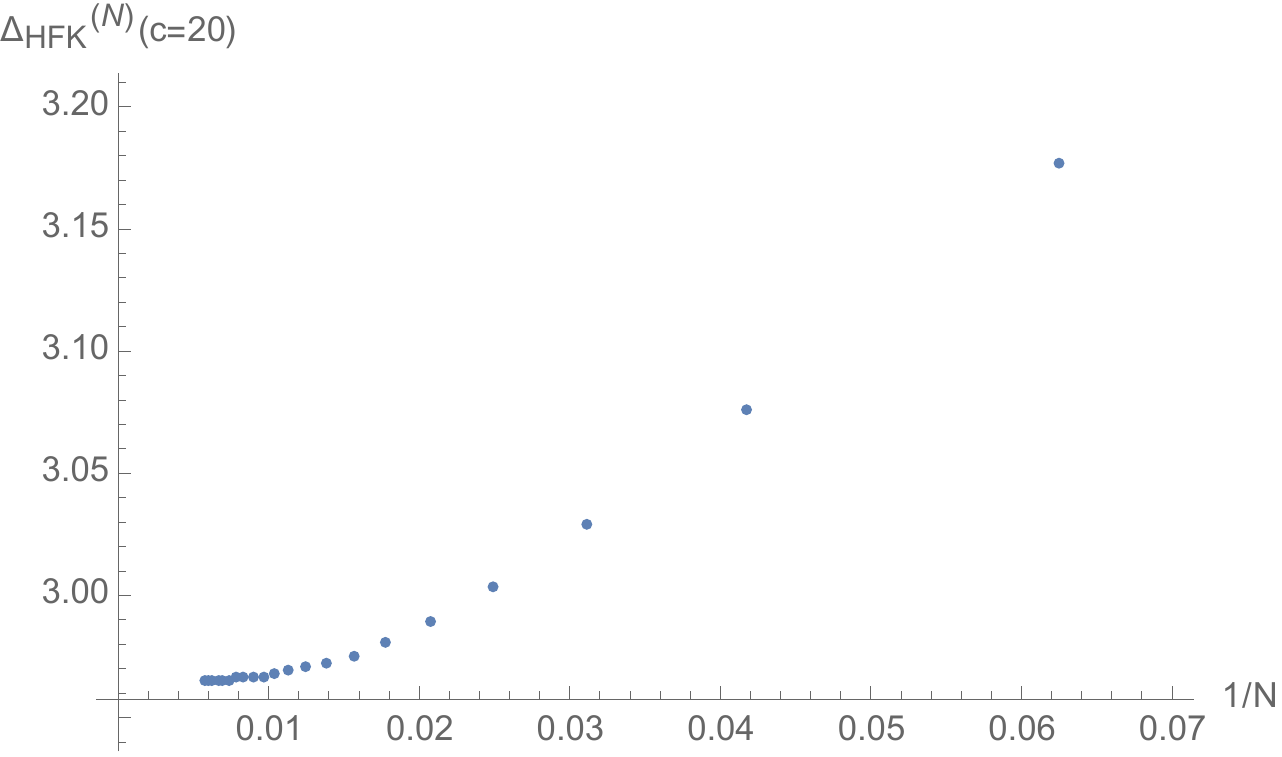}}~
\subfloat{\includegraphics[width=.49\textwidth]{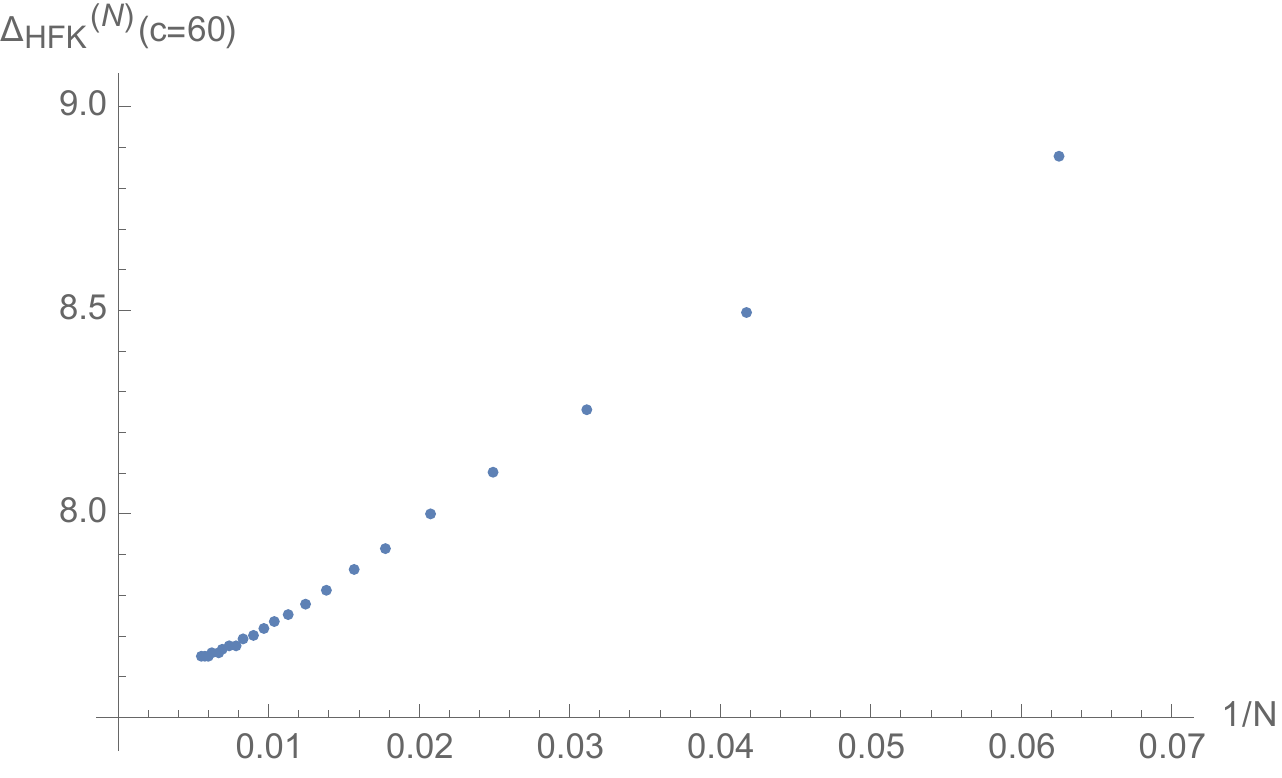}}
\caption{The gap bound $\Delta_{\rm HFK}^{(N)}(c)$ obtained using HFK functionals up to derivative order $N=175$ for central charge $c=20$ (left) and up to $N=183$ for $c=60$ (right). }\label{fig:HFK}
\end{figure}

\begin{figure}[h!]
\subfloat{\includegraphics[width=.49\textwidth]{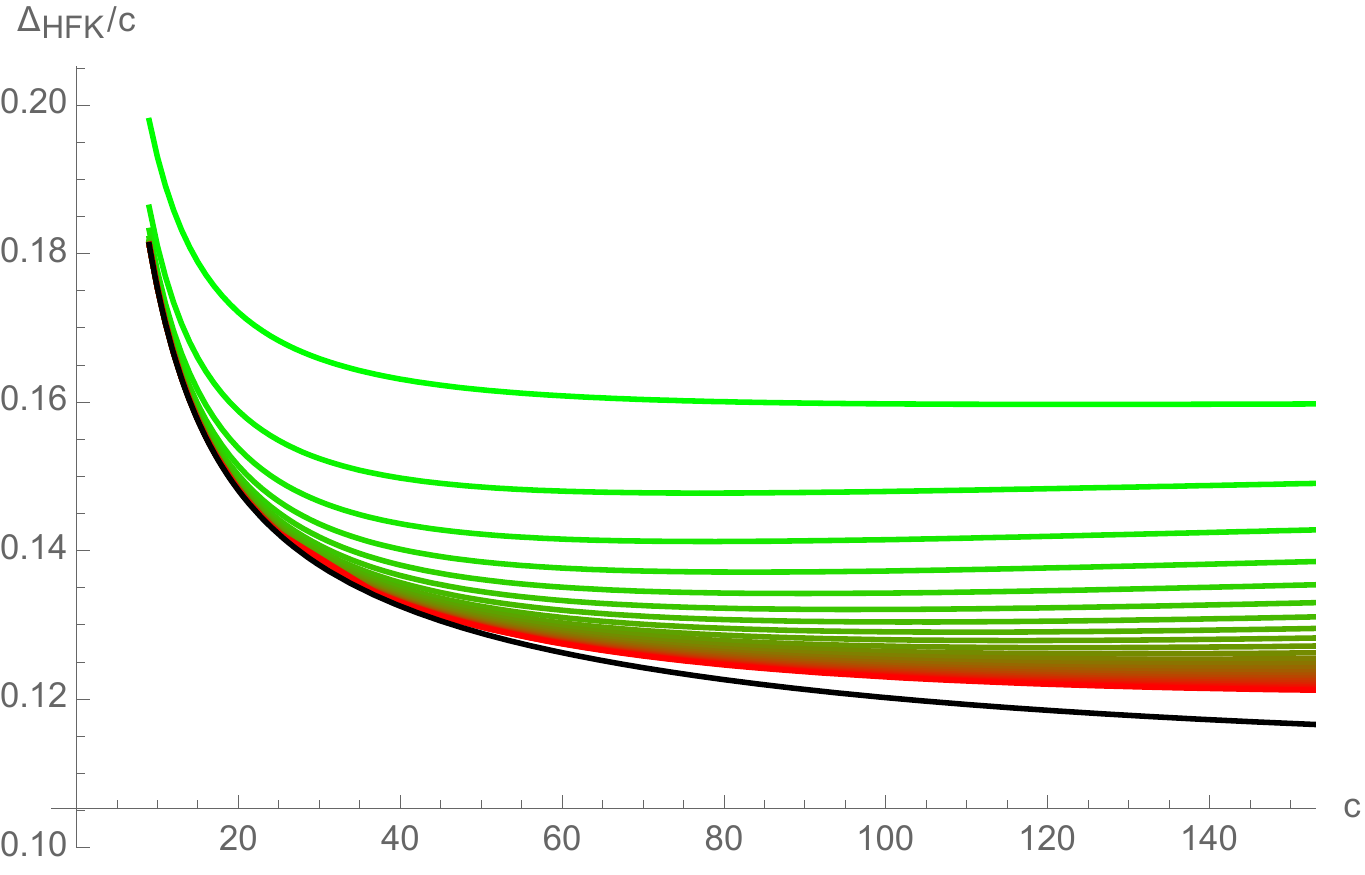}}~
\subfloat{\includegraphics[width=.49\textwidth]{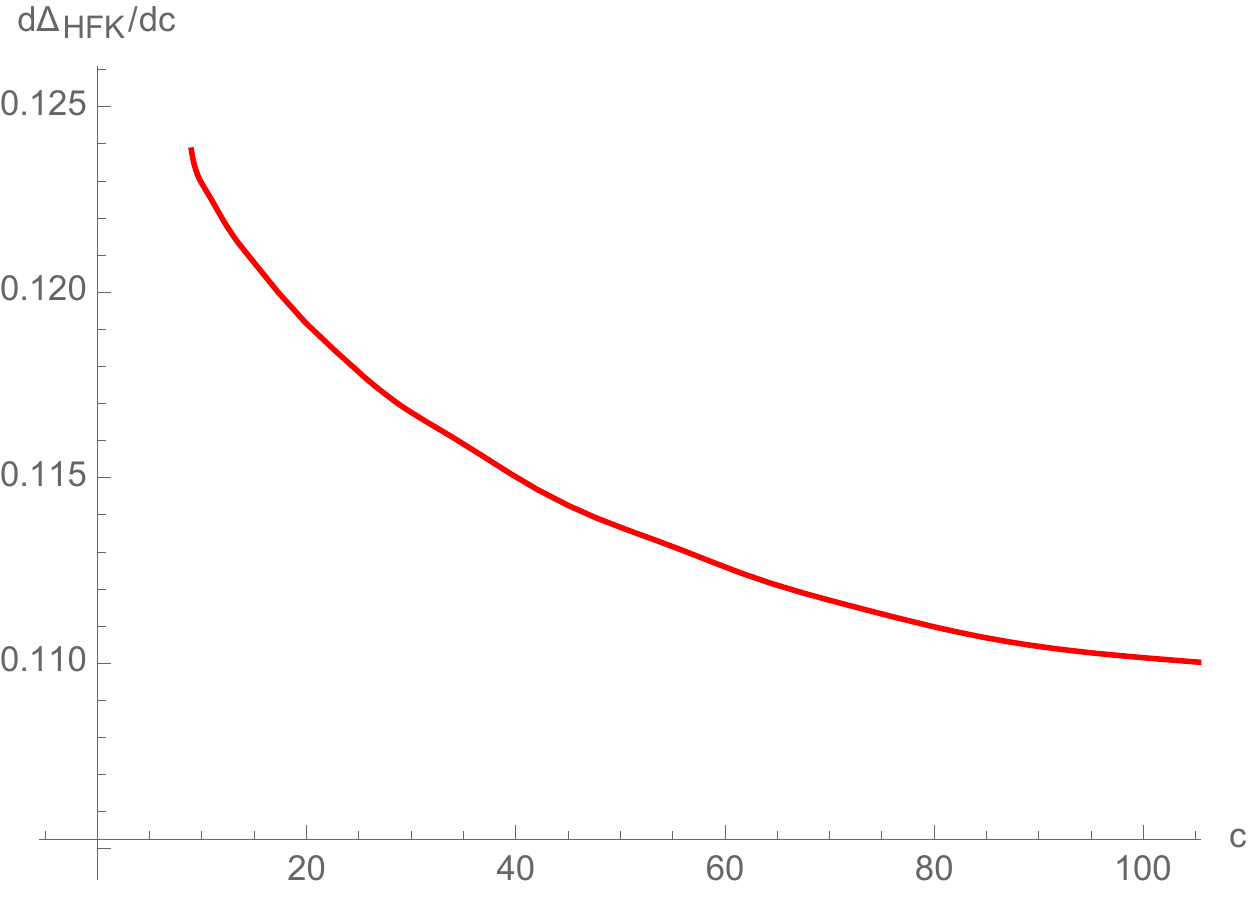}}
\caption{{\bf Left:} The green to red curves are plots of $\Delta_{\rm HFK}^{(N)}(c)$ with increasing $N$ ranging from 7 to 183, while the black curve represents the numerically extrapolated bound $\Delta_{\rm HFK}(c)=\Delta_{\rm HFK}^{(\infty)}(c)$. {\bf Right:} The slope of $\Delta_{\rm HFK}(c)$ as a function of $c$, obtained from taking the derivative of the Bezier fitting function of the extrapolated $\Delta_{\rm HFK}^{(\infty)}(c)$, over a range of $c$ where the numerical extrapolation appears reliable. The slope drops below ${1\over 9}$ for $c>75.5$. }\label{fig:SlopeDeltaGap}
\end{figure}

We employ the strategy of \cite{Friedan:2013cba} but consider the extrapolation $\lim_{N\to \infty}\Delta_{\rm HFK}^{(N)}(c)$ at each value of $c$, thereby exhausting the constraints from the HFK functionals (\ref{LinearFunctionalBeta}). Examples of data points used to obtain such extrapolations are shown in Figure~\ref{fig:HFK}. 
We observe numerically that as we increase $N$, the value of $N$ at which $\Delta_{\rm HFK}^{(N)}(c)$ stabilizes grows with the central charge $c$. Thus, it becomes increasingly difficult to extrapolate to $\Delta_{\rm HFK}^{(\infty)}(c)$ as we go to larger values of $c$, requiring that we work to larger derivative orders $N$ to obtain an accurate extrapolation. 
Figure~\ref{fig:SlopeDeltaGap} shows the result of an extrapolation of $\Delta_{\rm HFK}^{(\infty)}(c)/c$, as well as the slope $d\Delta_{\rm HFK}^{(\infty)}(c)/dc$ of the bound as a function of the central charge, over a range of the central charge where the numerical extrapolation appears reliable (using results up to $N=183$).

We observe that the slope of $\Delta_{\rm HFK}(c)$ {\it decreases monotonically}, and we conjecture that this property holds for all values of $c$. Note that this slope falls well below ${1\over 6}$, the large $c$ asymptotic slope of $\Delta_{\rm HFK}^{(N)}(c)$ for {\it fixed} $N$ (as was found in \cite{Friedan:2013cba}). Since a reliable extrapolation requires going to derivative order $N$ that grows with $c$, so far we have not been able to perform a reliable extrapolation for $c\gtrsim \mathcal{O}(10^2)$. Based on the numerical results, together with the known constraint $\Delta_{\rm HFK}(c)\geq t_{gap}(c)={c-1\over 12}$, we conjecture that in the large $c$ limit, 
\ie
\lim_{c\to \infty} {\Delta_{\rm HFK}(c)\over c} = b_{\rm HFK} , ~~~{\rm with}~~ {1\over 12}\leq b_{\rm HFK}< {1\over 9}.
\fe

The need for computing the bound using HFK functionals to very large derivative order $N$ suggests that $\{(\beta\partial_\beta)^n|_{\beta=1}\}_{{\rm odd} ~n}$ is in fact a poor choice of basis for the optimal linear functional at large $c$. We will examine such optimal functionals in detail in section \ref{optimal}.

\subsection{Bounds from full modular invariance}

We now consider the stronger constraints obtained by imposing modular invariance on $\hat Z(\tau,\bar\tau)$ on the entire upper half $\tau$-plane (as opposed to the imaginary $\tau$-axis), by applying to the modular crossing equation  linear functionals of the form
\ie
\A = \sum_{{\rm odd~}m+n\leq N} \A_{m, n} \partial_z^m \partial_{\bar z}^n|_{z = \bar z = 0}
\fe
that satisfy
\begin{align}
	\alpha\bigg[\hat{\chi}_0(\tau)\hat{\bar\chi}_0(\bar\tau)\bigg]>&0\nonumber\\
	\alpha\bigg[\hat{\chi}_{\Delta-s\over 2}(\tau)\hat{\bar\chi}_{\Delta+s\over 2}(\bar\tau) + \hat{\chi}_{\Delta+s\over 2}(\tau)\hat{\bar\chi}_{\Delta-s\over 2}(\bar\tau)\bigg]\ge& 0,\quad\Delta\geq\text{max}\big(\Delta^*_{\rm mod},s\big),
\end{align}
for some $\Delta_{\rm mod}^*$. If such a linear functional $\A$ is found, we learn that the gap in the dimension spectrum of all primaries is bounded from above by $\Delta_{\rm mod}^*$. The bound that results from the smallest such $\Delta_{\rm mod}^*$ will be denoted $\Delta_{\rm mod}^{(N)}(c)$. The optimal bound would be obtained in the infinite $N$ limit, namely,
\ie
\Delta_{\rm mod}(c) = \lim_{N\to \infty} \Delta_{\rm mod}^{(N)}(c).
\fe
Note that, just as with the HFK bound, it is important that we take the $N\to \infty$ limit at fixed $c$.

\begin{figure}[h!]
\centering
\includegraphics[width=.75\textwidth]{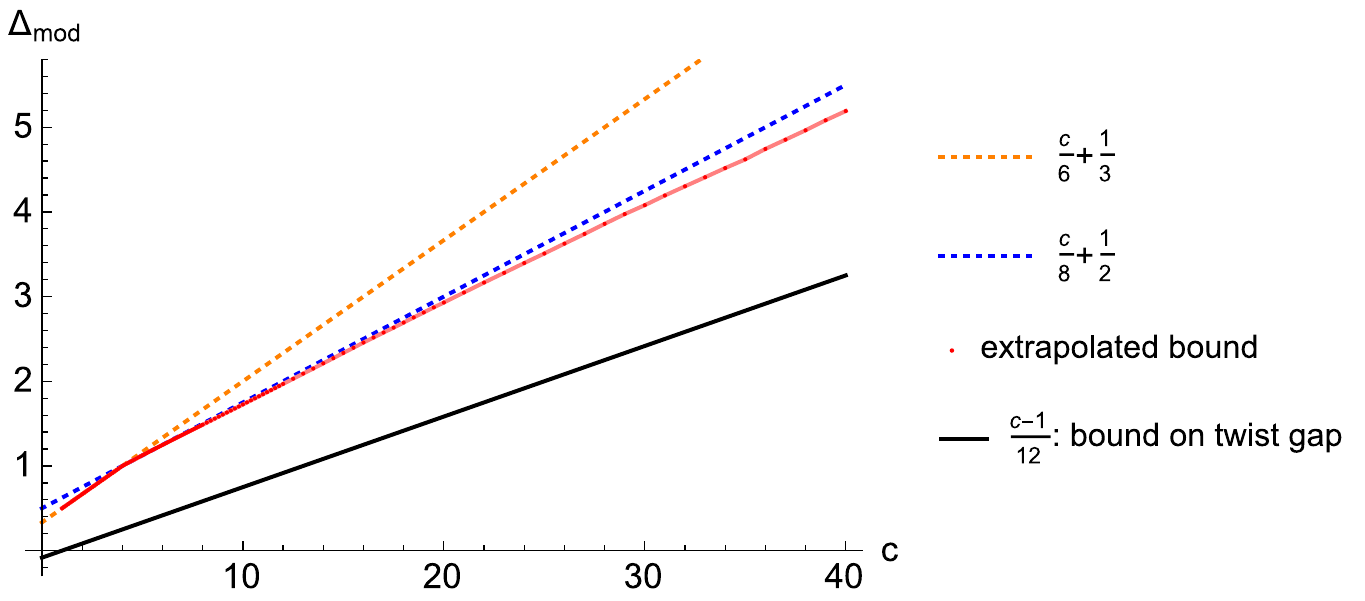}
\caption{The bound on the dimension gap $\Delta_{\rm mod}$ as a function of central charge $c$, obtained by extrapolating $\Delta_{\rm mod}^{(N)}$ to $N=\infty$. The numerical extrapolation is performed by fitting $31\leq N\leq 55$ bounds with a quadratic polynomial in $1/N$.}\label{fig:DeltaGapExtrapolated-TAU}
\end{figure}

Figure~\ref{fig:DeltaGapExtrapolated-TAU} shows the bound $\Delta_{\rm mod}(c)$ obtained by a numerical extrapolation of $\Delta_{\rm mod}^{(N)}$ to infinite derivative order $N$. 
We find with high numerical precision that 
\begin{equation}
	\Delta_{\rm mod}(c)={c\over 6}+{1\over 3}, ~~~~{\rm for}~~c\in [1,4].
\end{equation}
A kink appears at $c=4$ and $\Delta_{\rm mod}=1$, where the slope of $\Delta_{\rm mod}(c)$ jumps from ${1\over 6}$ to the left of the kink, to ${1\over 8}$ to the right of the kink. As the central charge is increased, the slope of $\Delta_{\rm mod}(c)$ appears to decrease monotonically, just as $\Delta_{\rm HFK}(c)$ seen in the previous subsection. For larger values of $c$, a numerical extrapolation to infinite derivative order $N$ is again needed. Based on the numerical results, we conclude that
\begin{equation}
	\Delta_{\rm mod}(c) < {c\over 8}+{1\over 2},~~~~c>4.
\end{equation}
Furthermore, we conjecture that the slope $\Delta_{\rm mod}(c)$ decreases monotonically, and asymptotic to a value
\ie
\lim_{c\to \infty} {\Delta_{\rm mod}(c)\over c} = b_{\rm mod},~~~{\rm with}~~{1\over 12}\leq b_{\rm mod}\leq b_{\rm HFK}< {1\over 9}.
\fe

\begin{figure}[h!]
\centering
\includegraphics[width=0.5\textwidth]{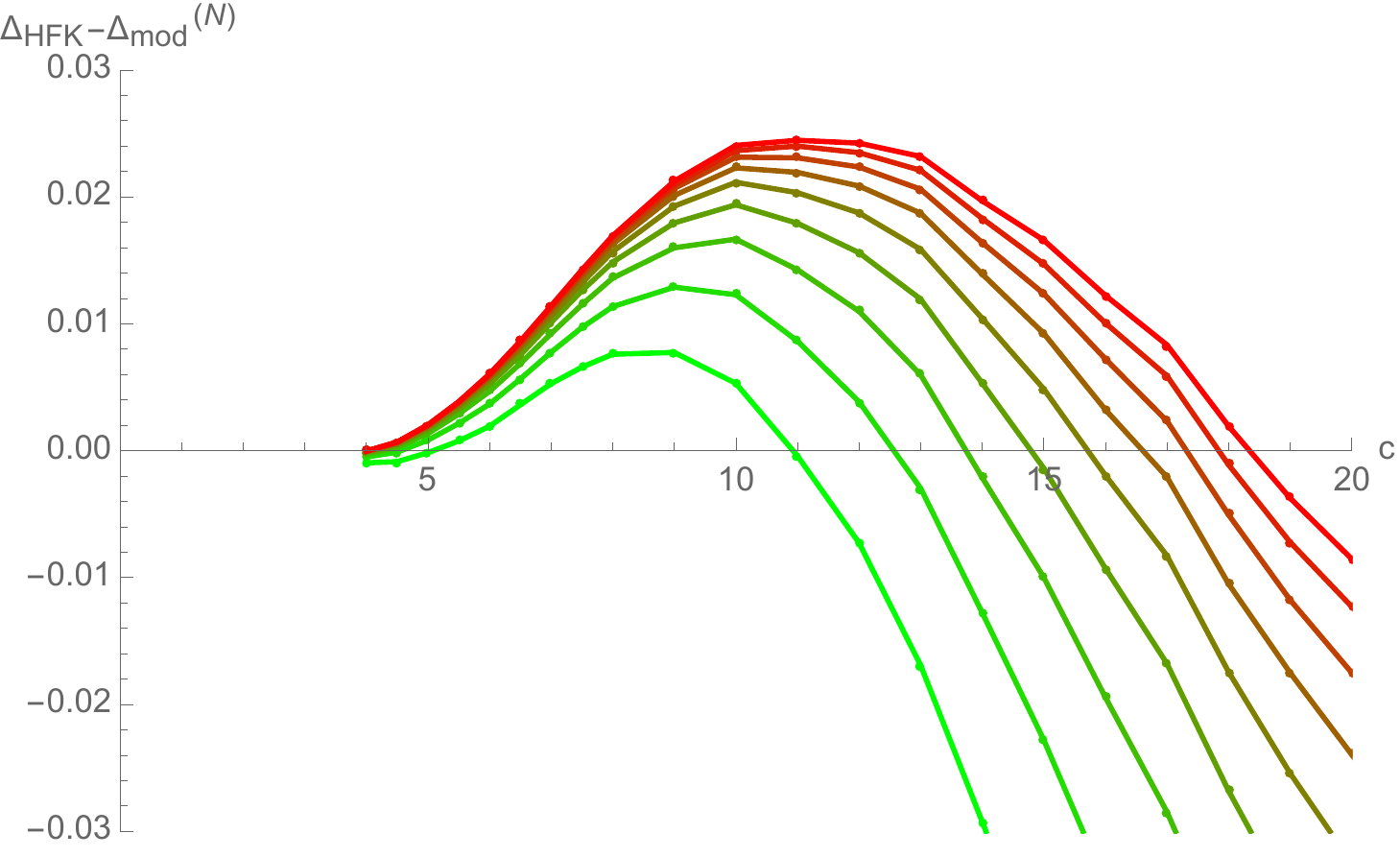}
\caption{$\Delta_{\rm HFK}-\Delta_{\rm mod}^{(N)}$ as a function of central charge $c$, for $c\in [4,22]$ and increasing $N$. The numerical values of the HFK bound are obtained by a linear extrapolation of $\Delta_{\rm HFK}^{(N)}$ to $1/N \to 0$ using bounds for $137\leq N\leq 175$.}\label{fig:differences}
\end{figure}

Interestingly, $\Delta_{\rm mod}(c)$ coincides with $\Delta_{\rm HFK}(c)$ at $c=4$ (where both are equal to 1), but the bounds do not agree for $c$ above or below 4.  The numerical evaluation of $\Delta_{\rm mod}^{(N)}(c)$ is more time consuming than that of $\Delta_{\rm HFK}^{(N)}(c)$, and we are unable to perform a reliable extrapolation of the large $c$ asymptotics of $\Delta_{\rm mod}(c)$ directly. We can nonetheless analyze the difference between the two bounds, $\Delta_{\rm HFK}(c)-\Delta^{(N)}_{\rm mod}(c)$, for moderate values of $c$, as shown in Figure~\ref{fig:differences}. We observe that at a fixed derivative order $N$ of the linear functional, the difference between $\Delta_{\rm HFK}(c)$ and $\Delta_{\rm mod}^{(N)}(c)$ will initially grow with the central charge until it eventually begins to decrease and becomes negative. This is related to the observed phenomenon that as the central charge is increased, one must use linear functionals of larger and larger derivative order to obtain stabilized bounds. While it is possible that the asymptotic slope $b_{\rm mod}$ is smaller than $b_{\rm HFK}$, we have not been able to resolve their difference numerically.

\subsection{The optimal linear functional}
\label{optimal}

It is somewhat unexpected that such delicate numerical analysis is required to extract bounds from a very simple form of the modular crossing equation for the reduced partition function, and we still do not know the value of the asymptotic slopes, $b_{\rm HFK}$ and $b_{\rm mod}$. As an analytic derivation of the optimal bounds is not yet available, we may look for hints in the optimal linear functional $\A$. At a given derivative order $N$, by minimizing the gap bound $\Delta_{\rm HFK}^*$ or $\Delta_{\rm mod}^*$, we can determine the optimal linear functional, which we denote by $\A_{\rm HFK}^{(N)}$ or $\A_{\rm mod}^{(N)}$. Numerically, it appears that there are indeed well defined $N\to \infty$ limits on the optimal linear functionals,
\ie
& \A_{\rm HFK} = \lim_{N\to \infty} \A_{\rm HFK}^{(N)} = F_{\rm HFK}(\beta\partial_\beta)|_{\beta=1},
\\
& \A_{\rm mod} = \lim_{N\to \infty} \A_{\rm mod}^{(N)} = F_{\rm mod}(\partial_z, \partial_{\bar z})|_{z=\bar z=0}.
\fe
Here $F_{\rm HFK}$ is a power series in $\beta\partial_\beta$ (to be evaluated at $\beta=1$), and $F_{\rm mod}$ is a power series in $\partial_z, \partial_{\bar z}$ (to be evaluated at $z=\bar z=0$). Both $F_{\rm HFK}$ and $F_{\rm mod}$ can be computed numerically.

\begin{figure}[h!]
\subfloat{\includegraphics[width=.33\textwidth]{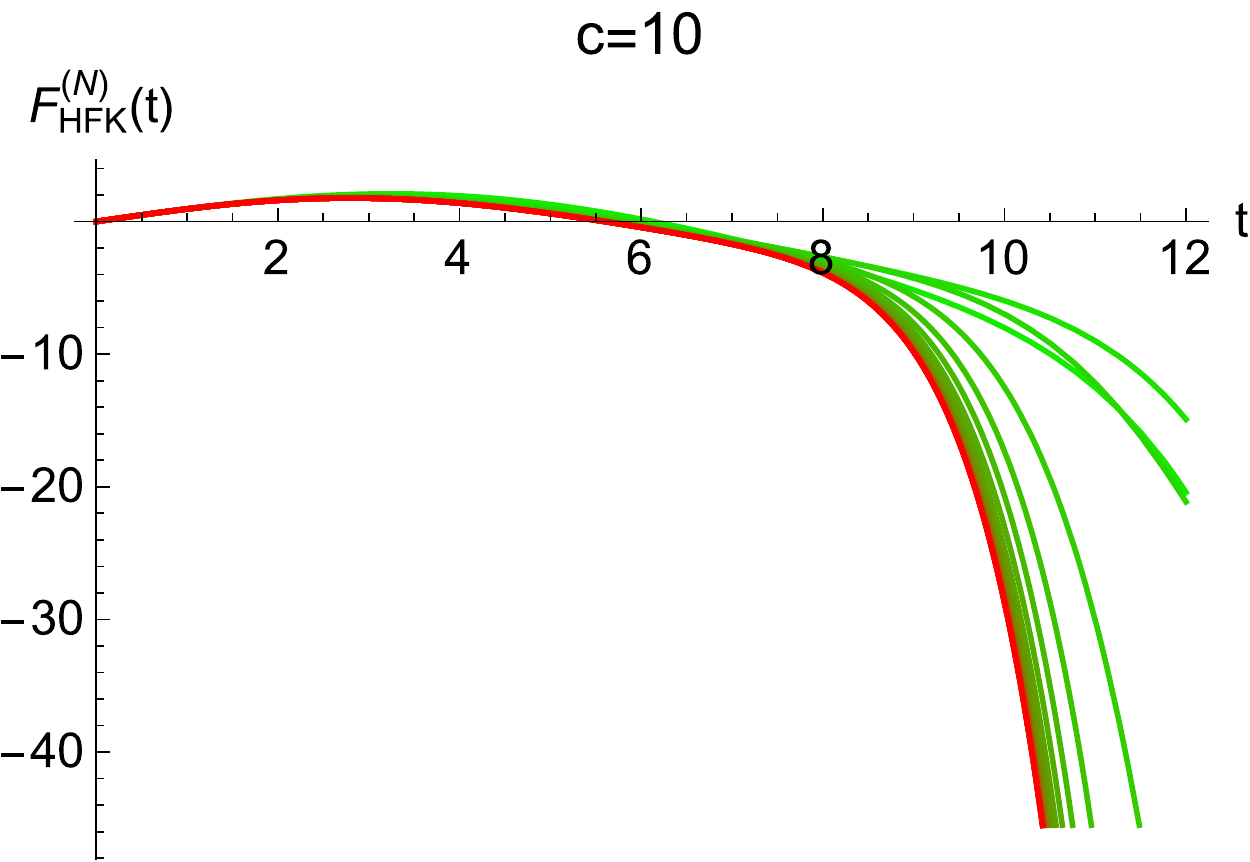}}~~
\subfloat{\includegraphics[width=.33\textwidth]{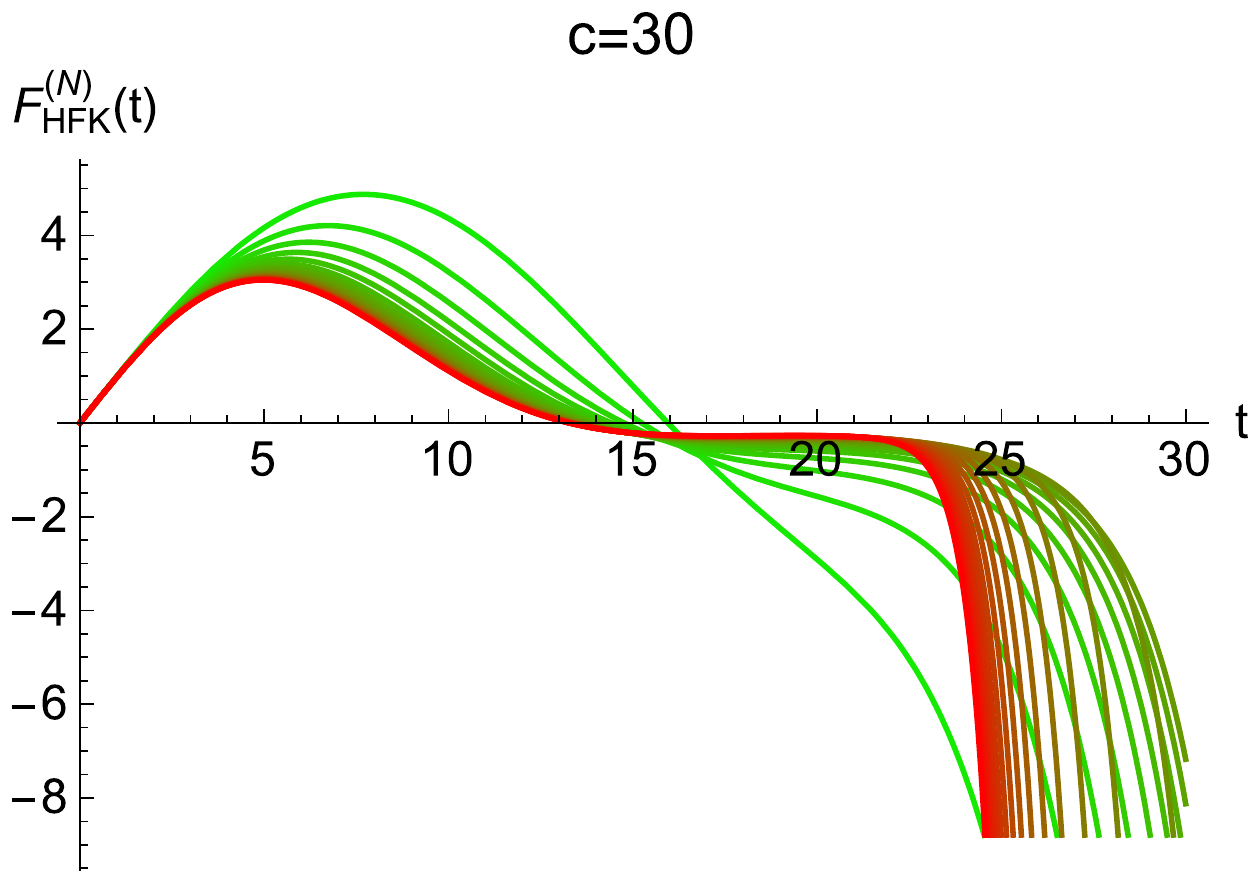}}~~
\subfloat{\includegraphics[width=.33\textwidth]{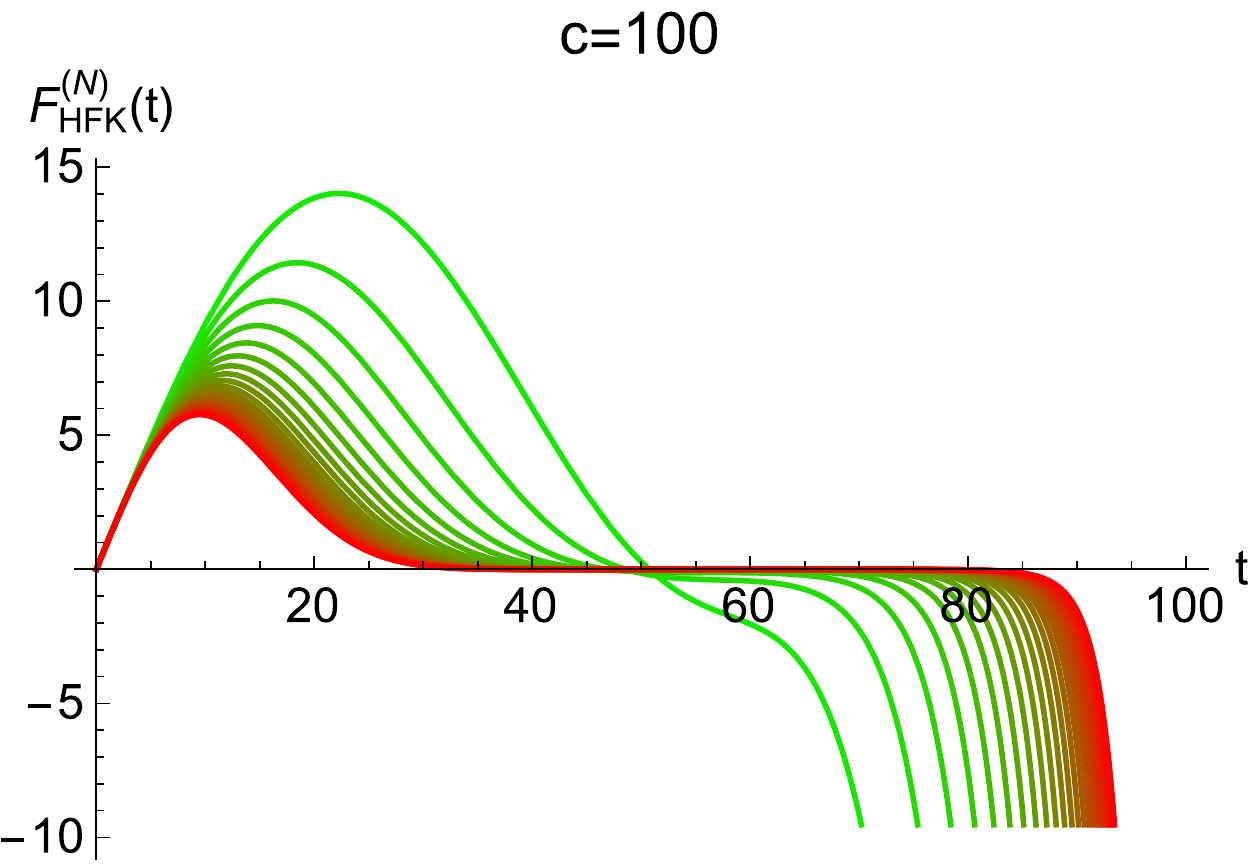}}
\caption{Plots of $F^{(N)}_{\rm HFK}(t)$ with increasing derivative orders $N$ (from green to red), at central charges $c=10, 30, 100$. }\label{fig:functional-HFK}
\end{figure}

Figure~\ref{fig:functional-HFK} shows a few examples of the polynomials $F_{\rm HFK}^{(N)}(t)$ that represent the optimal HFK functional up to derivative order $N$, that converge in the infinite $N$ limit. For small values of $c$, $F_{\rm HFK}(t)$ can be rather accurately fitted by a linear combination of $\sin(a t)$ and $\sinh(b t)$, for some constants $a, b$, suggesting that the optimal linear functional is well approximated by a linear combination of the modular crossing equation evaluated at two values of $\B$, one real and the other lying on the unit circle (analytically continued in $\B$). At large $c$, however, the behavior of $F_{\rm HFK}(t)$ changes: it can be approximately fitted by $t e^{- a t^2}$ over a large range of $t$, suggesting that the optimal linear functional is more appropriately represented by an integral transform in $\B$ rather than taking derivatives at $\B=1$.\footnote{It is tempting to suggest that the optimal functional in the large $c$ limit has the form $\partial_z e^{-a \partial_z^2}|_{z=0}$, for some $c$-dependent constant $a$, but the latter by itself does not take the same sign when acting on arbitrarily high dimension characters as on the vacuum character, and thus cannot be used to derive an analytic bound on the gap.}

\begin{figure}[h!]
\centering
\subfloat{\includegraphics[width=.4\textwidth]{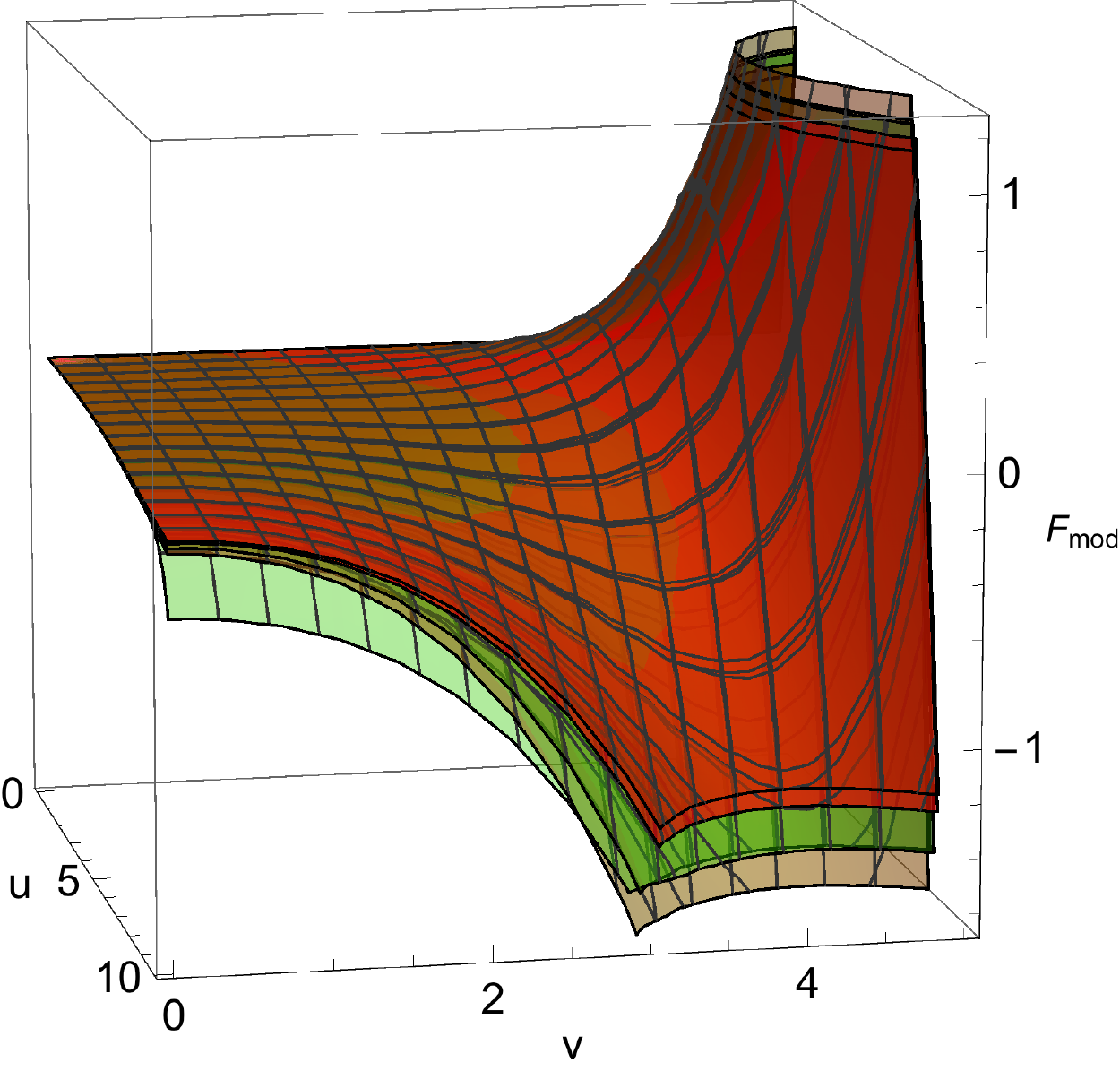}}
\caption{A plot of $F^{(N)}_{\rm mod}(u+iv,u-iv)$ with increasing derivative orders $N$ (as indicated by colors from green to red), at central charge $c=5$. }\label{fig:functional-mod}
\end{figure}

Similarly, we can obtain $F_{\rm mod}^{(N)}(w,\bar w)$ that represents the optimal functional obtained by imposing full modular invariance. It exhibits a nontrivial dependence on both the real and imaginary parts of $w$, refining the HFK functional. An example is shown in Figure~\ref{fig:functional-mod}.

\section{Spin-Dependent Bounds}

Using the complete basis of linear functionals (\ref{LinearFunctionalZ}), it is now possible to obtain more refined bounds on the spectrum that distinguish primaries of different spins. In this section, we present a number of results on the spin-dependent bounds.

\subsection{Gap in the spectrum of scalar primaries}

\begin{figure}[h!]
\centering
\subfloat{
\includegraphics[width=0.5\textwidth]{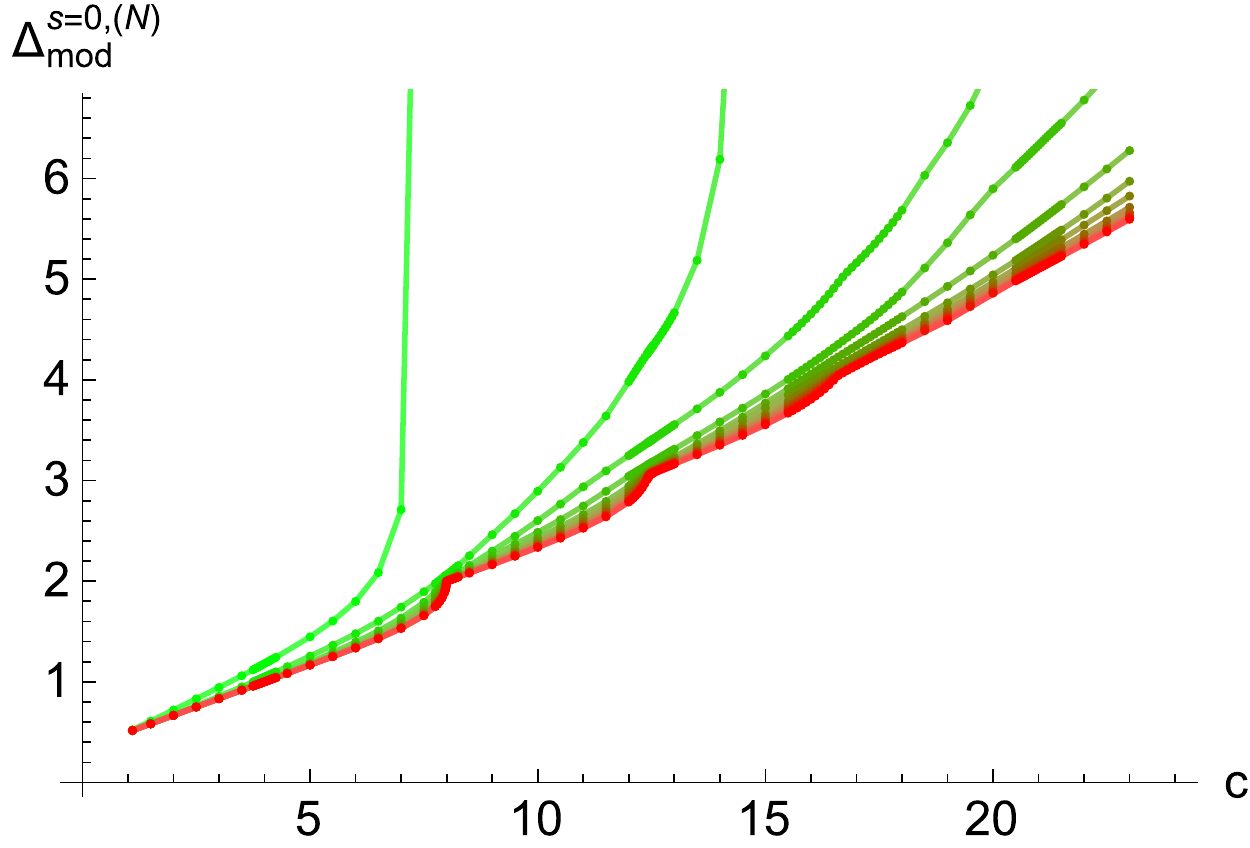}
}
\subfloat{
\includegraphics[width=0.5\textwidth]{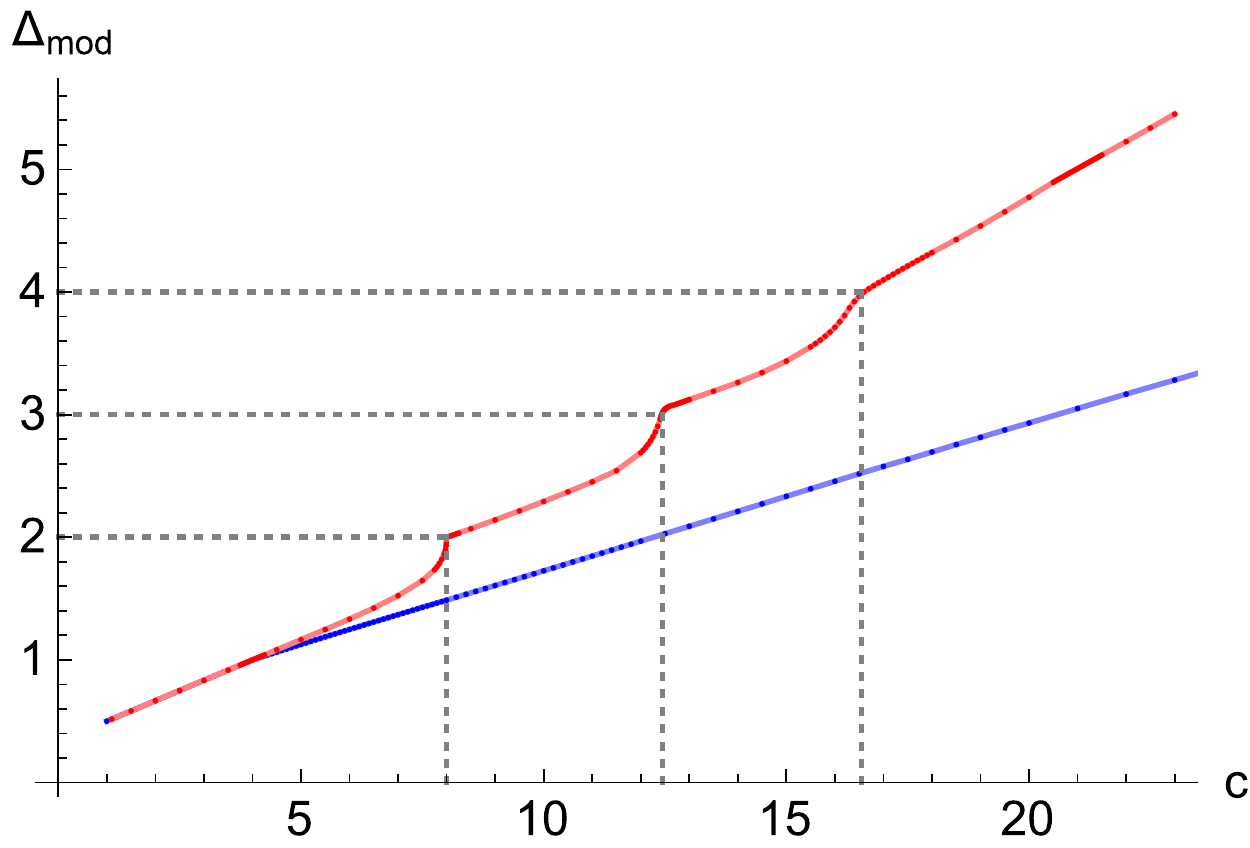}
}
\\
\subfloat{
\includegraphics[width=0.5\textwidth]{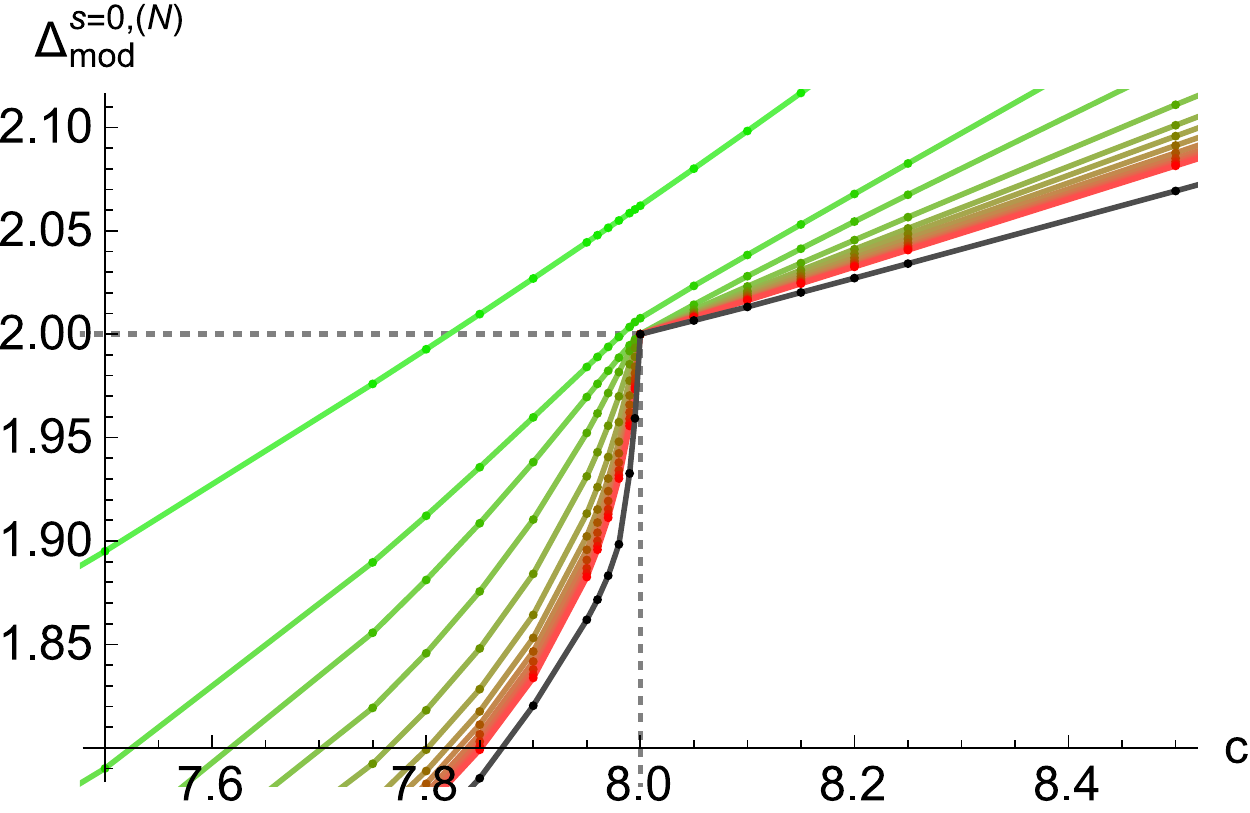}
}
\subfloat{
\includegraphics[width=0.5\textwidth]{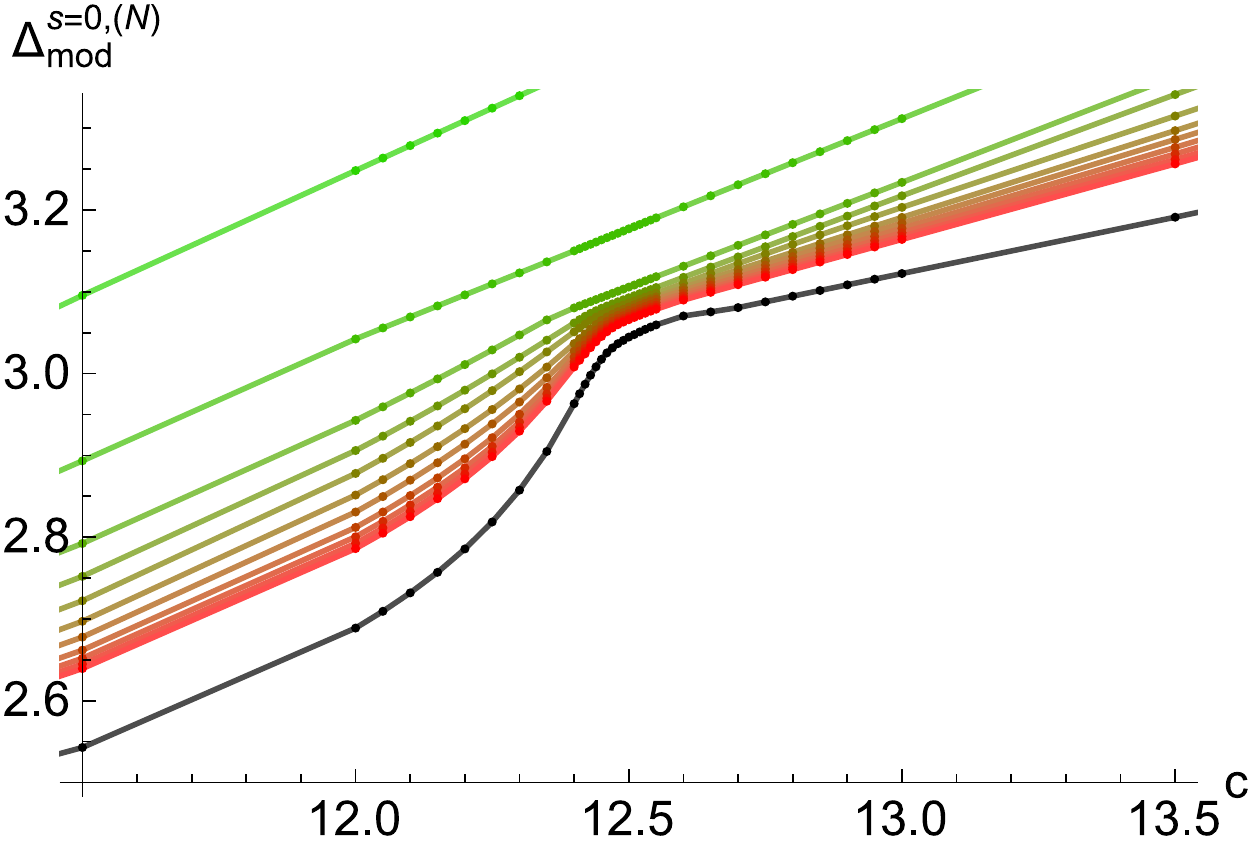}
}
\caption{{\bf Top-left:} The upper bound on the gap in the dimension of scalar primaries obtained at increasing derivative order of the linear functional (from green to red, up to $N=51$) as a function of the central charge. {\bf Top-right:} The extrapolated bounds on the dimension of the lightest scalar operator $\Delta_{\rm mod}^{s=0}$ (red) as a function of the central charge, superposed with the bound on the overall gap $\Delta_{\rm mod}$ (blue). For the bound on the scalar gap, the numerical extrapolation is performed by fitting $19\leq N\leq 51$ bounds with a quadratic polynomial in $1/N$. {\bf Bottom:} The extrapolated bound on the scalar gap (black) superposed with the bounds at fixed derivative orders (increasing from green to red) near the first two kinks.}\label{fig:ScalarDeltaGapExtrapolated}
\end{figure}

By searching for linear functionals that satisfy the following positivity conditions
\begin{align}
	\alpha\bigg[\hat{\chi}_0(\tau)\hat{\bar\chi}_0(\bar\tau)\bigg]>&0\nonumber\\
	\alpha\bigg[\hat{\chi}_{\Delta-s\over 2}(\tau)\hat{\bar\chi}_{\Delta+s\over 2}(\bar\tau) + \hat{\chi}_{\Delta+s\over 2}(\tau)\hat{\bar\chi}_{\Delta-s\over 2}(\bar\tau)\bigg]\ge& 0,\quad{\begin{cases}\Delta\ge\Delta_{s=0}^{*},~&s=0\\ \Delta\ge s,~&s>0\end{cases}}
\end{align}
we can place upper bounds on the dimension of the lightest scalar primary operator, with no assumption on the spectra of other spins. Namely, if such a linear functional is found, we would learn that the gap in the dimensions of scalar primaries must be bounded from above by $\Delta_{s=0}^{*}$. The smallest such $\Delta_{s=0}^{*}$, obtained using functionals up to derivative order $N$, will be denoted $\Delta_{\rm mod}^{s=0,(N)}$. We can numerically extrapolate to infinite $N$, which results in the optimal bound $\Delta_{\rm mod}^{s=0}$ on the scalar gap. Our results are shown in Figure~\ref{fig:ScalarDeltaGapExtrapolated}.

By definition, $\Delta_{\rm mod}^{s=0}(c)\geq \Delta_{\rm mod}(c)$, and the two agree when $c\leq 4$ (where the bound is less than 1 and only scalars could lie below the bound). It appears that the discontinuity in the slope at $c=4$ is absent in $\Delta_{\rm mod}^{s=0}(c)$. However, there appear to be new kinks\footnote{We would like to emphasize, however, that the extrapolation to infinite $N$ is not accurate enough to determine whether the apparent kinks at $\Delta_{\rm mod}^{s=0}=3,4,\ldots$ represent genuine discontinuities in the derivative of the $\Delta_{\rm mod}^{s=0}(c)$ curve.} in the $\Delta_{\rm mod}^{s=0}(c)$ curve when the bound attains integer values $2,3,4,\cdots$. In particular, $\Delta_{\rm mod}^{s=0}(c)<2$ for $c<8$, which implies that unitary CFTs with no conserved currents and $c<8$ must admit relevant deformations. The kink at $c=8$ and $\Delta_{\rm mod}^{s=0}=2$ is in fact realized by a Narain lattice CFT of 8 free compact bosons (even though this CFT contains conserved currents, its partition function is of generic type), as will be discussed, among other examples, in the next subsection.

\begin{figure}[h!]
\centering
\subfloat{
\includegraphics[width=0.53\textwidth]{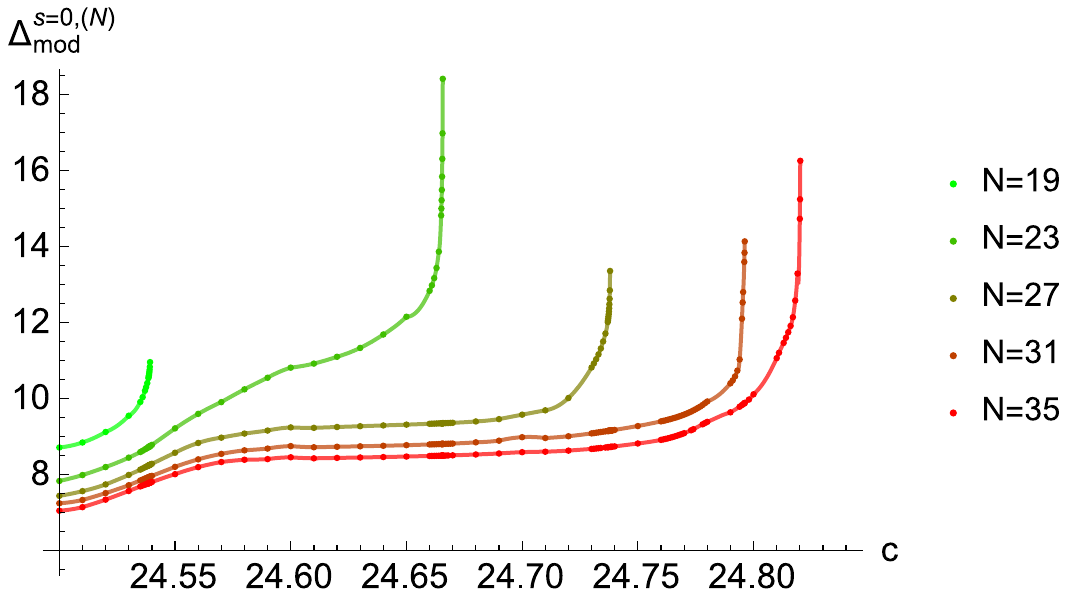}
}
\subfloat{
\includegraphics[width=0.46\textwidth]{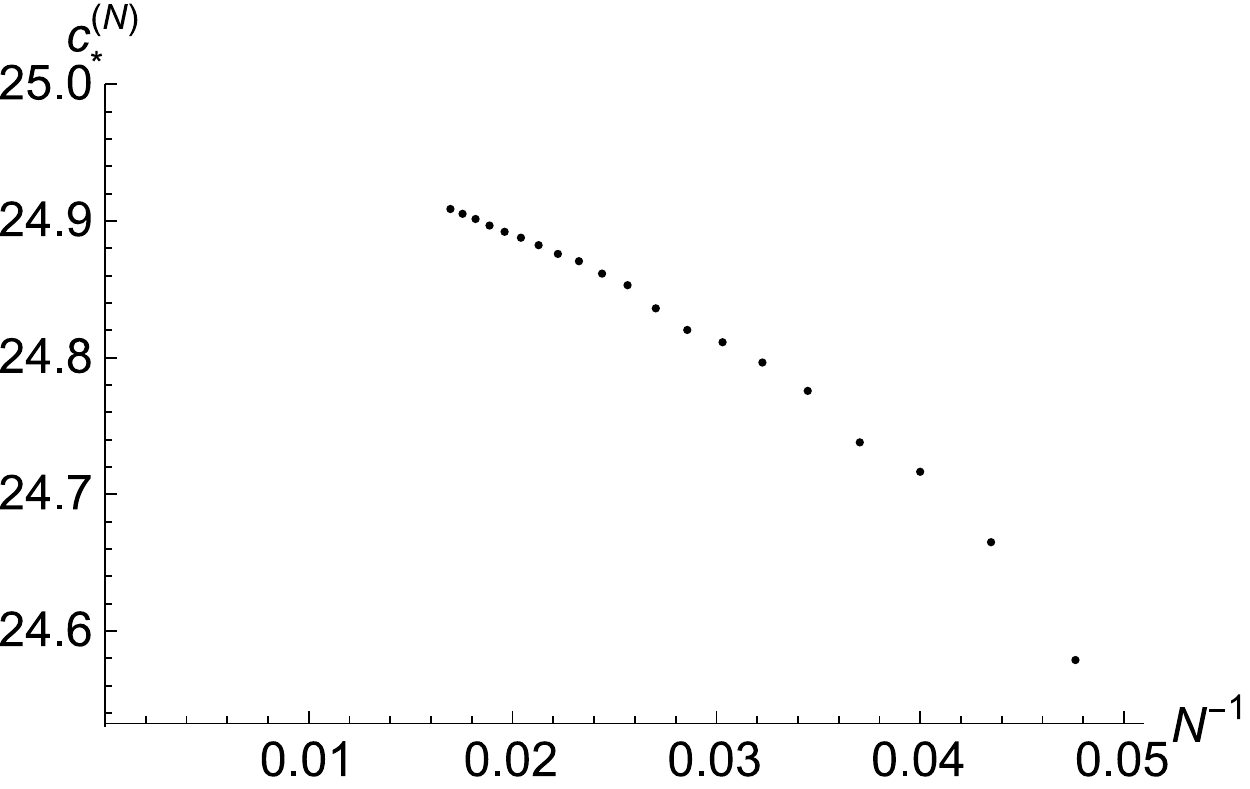}
}
\caption{{\bf Left:} The upper bound on the dimension of the lightest scalar primary operator as a function of central charge at fixed derivative order of the linear functional. {\bf Right:} The central charge at which a bound on the scalar dimension gap can no longer be found as a function of inverse derivative order of the linear functional. }\label{fig:ScalarResolve}
\end{figure}

We find no bound on the dimension of the lightest scalar operator for $c\ge 25$. At a given derivative order $N$, we denote by $c_*^{(N)}$ the central charge at which the scalar gap bound $\Delta_{\rm mod}^{s=0,(N)}$ ceases to exist. As shown in Figure~\ref{fig:ScalarResolve}, $c_*^{(N)}$ approaches 25 from below as $N\rightarrow \infty$. 

The disappearance of an upper bound on the scalar gap for $c\geq 25$ has a very simple explanation. Our positivity criteria on the degeneracy of the primaries do not exclude the limit where the degeneracies of the primaries diverge, and the partition function becomes a divergent factor multiplied by the partition function of a noncompact CFT spectrum, namely that of a continuous spectrum with a finite density of states, with no $SL(2,\mathbb{R})\times SL(2,\mathbb{R})$ invariant vacuum. Indeed, consider the modular invariant function 
\begin{equation}\label{PartitionFunction-c=25}
Z(\tau,\bar\tau) = {J(\tau)+\bar J(\bar \tau)\over \tau_2^{1\over 2}|\eta(\tau)|^2} ,
\end{equation}
where $J(\tau)$ is related to the elliptic $j$-invariant $j(\tau)$ by $J(\tau) = j(\tau) - 744 =  q^{-1} + 196884 q + \mathcal{O}(q^2)$.
When interpreted as the partition function of a noncompact CFT of central charge $c$, it admits a decomposition in terms of Virasoro characters associated with primaries of {\it nonzero} spin and twist $\geq {c-25\over 12}$. Thus, the absence of scalar primaries is consistent with modular invariance for $c\geq 25$.

\subsection{Kinks and CFTs with generic type partition function}\label{Sec:Kinks}

An obvious question is whether there are CFTs whose spectra saturate the dimension gap bounds. This turns out to be the case for a few values of the central charge, where the bound is saturated by a rational CFT whose partition function is of the generic type. These examples provide good consistency checks of our numerical methods.

\noindent $\bullet$ At $c=1$, the bound $\Delta_{\rm mod}={1\over 2}$ is saturated by the gap of the $SU(2)$ WZW model at level $1$, otherwise known as the free compact boson at the self-dual radius.

\noindent $\bullet$ At $c=2$, the bound $\Delta_{\rm mod}={2\over 3}$ is saturated by the $SU(3)$ WZW model at level $1$. This theory admits a description in terms of free bosons with target space $T^2$ at the $\mathbb{Z}_3$-invariant point in both its K\"ahler and complex structure moduli spaces.

\noindent $\bullet$ At $c={14\over 5}$, the bound $\Delta_{\rm mod} = {4\over 5}$ is saturated by the $G_2$ WZW model at level $1$.

\noindent $\bullet$ At $c=4$, the bound $\Delta_{\rm mod}=1$ is saturated by the $SO(8)$ WZW model at level $1$. This theory also admits a description in terms of 8 free fermions with diagonal GSO projection. Note that this is the kink on the curve $\Delta_{\rm mod}(c)$.

\noindent $\bullet$ At $c=8$, the bound $\Delta_{\rm mod}^{s=0}=2$ is saturated by the $E_8$ WZW model at level $1$. This theory also admits a description in terms of 8 compact bosons at the holomorphically factorized point in its moduli space, where the holomorphic factor can be described as the Narain compactification on the root lattice of $E_8$. This is the first kink on the curve for the upper bound on the scalar dimension gap $\Delta_{\rm mod}^{s=0}(c)$.

In section 4, we will show that the spectra of these theories are in fact completely determined by the saturation of the gap bound.

\subsection{Allowing for primary conserved currents}

So far, we have focused on spectra with no conserved current primaries, together with the limiting case of generic type spectra, where conserved currents were allowed as long as their contributions to the partition function were combined with twist-2 primaries to give non-degenerate Virasoro characters. It is straightforward to relax this assumption by including degenerate characters of the form $\chi_j(\tau) \bar\chi_0(\bar\tau)$ and $\chi_0(\tau) \bar\chi_j(\bar\tau)$ in the partition function, as in (\ref{genchar}), and try to rule out hypothetical spectra by seeking linear functionals that act non-negatively on the degenerate Virasoro characters as well as the non-degenerate characters present in the spectrum. 

Before doing so, let us note that there are many constraints on the spectrum due to the associativity of the OPE that are not taken into account by modular invariance of the partition function alone. This is apparent in the presence of conserved current primaries\footnote{When the currents are not conserved, the constraints on their OPE are much more delicate.}: operators must form representations of an extended chiral algebra, and in particular, operators formed by taking product of left and right moving currents (e.g. of the form $J^a(z) \tilde J^b(\bar z)$) are part of the spectrum. 

In the discussion that follows, we will again assume that the spectrum is parity-invariant. If there are holomorphic and anti-holomorphic spin-1 currents in the CFT, then the spectrum consists of representations of a current algebra, whose characters always admit non-negative decompositions into {\it non-degenerate} Virasoro characters. In other words, if conserved spin-1 currents are present in a parity invariant CFT, then the partition function is necessarily of the generic type, except the case where there is a single $U(1)$ current algebra, which can easily be taken into account by replacing the vacuum Virasoro character by the $U(1)$ current algebra character.

We shall study the numerical bounds on the scalar dimension gap in the following three cases:

\noindent {\bf (I)} Conserved current primaries of all spins are allowed. 
Due to the basic OPE constraints discussed above (under the assumption of a parity-invariant spectrum), it suffices to assume either there are no conserved spin-1 currents (i.e. case (II) below), or there is a $U(1)$ current (in both the left and right sector) while higher-spin primaries come with non-degenerate characters.

\noindent {\bf (II)} Conserved current primaries of spins $j\geq 2$ are allowed.

\noindent {\bf (III)} Conserved current primaries of spins $j\geq 3$ are allowed, i.e., we consider CFTs with a unique stress-energy tensor and no spin-1 currents.

\noindent The resulting upper bound on the scalar gap will be denoted $\Delta_{{\rm mod},j\geq 1}^{s=0,(N)}(c)$, $\Delta_{{\rm mod},j\geq 2}^{s=0,(N)}(c)$, and $\Delta_{{\rm mod},j\geq 3}^{s=0,(N)}(c)$ respectively.

In case {\bf (III)}, we find essentially no difference between $\Delta_{{\rm mod},j\geq 3}^{s=0,(N)}(c)$ and $\Delta_{{\rm mod}}^{s=0,(N)}(c)$, up to the numerical error due to the finite resolution $\epsilon$ of the binary search implemented for determining the optimal bound,\footnote{Here the resolution of our binary search for the optimal bound is taken to be $\epsilon = {c-1\over 48000}$.} except for a tiny peak in their difference that is localized near $c\sim 12.5$ and narrows with increasing derivative truncation order $N$. 

\begin{figure}[h!]
\centering
\subfloat{\includegraphics[width=0.5\textwidth]{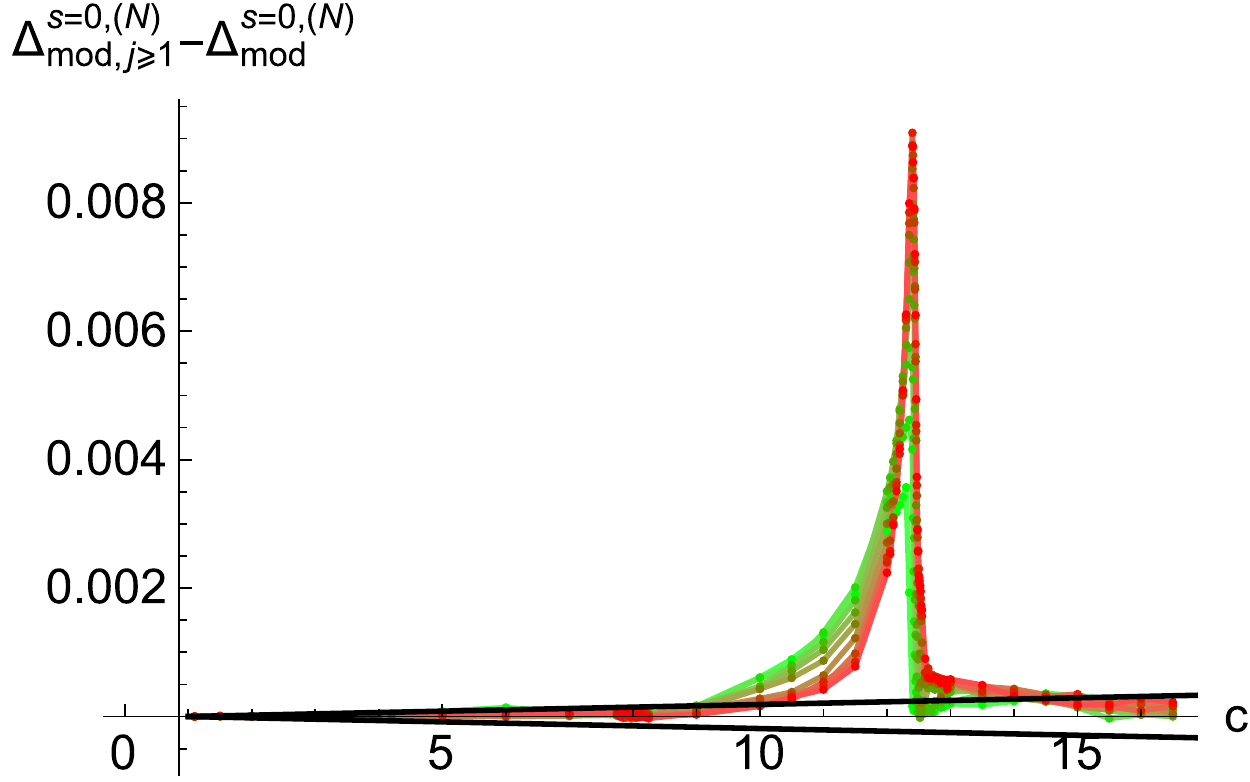}}
\subfloat{\includegraphics[width=0.51\textwidth]{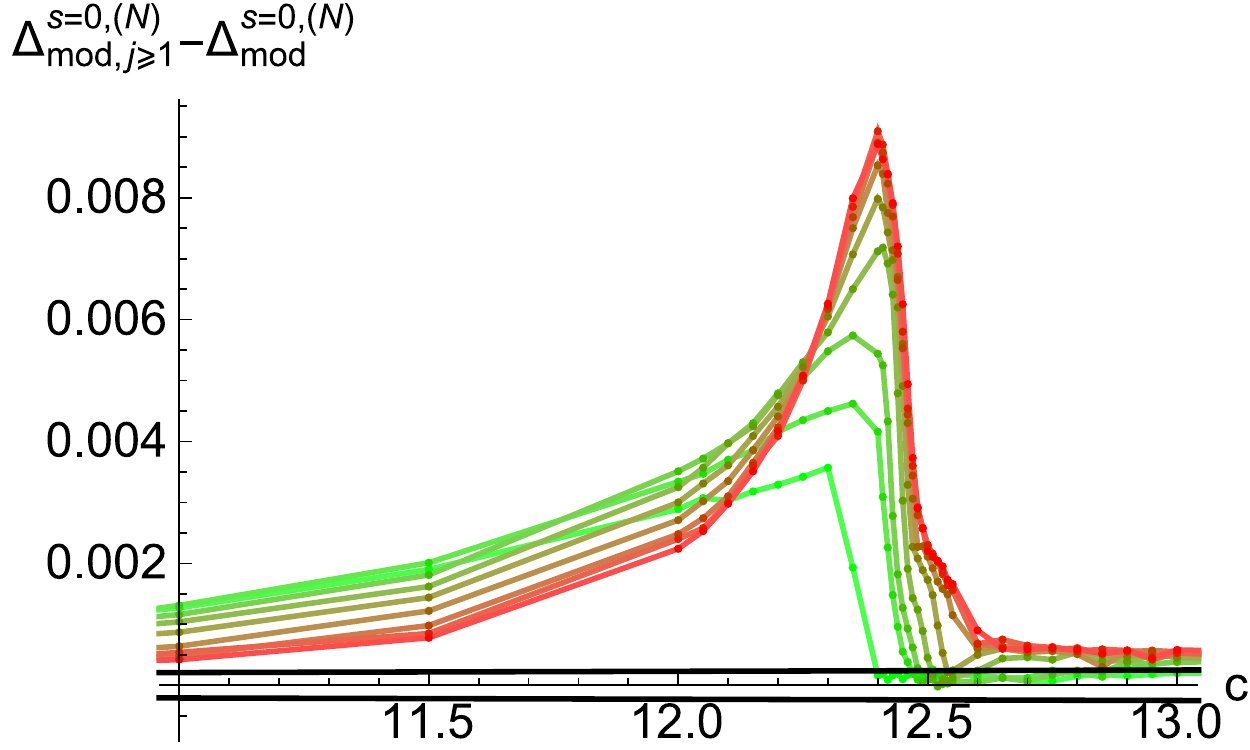}}
\\
\subfloat{\includegraphics[width=0.6\textwidth]{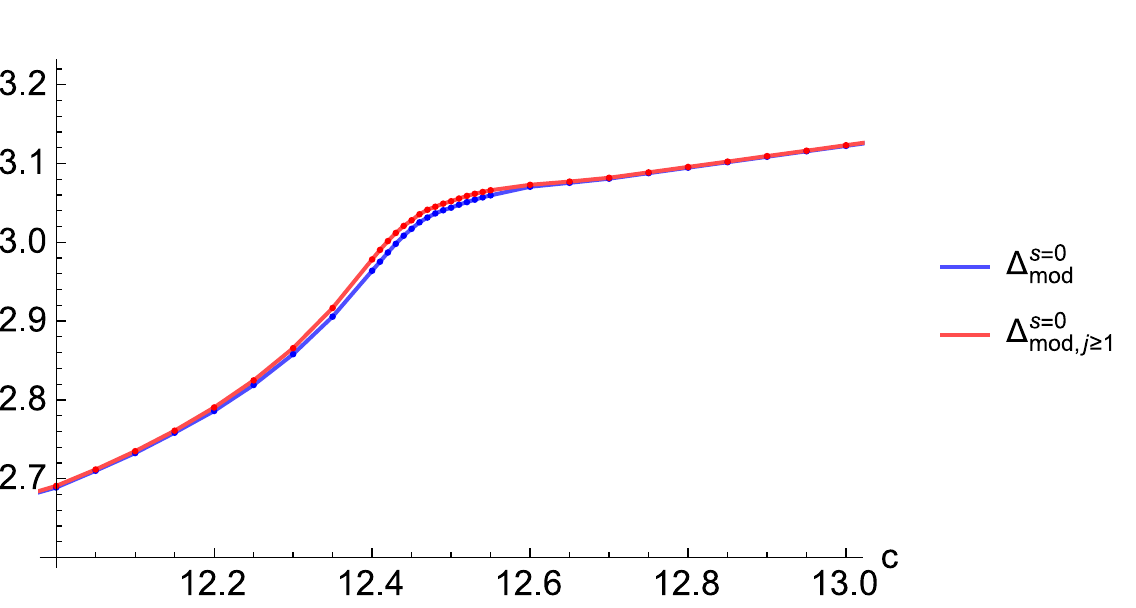}}
\caption{{\bf Top-left:} The difference between the bounds $\Delta_{\rm mod}^{s=0}(c)$ obtained with and without allowing extra conserved currents, computed using linear functionals up to a given derivative order $N$. The colors from green to red indicate bounds at increasing values of $N$, up to $N=51$. The black lines indicate the error threshold due to our binary search for the bound. {\bf Top-right:} The same plot zoomed into the range $11\leq c\leq 13$. {\bf Bottom:} The extrapolated bounds $\Delta_{\rm mod}^{s=0}(c)$ obtained with and without allowing extra conserved currents, plotted in the vicinity of the second kink. The numerical extrapolation is performed by fitting $19\leq N\leq 51$ bounds with a quadratic polynomial in $1/N$.}\label{fig:ScalarDeltaGapDiff}
\end{figure}

In cases ({\bf I}) and ({\bf II}), we find the same bounds $\Delta_{{\rm mod},j\geq 1}^{s=0,(N)}(c)$ and $\Delta_{{\rm mod},j\geq 2}^{s=0,(N)}(c)$ for the scalar gap, at sufficiently high derivative order $N$. For $N\geq 19$, we find no difference between $\Delta_{{\rm mod},j\geq 1}^{s=0,(N)}(c)$ and $\Delta_{\rm mod}^{s=0,(N)}(c)$ (the latter obtained assuming the absence of conserved current primaries) for $c\leq 8$, up to our numerical resolution. A small difference $\Delta_{{\rm mod},j\geq 1}^{s=0,(N)}(c)-\Delta_{\rm mod}^{s=0,(N)}(c)$ is found for $9\lesssim c\lesssim 15$, as shown in Figure~\ref{fig:ScalarDeltaGapDiff}, for $N$ up to 51. 
The numerics suggests that when conserved current primaries of all spins are allowed, the second kink near $c\sim 12.5$ for $\Delta_{{\rm mod}}^{s=0}(c)$ may be shifted slightly to the left in the curve for $\Delta_{{\rm mod},j\geq 1}^{s=0}(c)$.
Note that unlike the first kink in Figure~\ref{fig:ScalarDeltaGapExtrapolated} at $c=8$, we do not know of a candidate CFT that resides at this second kink. 

In conclusion, apart from a small shift in the position of the second kink near $c\sim 12.5$, we do not find any significant weakening of the scalar gap bound $\Delta_{\rm mod}^{s=0}(c)$ when conserved current primaries are allowed.

Unitary CFTs that admit only irrelevant deformations describe what are known as ``perfect metals" \cite{Plamadeala:2014roa}. It follows from our bound on the scalar dimension gap that perfect metals do not exist when the central charge $c$ is less than or equal to 8.

\subsection{Turning on a twist gap}

As remarked in the introduction, while there is a nontrivial upper bound on the twist gap $t_{\rm mod} = {c-1\over 12}$, we do not know any explicit construction of unitary, compact CFTs with nonzero twist gap.\footnote{Nor do we know one with zero twist gap but no conserved primary currents, i.e., a unitary, compact CFT with infinitely many non-conserved higher spin primaries whose twists accumulate to zero.} Obviously it would be of interest to exhibit such theories, if they are indeed as ubiquitous as one might expect (some candidates for irrational CFTs with no extra conserved currents have been considered in \cite{Halpern:1995js, Dotsenko:aa}). Here we study the upper bound on the gap in the dimension of scalar primaries while imposing a nonzero gap in the twist of all primary operators in the spectrum. This would in particular exclude theories that contain conserved current primaries (regardless of whether their partition functions are of the generic type).

\begin{figure}[h!]
\subfloat{
\includegraphics[width=.33\textwidth]{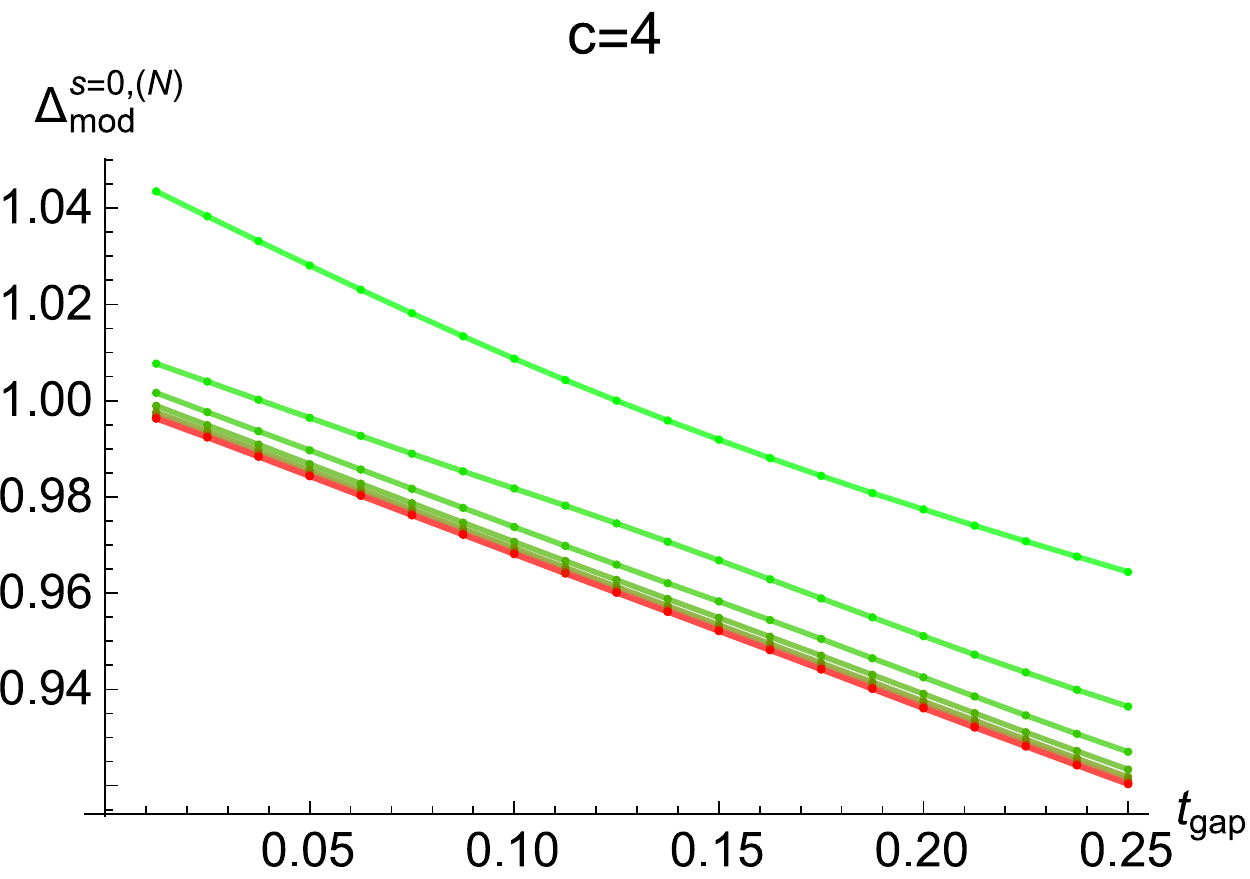}
}
\subfloat{
\includegraphics[width=.33\textwidth]{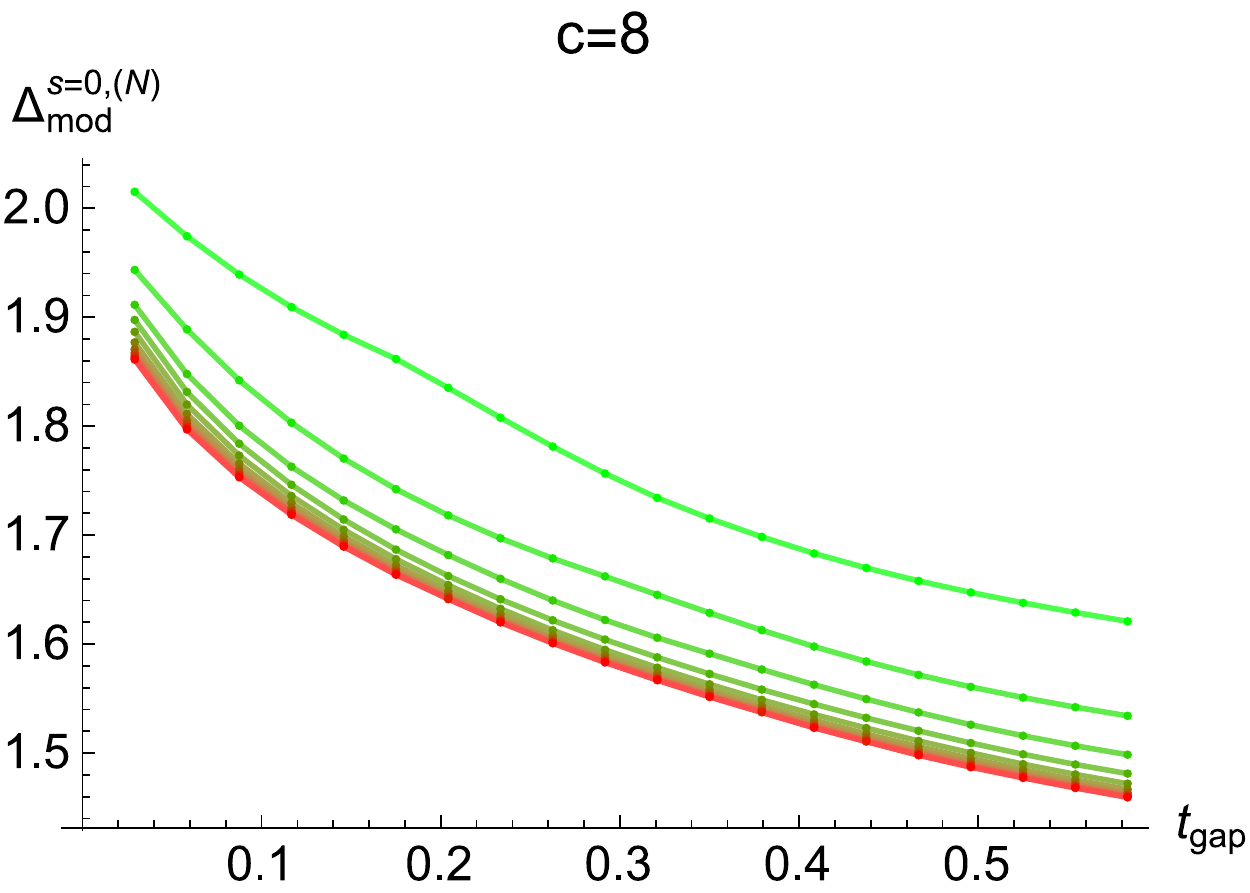}
}
\subfloat{
\includegraphics[width=.33\textwidth]{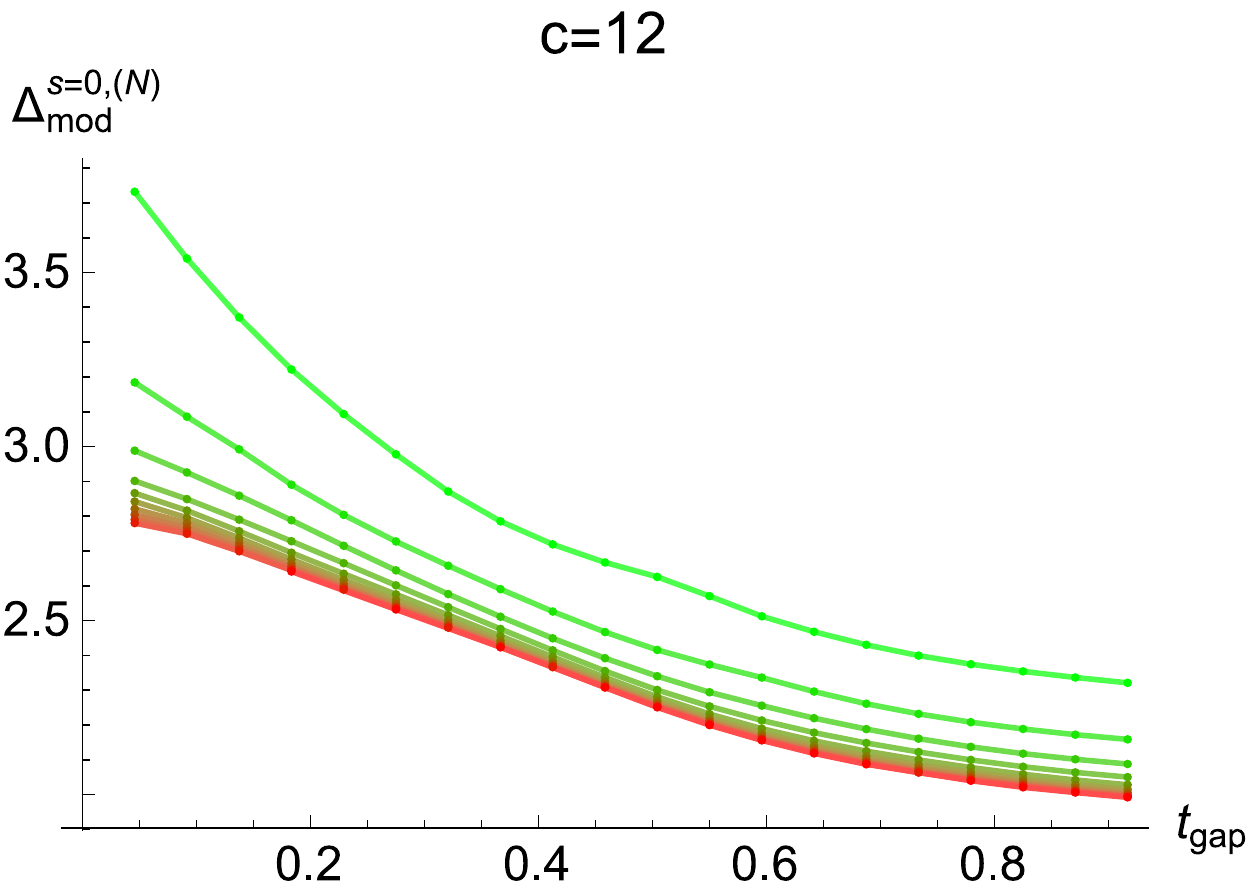}
}
\caption{The gap in the dimension of scalar primaries as a function of the gap in the twist at increasing derivative order (from green to red, up to $N=47$) for $c=4,\,8,\,12$. }\label{fig:AllSpinTwistGap}
\end{figure}

Figure~\ref{fig:AllSpinTwistGap} shows the bound on the scalar dimension gap $\Delta^{(N)}_{\rm mod}(t_{\rm gap})$ as a function of the twist gap $t_{\rm gap}$, for various values of central charge $c$. As $t_{\rm gap}$ is increased from $0$ to its upper bound $t_{\rm mod}={c-1\over 12}$, $\Delta_{\rm mod}(t_{\rm gap})$ decreases smoothly (but interestingly, it never approaches $c-1\over 12$). 

\begin{figure}[h!]
\centering
\subfloat{
\includegraphics[width=.5\textwidth]{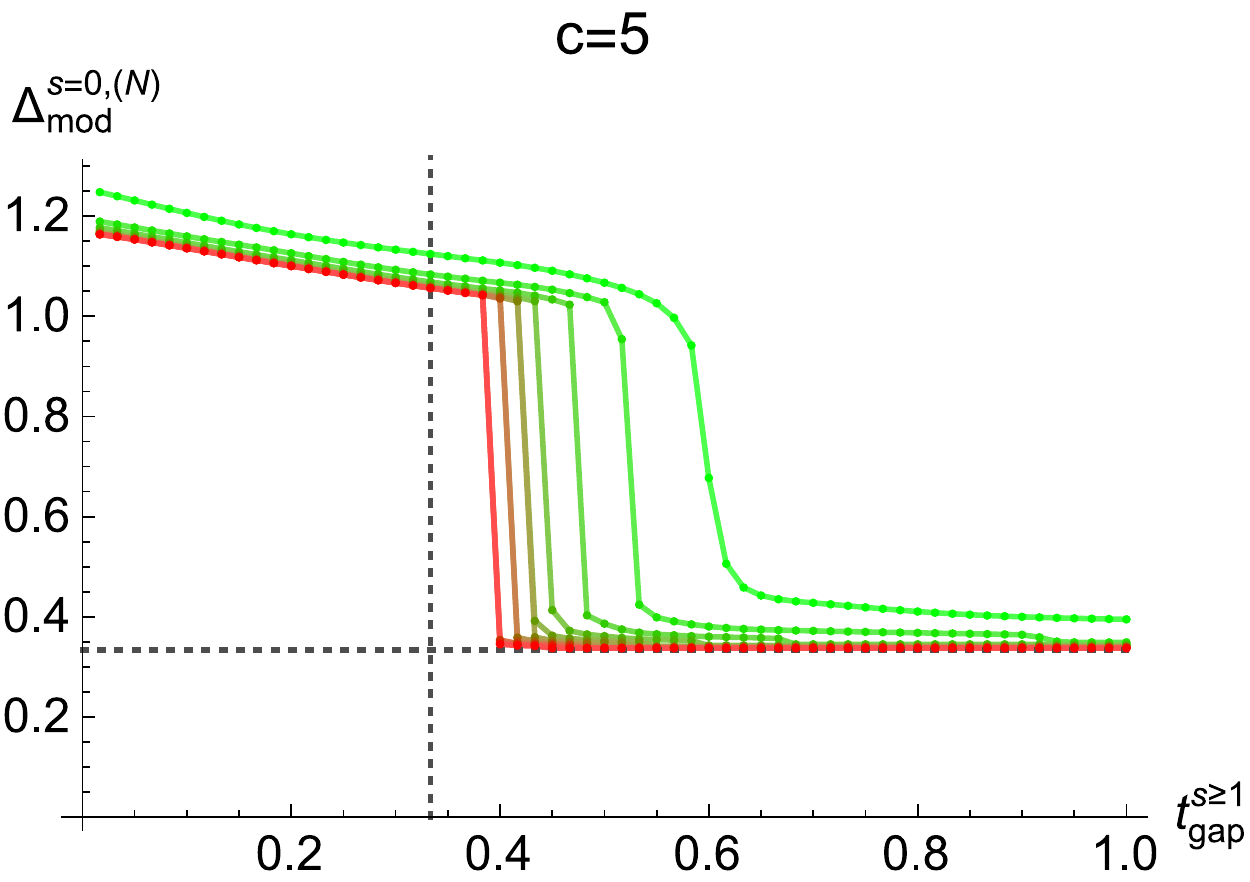}
}
\caption{The gap in the dimension of scalar primaries as a function of the gap in the twist of all {\it nonzero-spin} primaries at increasing derivative order (from green to red, up to $N=47$) for $c=5$. The dotted lines denote $t_{\rm gap}^{s\ge 1}={c-1\over 12}$ and $\Delta_{\rm mod}^{s=0,(N)}={c-1\over 12}$.}\label{fig:HigherSpinTwistGapDrop}
\end{figure}

If we further increase the twist gap for {\it nonzero-spin} primaries beyond $c-1\over 12$, the bound on scalar dimension gap drops to $c-1\over 12$ as expected. An example of this is shown in Figure~\ref{fig:HigherSpinTwistGapDrop}.

\begin{figure}[h!]
\centering
\subfloat{
\includegraphics[width=.45\textwidth]{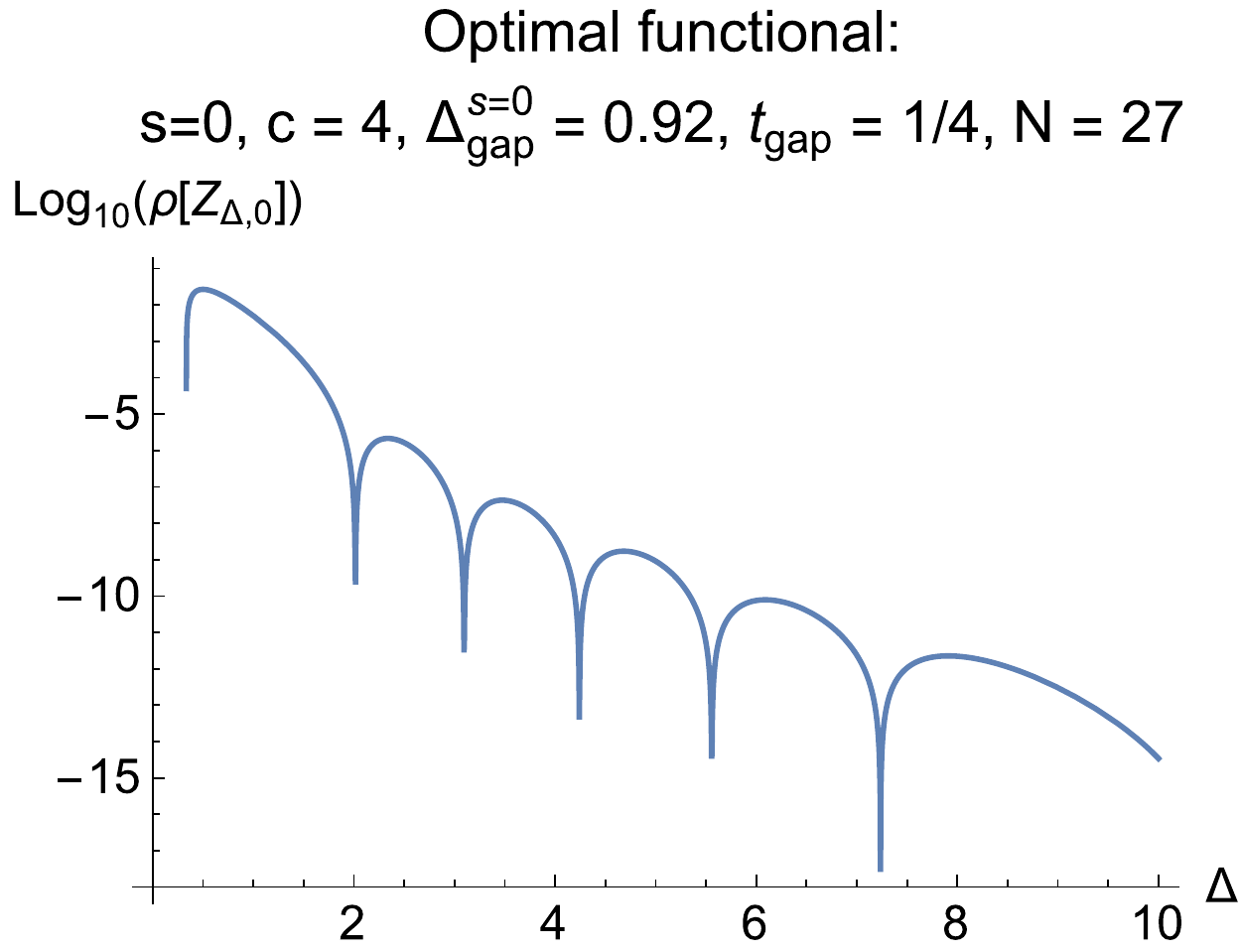}
}
\subfloat{
\includegraphics[width=.45\textwidth]{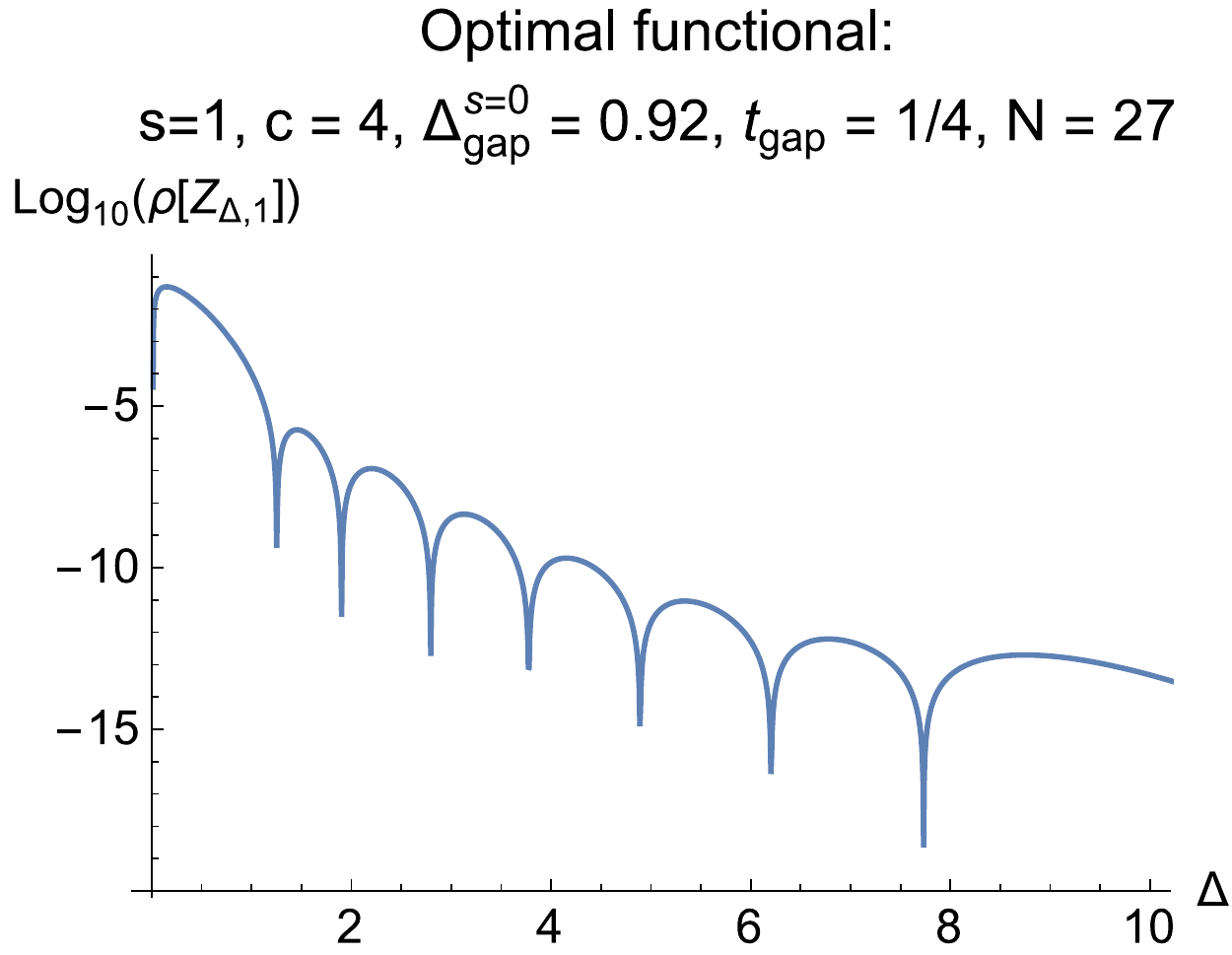}
}
\caption{The optimal linear functional acting on spin-0 and spin-1 reduced characters with the maximal twist gap imposed. The zeroes of this functional determine the dimensions of operators in the extremal spectrum.}\label{fig:AllSpinTwistGapOptimalZeroes}
\end{figure}

As we will explain in sections \ref{Sec:Degeneracy} and \ref{Sec:ExtremalSpectrum}, when the upper bound on the degeneracy of the lightest operator is saturated, the rest of the spectrum (which we refer to as the ``extremal spectrum'') is uniquely determined by the zeroes of the optimal linear functional acting on the characters as a function of the dimension. This provides a procedure to explicitly construct a spectrum for which the twist gap is nonzero.\footnote{However, this procedure does not on its own determine the degeneracies of the higher-dimension operators in the extremal spectrum.} As a proof of principle, in Figure~\ref{fig:AllSpinTwistGapOptimalZeroes} we plot the optimal functional acting on spin-0 and spin-1 characters for $c=4$ with the maximal twist gap and a value of the dimension gap close to the upper bound. The resulting spectra with finite twist gap and maximal degeneracy of the lightest primary appear to be discrete and are not obviously inconsistent.

It is interesting to compare $\Delta^{s=0}_{\rm mod}(t_{\rm gap})$ with the bound on scalar dimension gap when a nonzero twist gap {\it only} for spin-1 primaries is introduced. We denote the latter bound by $\widetilde\Delta^{s=0}_{\rm mod}(t_{\rm gap}^{s=1})$. Obviously, by definition $\widetilde\Delta^{s=0}_{\rm mod}(t)\geq\Delta^{s=0}_{\rm mod}(t)$. We find numerically that $\widetilde\Delta^{s=0}_{\rm mod}(t)$ coincides with $\Delta^{s=0}_{\rm mod}(t)$ for $t\leq {c-1\over 12}$. A transition occurs at a larger value of $t$, after which $\widetilde\Delta^{s=0}_{\rm mod}(t)$ decays smoothly with $t$ (potentially exponentially fast) towards $c-1\over12$. We have not managed to obtain a reliable plot of the full curve of $\widetilde\Delta^{s=0}_{\rm mod}(t_{\rm gap}^{s=1})$ as the numerics stabilize slowly with the truncation on derivative order $N$, for an intermediate range of $t_{\rm gap}^{s=1}$.

\section{Operator Degeneracies and Extremal Spectra}

\subsection{Bounds on the degeneracy at the gap}
\label{Sec:Degeneracy}

If we impose a dimension gap $\Delta_{\rm gap}$ on the spectrum (not to be confused with the upper bound on such a gap, which we denoted by $\Delta_{\rm mod}$), we can use semi-definite programming to place universal bounds on the degeneracies of primary operators in such a CFT. In particular, if $\Delta_{\rm gap}$ lies between the twist gap bound ${c-1\over 12}$ and the upper bound on the dimension gap (that follows from modular invariance) $\Delta_{\rm mod}$, one can place upper bounds on the degeneracies of primaries, as follows.

Fixing the dimension $\Delta_{\text{gap}}$($\leq \Delta_{\rm mod}$) of the lowest primary operator, consider all linear functionals $\rho$ of the form (\ref{LinearFunctionalZ}) such that 
\begin{align}\label{PositivityAlphaDeg}
	&\rho\bigg[\hat{\chi}_{\Delta-s\over 2}(\tau)\hat{\bar\chi}_{\Delta+s\over 2}(\bar\tau) + \hat{\chi}_{\Delta+s\over 2}(\tau)\hat{\bar\chi}_{\Delta-s\over 2}(\bar\tau)\bigg]\ge 0,\quad \Delta \ge \text{max}\bigg(\Delta_{\text{gap}},s\bigg).
\end{align}
Since the gap $\Delta_{\rm gap}$ is allowed by the modular crossing equation, $\rho$ must be negative when acting on the vacuum character. We will normalize $\rho$ so that
\begin{align}\label{condb}
	\rho\bigg[\hat{\chi}_0(\tau)\hat{\bar\chi}_0(\bar\tau)\bigg] = -1.
\end{align}
The degeneracy\footnote{Of course, the coefficients $\{d_{\Delta,s}\}$ only have an interpretations as degeneracies of primary operators in the absence of conserved currents. In the presence of conserved currents, the $\{d_{\Delta,s}\}$ are simply the coefficients in the decomposition of the partition function into non-degenerate characters.} $d_{\Delta,s}$ of a primary of dimension $\Delta$ and spin $s$ is then subject to the following upper bound (this mirrors the upper bound on the squared OPE coefficients derived in the context of the four-point function bootstrap in \cite{Caracciolo:2009bx})
\begin{align}\label{DegUpperBound}
	d_{\Delta,s} \le \bigg(\rho\bigg[\hat{\chi}_{\Delta-s\over 2}(\tau)\hat{\bar\chi}_{\Delta+s\over 2}(\bar\tau) + \hat{\chi}_{\Delta+s\over 2}(\tau)\hat{\bar\chi}_{\Delta-s\over 2}(\bar\tau)\bigg]\bigg)^{-1}.
\end{align}
Obviously, the optimal bound on the degeneracy would be obtained using the functional that maximizes $\rho\bigg[\hat{\chi}_{\Delta-s\over 2}(\tau)\hat{\bar\chi}_{\Delta+s\over 2}(\bar\tau) + \hat{\chi}_{\Delta+s\over 2}(\tau)\hat{\bar\chi}_{\Delta-s\over 2}(\bar\tau)\bigg]$, subject to the conditions (\ref{PositivityAlphaDeg}) and (\ref{condb}). 

\begin{figure}[h!]
\centering
\includegraphics{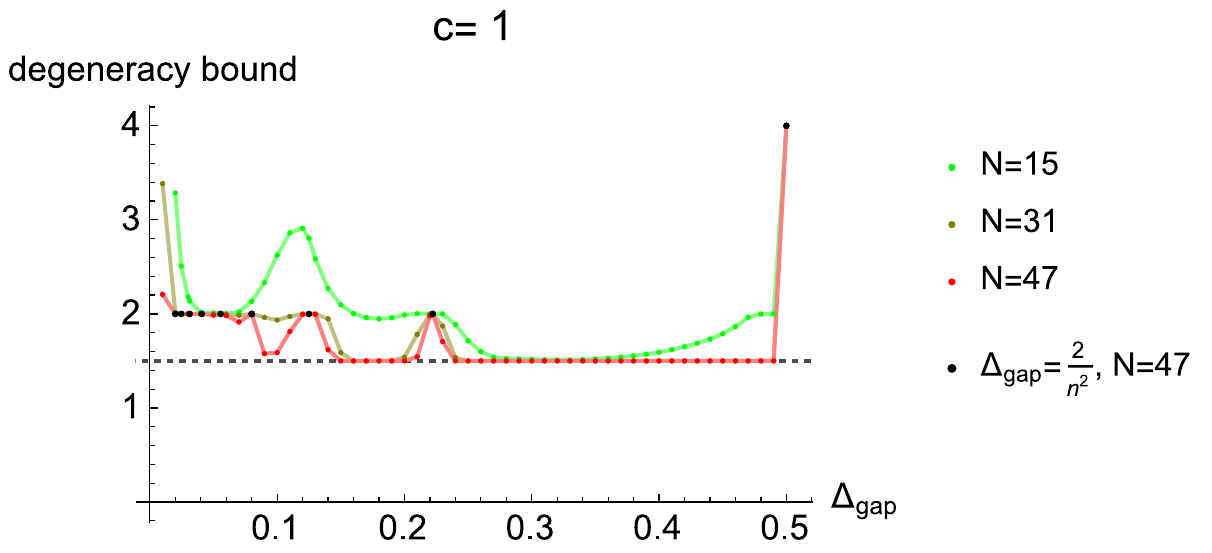}
\caption{The upper bound on the degeneracy of the lowest-lying operator for $c=1$ as a function of the assumed gap in the spectrum for derivative orders $N=15,31,47$.}\label{fig:deg-c=1}
\end{figure}

We will illustrate this method starting with the special case of $c=1$. Figure~\ref{fig:deg-c=1} shows the upper bound on the degeneracy of the lowest dimension scalar primaries above the vacuum as a function of the dimension gap $\Delta_{gap}$ in the spectrum. As the derivative order $N$ is increased, the degeneracy bound converges to ${3\over 2}$ except for a sequence of peaks located at $\Delta_{gap}={1\over 2}, {2\over 9}, {1\over 8},\cdots$ where the degeneracy bound is 4 or 2. This is in fact precisely consistent with what we know about the partition function of $c=1$ CFTs, as we now explain.

The only unitary, compact $c=1$ CFTs are the compact boson and its orbifolds. The $S^1/\mathbb{Z}_2$ orbifold partition function is not of the generic type, due to the degenerate character of a conserved spin-4 current, except at the self-dual radius $R=1$, where the CFT is equivalent to a compact boson at $R=2$. The compact boson CFT at radius $R$ has a reduced partition function of the form
\begin{align}
	\hat{Z}_{\rm CB}(R) = |\tau|^{1\over 2}\sum_{a,b\in\mathbb{Z}}q^{{1\over 4}({a\over R}+bR)^2}\bar{q}^{{1\over 4}({a\over R}-bR)^2}. 
\end{align} 
For $R\ge 1$, the gap in the spectrum is $\Delta_{\text{gap}}(R) = {1\over 2 R^2}$.
If we decompose the reduced partition function in terms of {\it non-degenerate} reduced characters, $\hat{Z}_{\rm CB}(R)$ generically has a single negative coefficient of $-1$ at weight $(1,1)$. When the radius $R$ is a half-integer $R={n\over 2}, ~n\in\mathbb{Z}_{\ge 0}$ (or a $T$-dual equivalent thereof), however, the weight $(1,1)$ coefficient becomes positive due to the appearance of extra marginal primaries, and thus the corresponding partition function is of the generic type to which our bounds apply despite the presence of conserved currents. When $R=1$, it is easy to see that the reduced partition function has degeneracy 4 at the gap:
\begin{align}\label{SelfDualZ}
	\hat{Z}_{\rm CB}(1)=\hat{\chi}_0(\tau)\hat{\chi}_0(\bar\tau) + |\tau|^{1\over 2}\bigg[4(q \bar q)^{1\over 4}+3(q+\bar q) + 3 q \bar q+\ldots\bigg],
\end{align}
which corresponds to the peak at $\Delta_{\text{gap}} = {1\over 2}$ in Figure~\ref{fig:deg-c=1}. When the compactification radius takes another half-integer value, the degeneracy of the lowest-lying primary is 2:
\begin{align}\label{CompactBosonHalfInteger}
	\hat{Z}_{\rm CB}\left({n\over 2}\right)=\hat{\chi}_0(\tau)\hat{\bar\chi}_0(\bar\tau) + |\tau|^{1\over 2}\bigg[2 (q\bar{q})^{1\over n^2}+\ldots\bigg],~n\ge 3,
\end{align}
which corresponds to peaks of the degeneracy bound at $\Delta_{\text{gap}} = {2\over n^2}$ for $n\ge 3$, as seen in Figure~\ref{fig:deg-c=1}.

At first sight, it might seem odd that the degeneracy bounds in Figure~\ref{fig:deg-c=1}  approach ${3\over 2}$ at a generic value of $\Delta_{gap}$. Recall that the compact boson reduced partition function will generically have a negative coefficient $-1$ at weight $(1,1)$ in its decomposition into non-degenerate reduced characters. On the other hand, from (\ref{SelfDualZ}), the reduced partition function at the self-dual radius $R=1$ has a coefficient $3$ at weight $(1,1)$, and so the linear combination
\begin{align}\label{CompactBosonGenericType}
	{3\over 4}\hat{Z}_{\rm CB}(R)+{1\over 4}\hat{Z}_{\rm CB}(1) = \hat{\chi}_0(\tau)\hat{\bar\chi}_0(\bar\tau) + |\tau|^{1\over 2}\bigg[{3\over 2}(q\bar q)^{1\over 4R^2}+\ldots\bigg]
\end{align}
has non-negative coefficients; it is a partition function of the generic type. Furthermore, we see that the ``degeneracy" at the gap $\Delta_{\text{gap}} = {1\over 2R^2}$ is $3\over 2$, as seen in Figure~\ref{fig:deg-c=1}.\footnote{One could also have considered the linear combinations ${3\over4}\hat{Z}_{\rm CB}(R) + {1\over 4}\hat{Z}_{\rm CB}(2)$ (for $R>2$) or ${1\over 2}\hat{Z}_{\rm CB}(R) +{1\over 2} \hat{Z}_{\rm CB}\left({n\over 2}\right)$ for $n=3,5,6,7,\ldots$ (for $R>{n\over 2}$), but the conclusion that $3\over 2$ is the maximal leading coefficient is unchanged.} In this case, a more refined bound would be obtained if we demand that the degeneracies are integers.

\begin{figure}[h!]
\centering
\subfloat{
\includegraphics[width=.45\textwidth]{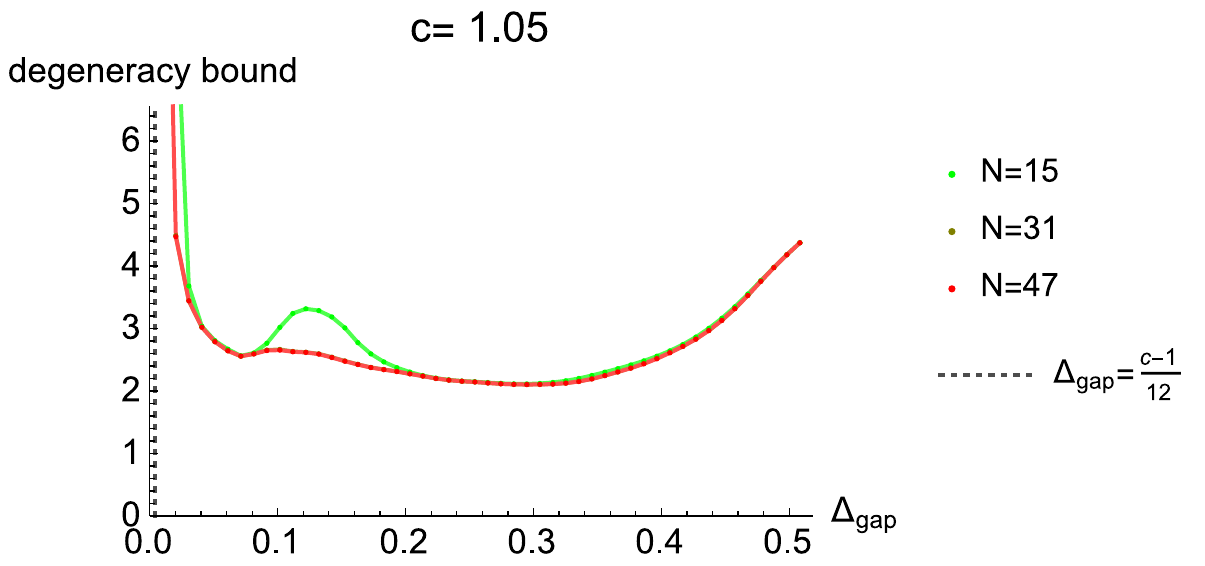}
}
\subfloat{
\includegraphics[width=.45\textwidth]{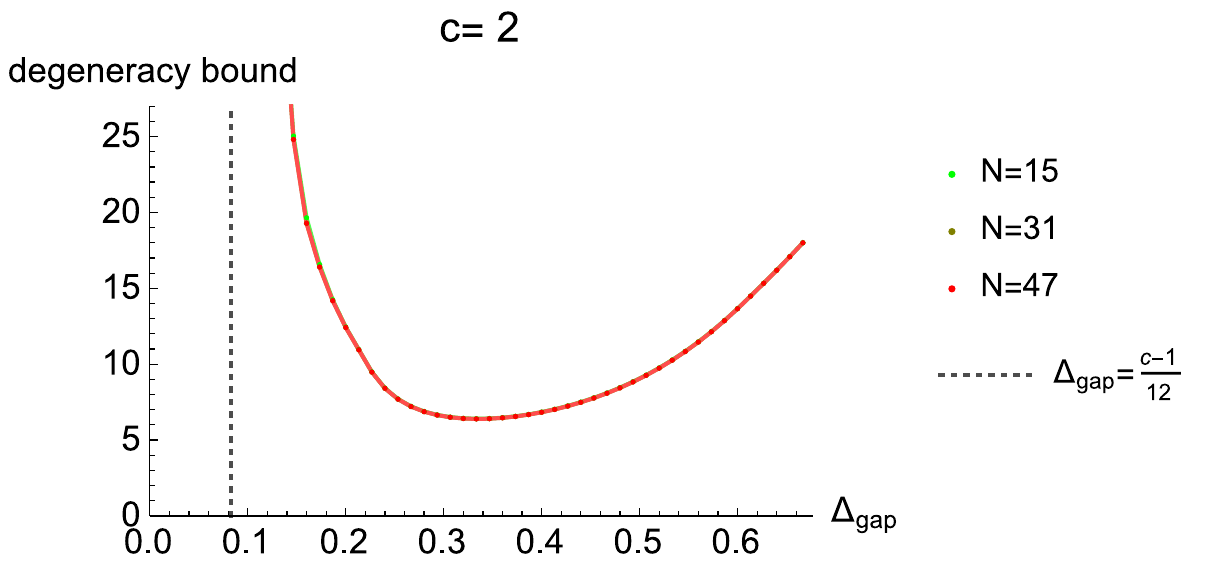}
}
\\
\subfloat{
\includegraphics[width=.45\textwidth]{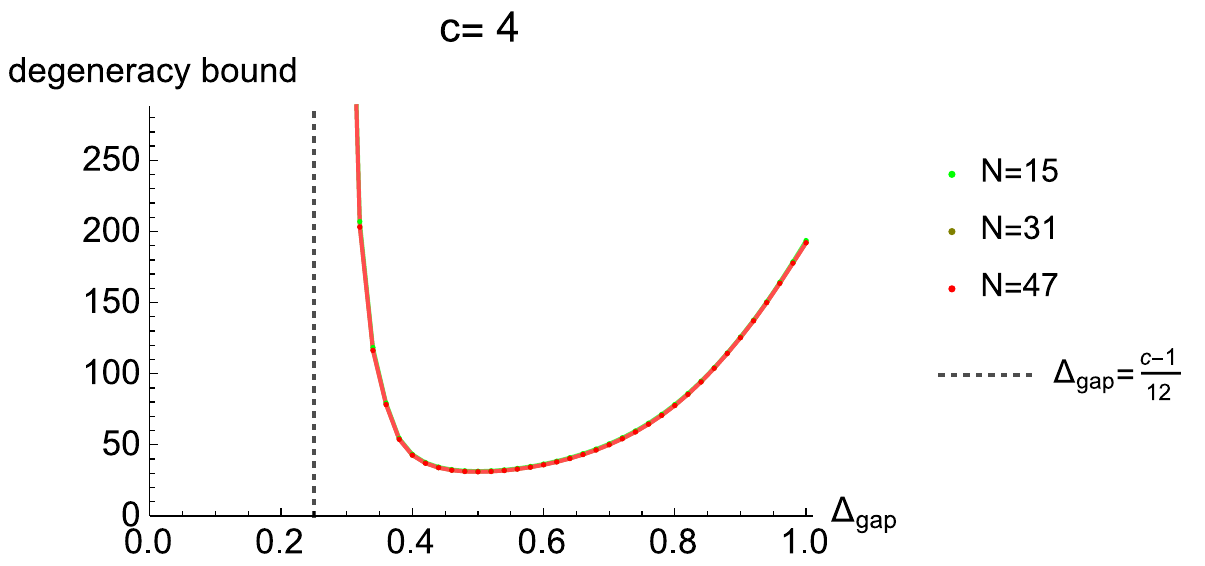}
}
\subfloat{
\includegraphics[width=.4\textwidth]{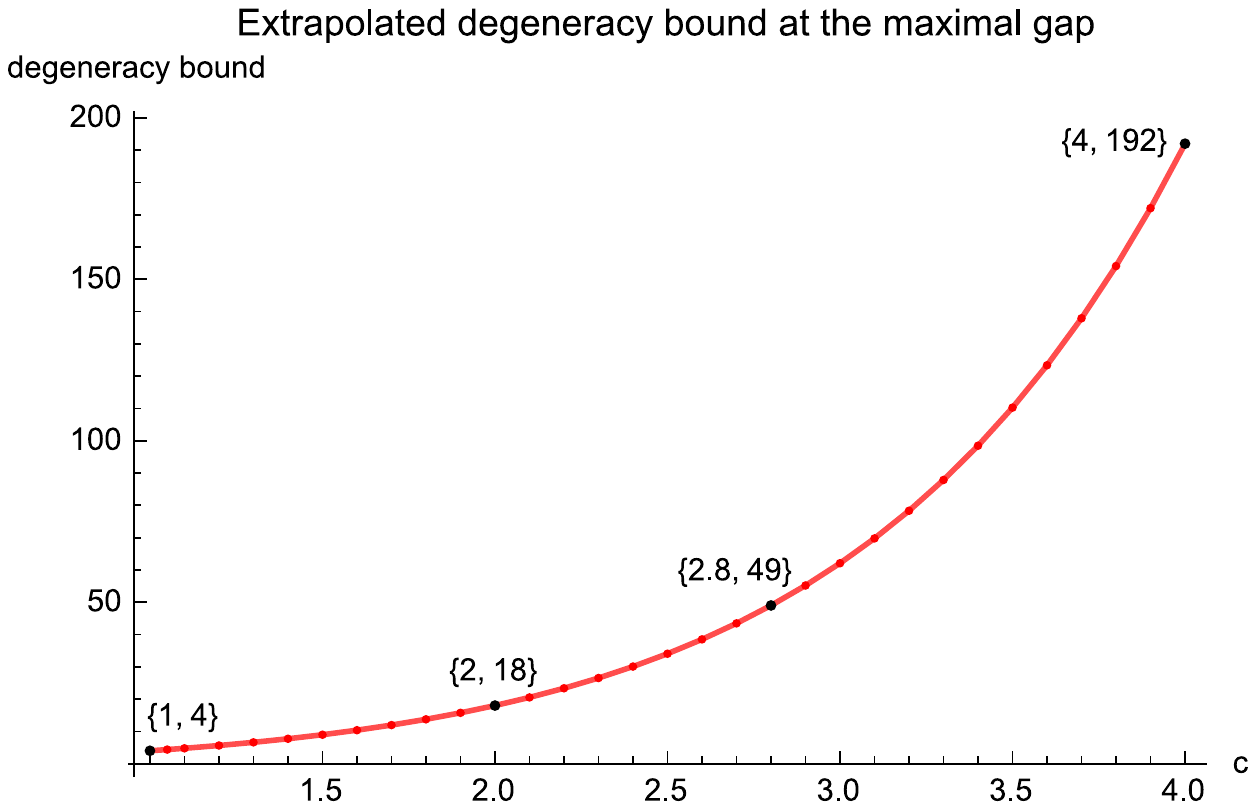}
}
\caption{The upper bound on the degeneracy of the lowest-lying operators as a function of $\Delta_{\rm gap}$ for a few values of the central charge, and the growth of the degeneracy at the maximal gap $\Delta_{\rm mod}$ as a function of the central charge. In the bottom-right plot, the black points denote the special theories discussed in section \ref{Sec:Kinks}.}\label{fig:deg}
\end{figure}

The bounding curves on the degeneracies of the lightest primary operators at larger values of the central charge are shown in Figure~\ref{fig:deg}.
Note that the bound diverges as $\Delta_{\rm gap}$ approaches $c-1\over 12$, as expected.\footnote{This can be understood by noting that the partition function of Liouville theory, with gap $c-1\over 12$, may be viewed as an infinite degeneracy limit of a compact CFT partition function.}

\subsection{Extremal spectrum from the optimal linear functional}
\label{Sec:ExtremalSpectrum}

When the degeneracy bound at the gap $\Delta_{\rm gap}$ is saturated (for ${c-1\over 12}<\Delta_{\rm gap}\leq \Delta_{\rm mod}$), the entire spectrum of the CFT is in fact determined \cite{ElShowk:2012hu, El-Showk:2014dwa}, for reasons we explain below. Such a spectrum will be called ``extremal", and the corresponding reduced partition function will be denoted $\hat{Z}_{\text{ext}}(c,\Delta_{\text{gap}})$.\footnote{This definition of the extremal spectrum does not guarantee that the degeneracies of operators are integers; in fact, the latter occurs only for a discrete set of values of $c$ and $\Delta_{\rm gap}$, and it is obviously only in these cases that the extremal spectrum could potentially be realized by a physical CFT.} The spectrum of spin-$s$ primaries will be denoted ${\cal I}_s^{\rm ext}$. The optimal functional $\rho$ we use to determine the degeneracy bound, as in (\ref{DegUpperBound}), satisfies
\begin{align}
	0 &= \rho\bigg[\hat{\chi}_0(\tau)\hat{\bar\chi}_0(\bar\tau) + \sum_s\sum_{\Delta\in\mathcal{I}_s^{\text{ext}}}d_{\Delta,s}\bigg(\hat{\chi}_{\Delta+s\over 2}(\tau)\hat{\bar\chi}_{\Delta-s\over 2}(\bar\tau) + \hat{\chi}_{\Delta-s\over 2}(\tau)\hat{\bar\chi}_{\Delta+s\over 2}(\bar\tau)\bigg)\bigg]\nonumber\\
	&= \sum_{s}\sum_{\Delta\in\mathcal{I}^\text{ext}_s,\Delta>\Delta_{\text{gap}}}d_{\Delta,s}\rho\bigg[\hat{\chi}_{\Delta+s\over 2}(\tau)\hat{\bar\chi}_{\Delta-s\over 2}(\bar\tau) + \hat{\chi}_{\Delta-s\over 2}(\tau)\hat{\bar\chi}_{\Delta+s\over 2}(\bar\tau)\bigg].
\end{align}
Here, the contribution of the vacuum and that of the primaries at the gap cancel, due to the saturation of the degeneracy bound at the gap.
Positivity of the coefficients $d_{\Delta,s}$ and (\ref{PositivityAlphaDeg}) then constrain the extremal spectrum to be such that the corresponding characters are annihilated by the linear functional $\rho$, namely
\begin{equation}
	\rho\bigg[\hat{\chi}_{\Delta+s\over 2}(\tau)\hat{\bar\chi}_{\Delta-s\over 2}(\bar\tau) + \hat{\chi}_{\Delta-s\over 2}(\tau)\hat{\bar\chi}_{\Delta+s\over 2}(\bar\tau)\bigg]=0,~~~\Delta\in\mathcal{I}^\text{ext}_s,~\Delta> \Delta_{\text{gap}}.
\end{equation}

\begin{figure}[h!]{
\centering
\subfloat{
\includegraphics[width=.45\textwidth]{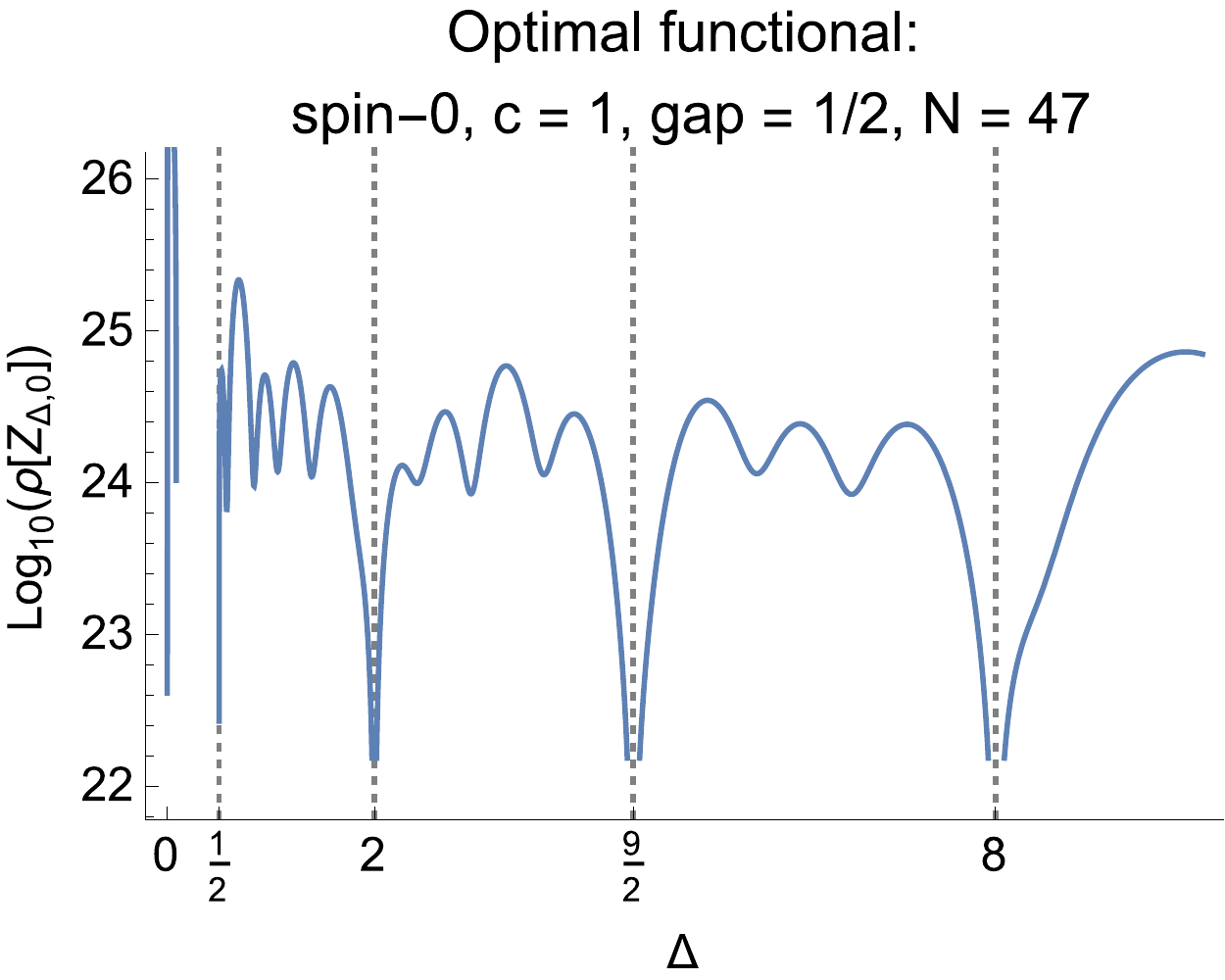}
}
\subfloat{
\includegraphics[width=.45\textwidth]{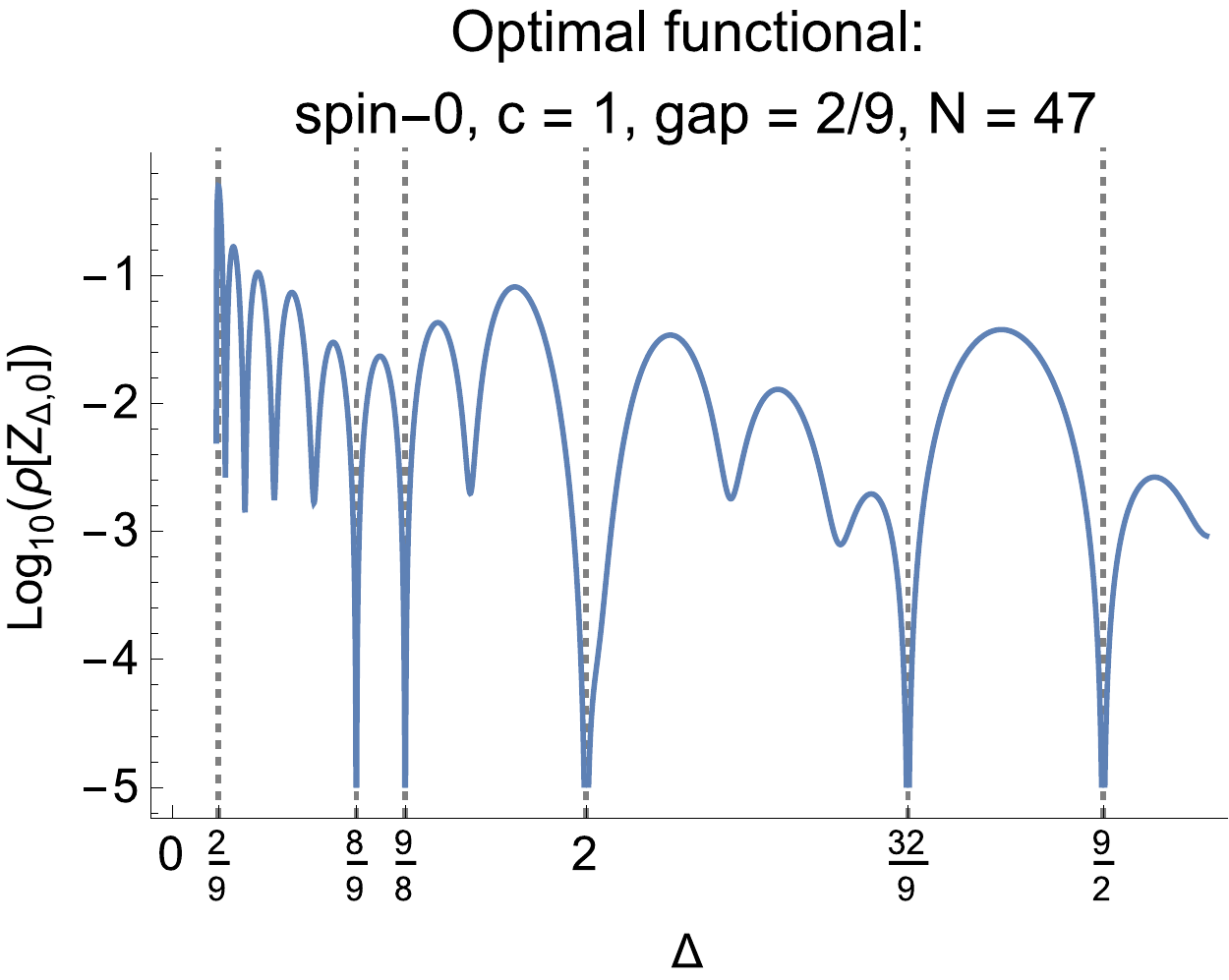}
}
\\
\subfloat{
\includegraphics[width=.45\textwidth]{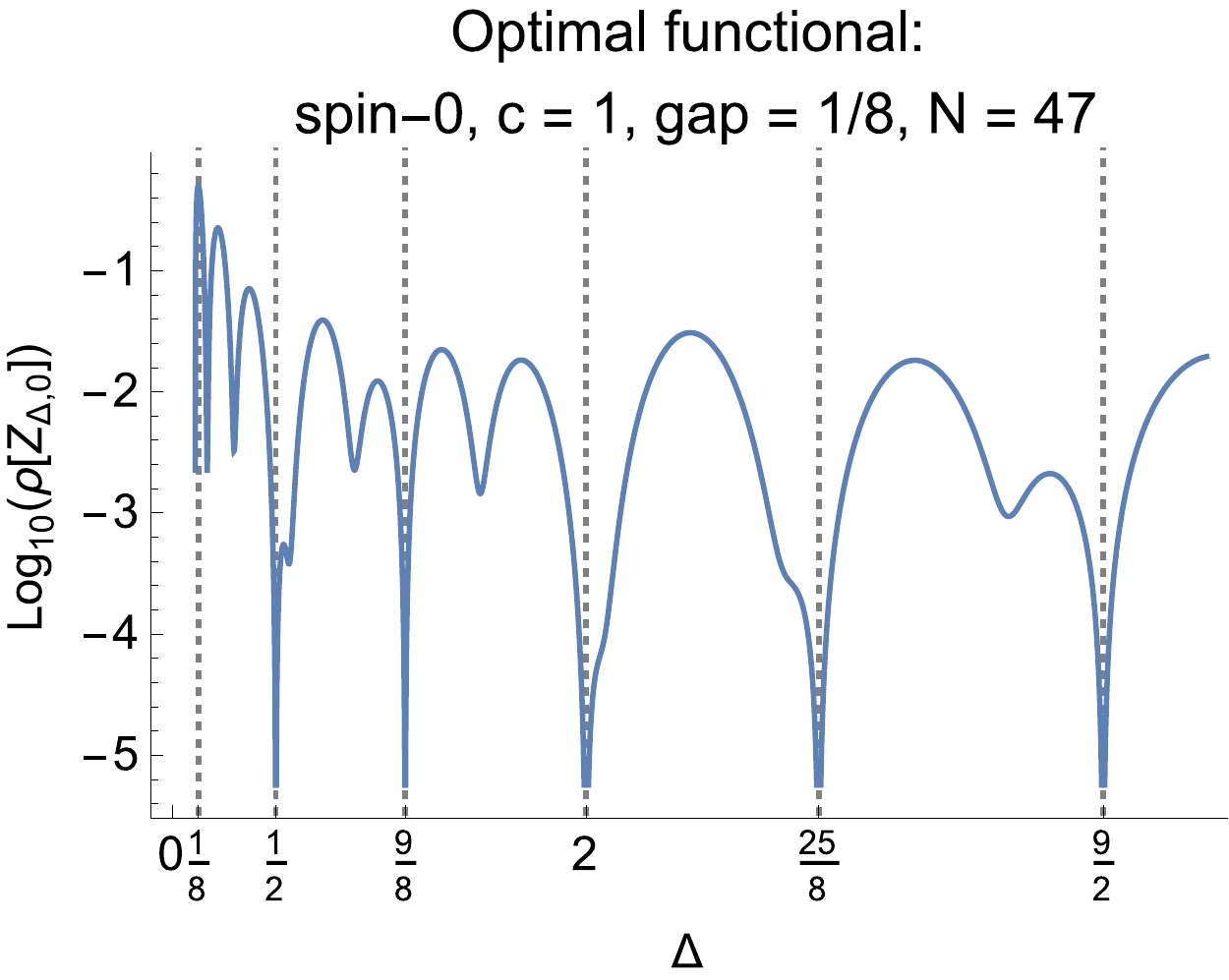}
}
\subfloat{
\includegraphics[width=.45\textwidth]{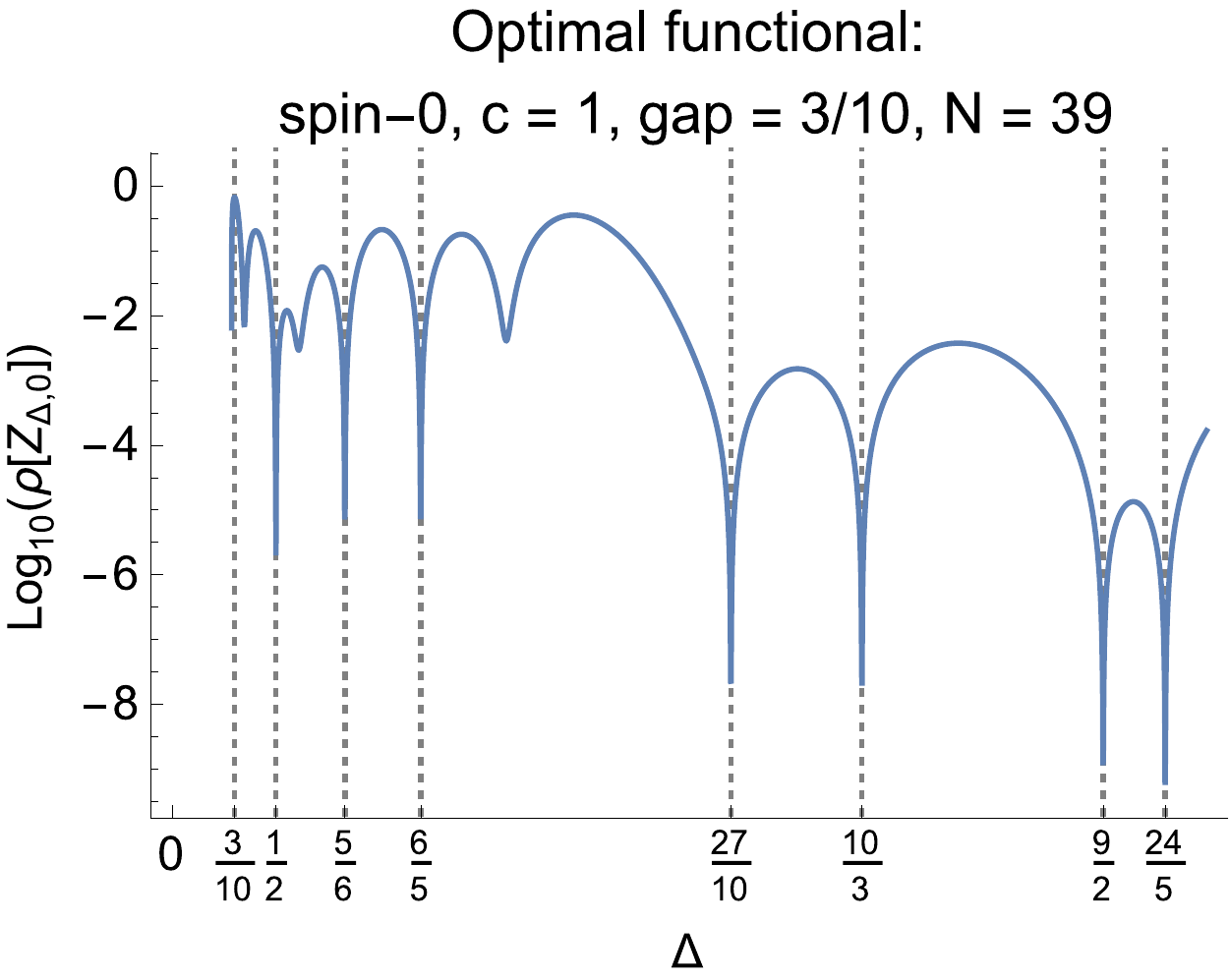}
}
\caption{The optimal linear functional that maximizes the degeneracy at the gap, acting on the spin-0 reduced characters. The zeroes of the optimal functional correspond to scalar operators in the extremal spectrum. The dotted lines correspond to the ``dimensions of scalar operators'' in the corresponding generic type compact boson reduced partition functions.}\label{fig:OptimalZeroesSpin0}}
\end{figure}

\begin{figure}[h!]
\centering
\subfloat{
\includegraphics[width=.45\textwidth]{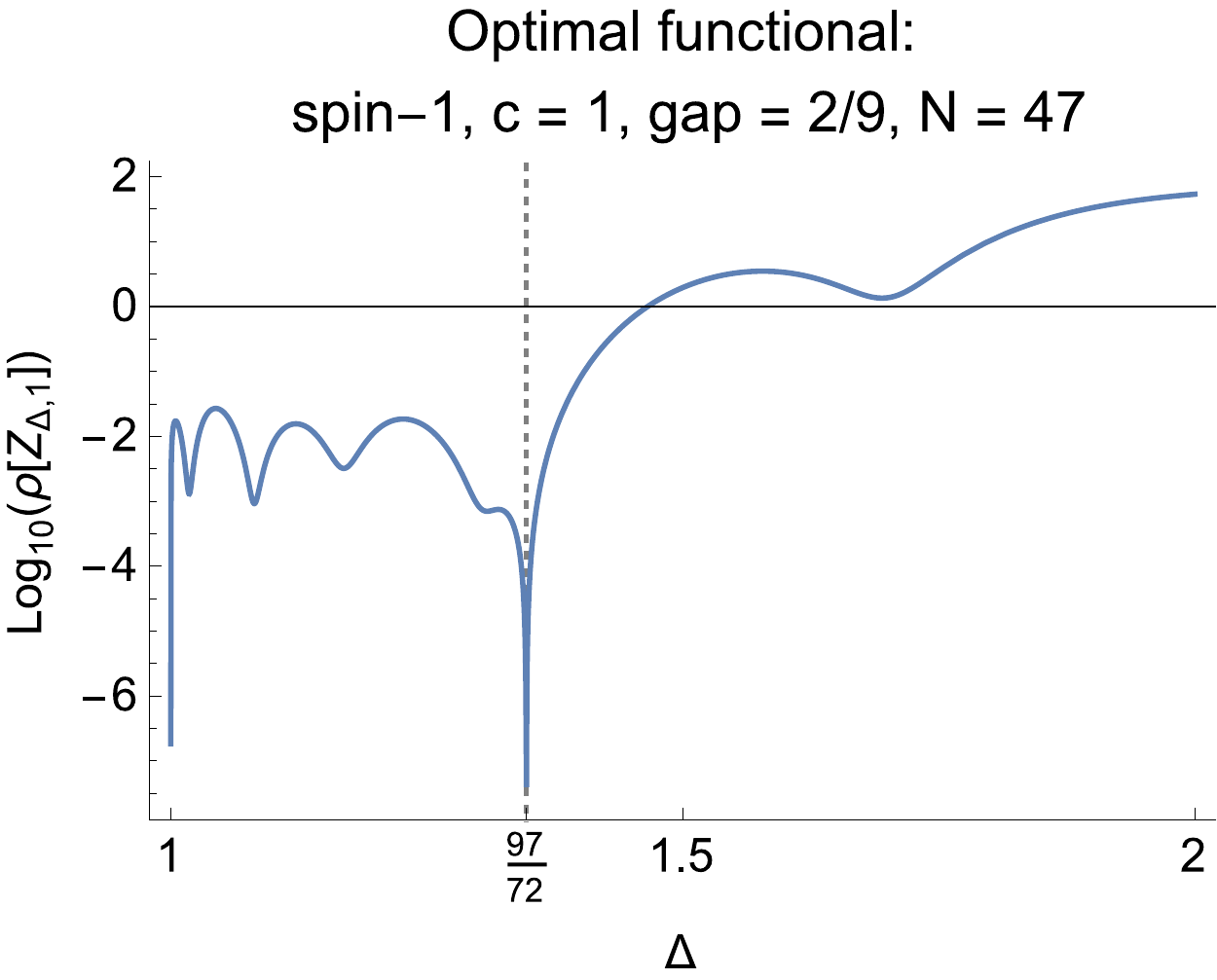}
}
\subfloat{
\includegraphics[width=.45\textwidth]{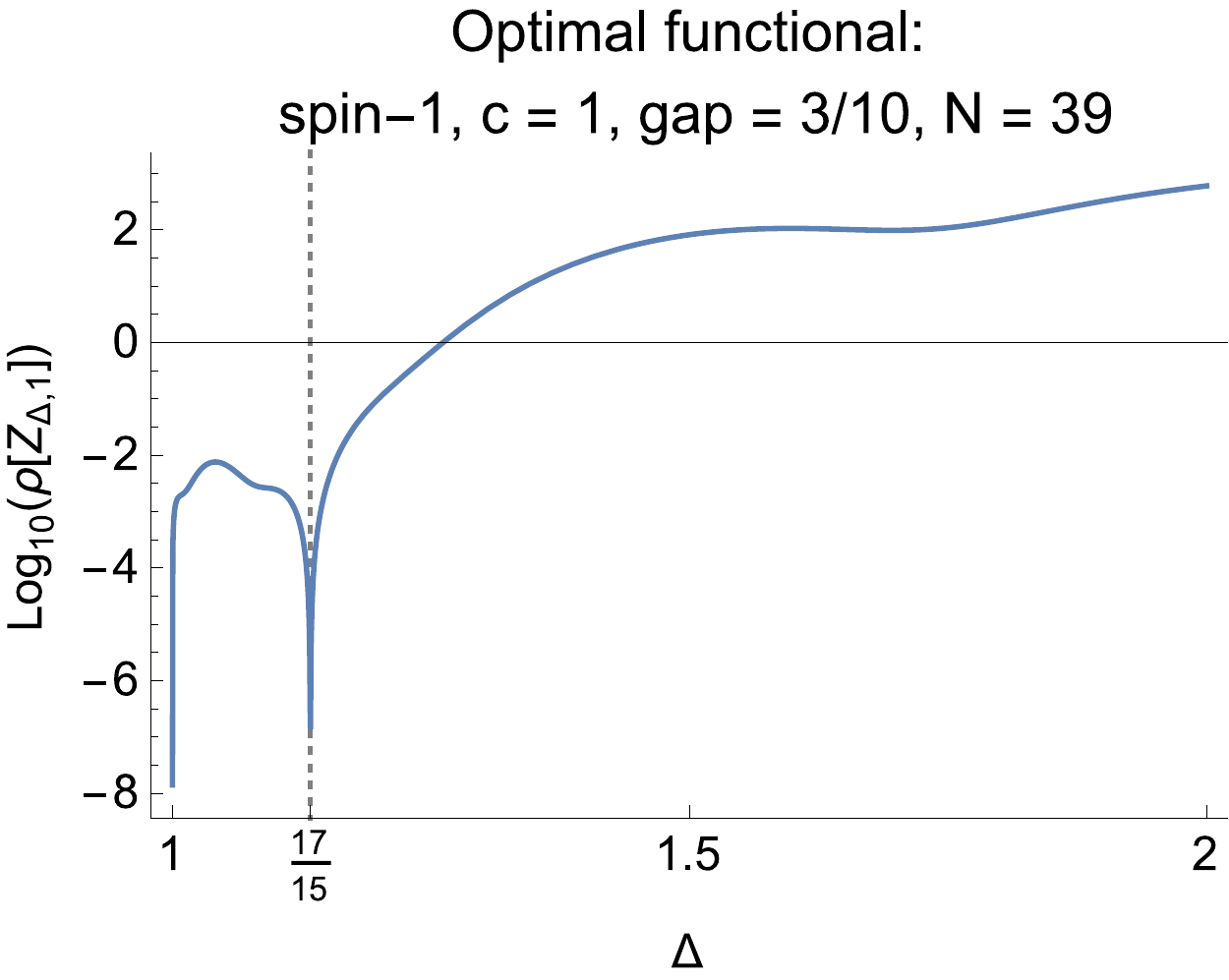}
}
\caption{The action of the optimal linear functional that maximizes the degeneracy of the lowest-lying operator on the spin-1 reduced characters. Again, the locations of the zeroes agree with the dimensions of spin-1 operators in the corresponding generic type compact boson reduced partition functions.}\label{fig:OptimalZeroesSpin1}
\end{figure}

Indeed, the extremal spectrum can be efficiently computed in this way. We will begin with the $c=1$ example. Figure~\ref{fig:OptimalZeroesSpin0} shows the value of the optimal functional acting on spin-0 reduced characters. 
For $\Delta_{\text{gap}} = {2\over n^2}$, the zeroes of the optimal functional correspond exactly to the scaling dimensions of scalar operators in the reduced compact boson partition function at radius $R={n\over 2}$ (\ref{CompactBosonHalfInteger}). On the other hand, at generic values of $\Delta_{\rm gap}$, the zeroes of the optimal functional precisely correspond to the scaling dimensions of scalar primaries in two different compact boson CFTs, whose partition functions combine to give one of generic type (but with non-integer coefficients) as in (\ref{CompactBosonGenericType}). Repeating this exercise with the optimal linear functional acting on nonzero spin characters also reveals zeroes at the locations predicted by the corresponding reduced compact boson partition functions (\ref{CompactBosonHalfInteger},\ref{CompactBosonGenericType}). See Figure~\ref{fig:OptimalZeroesSpin1} for some spin-1 examples.

One can also investigate the upper bound on the degeneracies of operators of higher dimensions that appear in the extremal spectrum. To do this, one fixes the gap and maximizes the action of the linear functional subject to (\ref{PositivityAlphaDeg},\ref{condb}) on the character corresponding to the higher-dimension operator of interest (rather than that of the operator whose dimension saturates the assumed gap). A priori, it need not be the case that a CFT that realizes the extremal spectrum also maximizes the degeneracies of the operators other than the lowest-lying (subject to the assumption that the gap in the spectrum is $\Delta_{\rm gap}$). However, for $c=1$, fixing the gap and maximizing the degeneracies of other operators in the extremal spectra, we find upper bounds that agree with the corresponding coefficients in the reduced compact boson partition functions of generic type (\ref{CompactBosonHalfInteger},\ref{CompactBosonGenericType}).

\subsection{Extremal spectra with maximal gap}
\begin{figure}[h!]{
\centering
\subfloat{
\includegraphics[width=.33\textwidth]{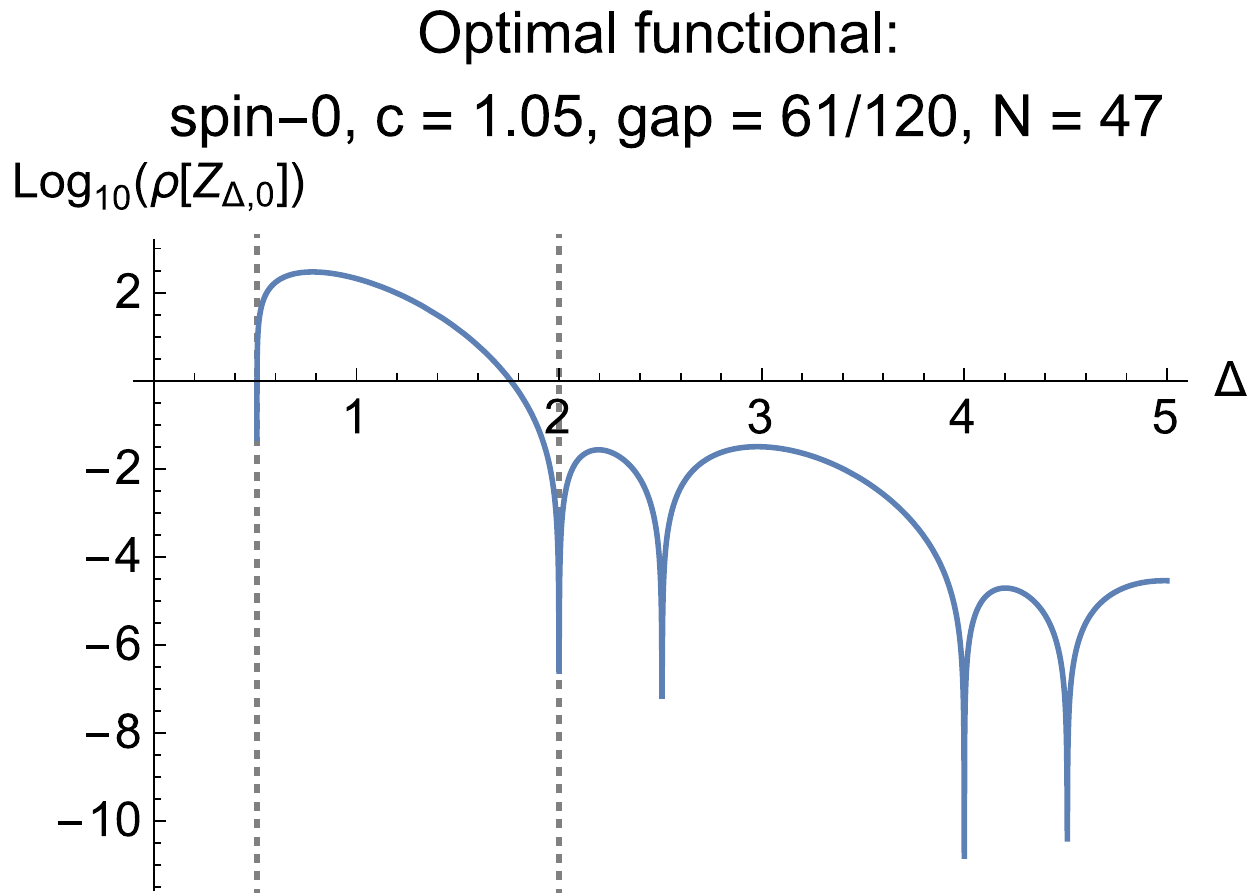}
}
\subfloat{
\includegraphics[width=.33\textwidth]{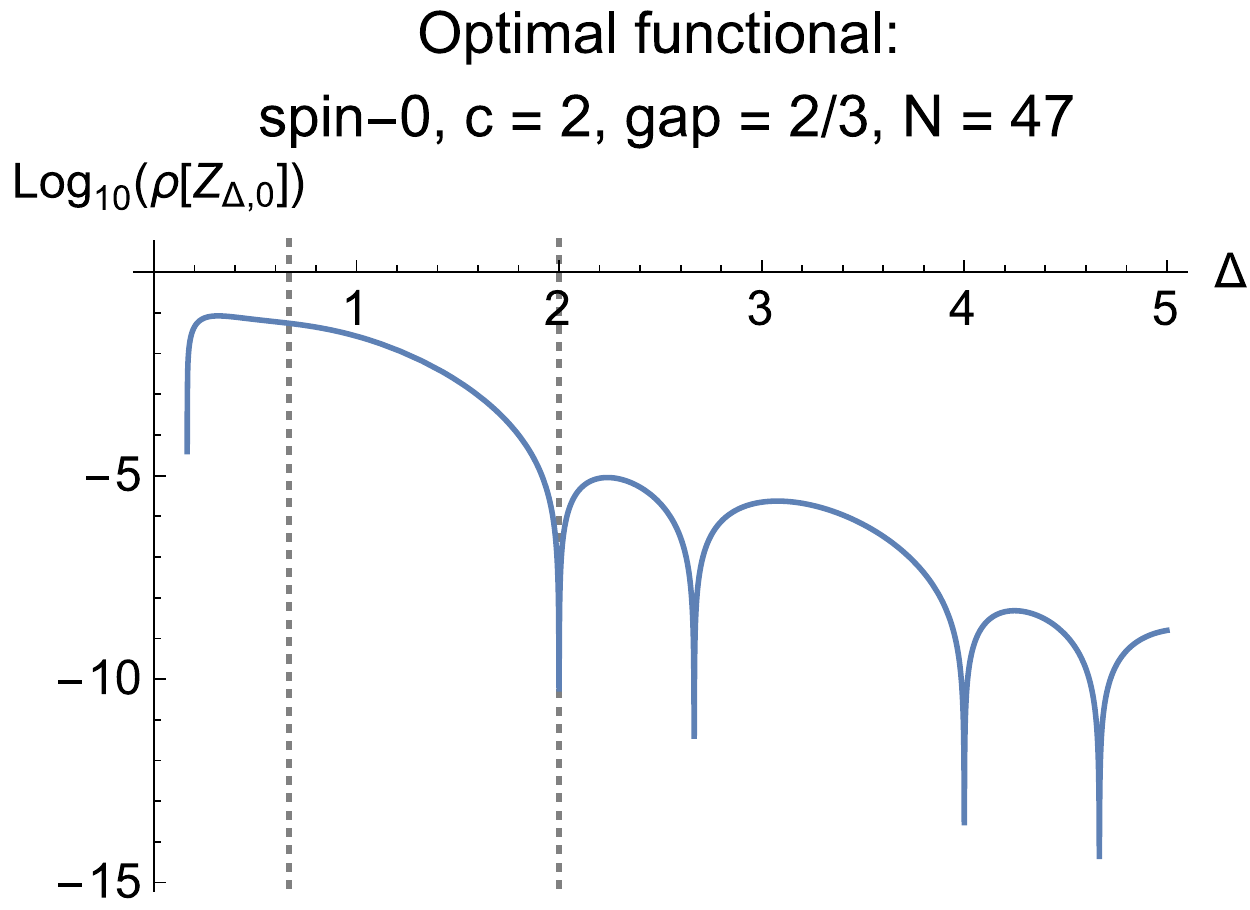}
}
\subfloat{
\includegraphics[width=.33\textwidth]{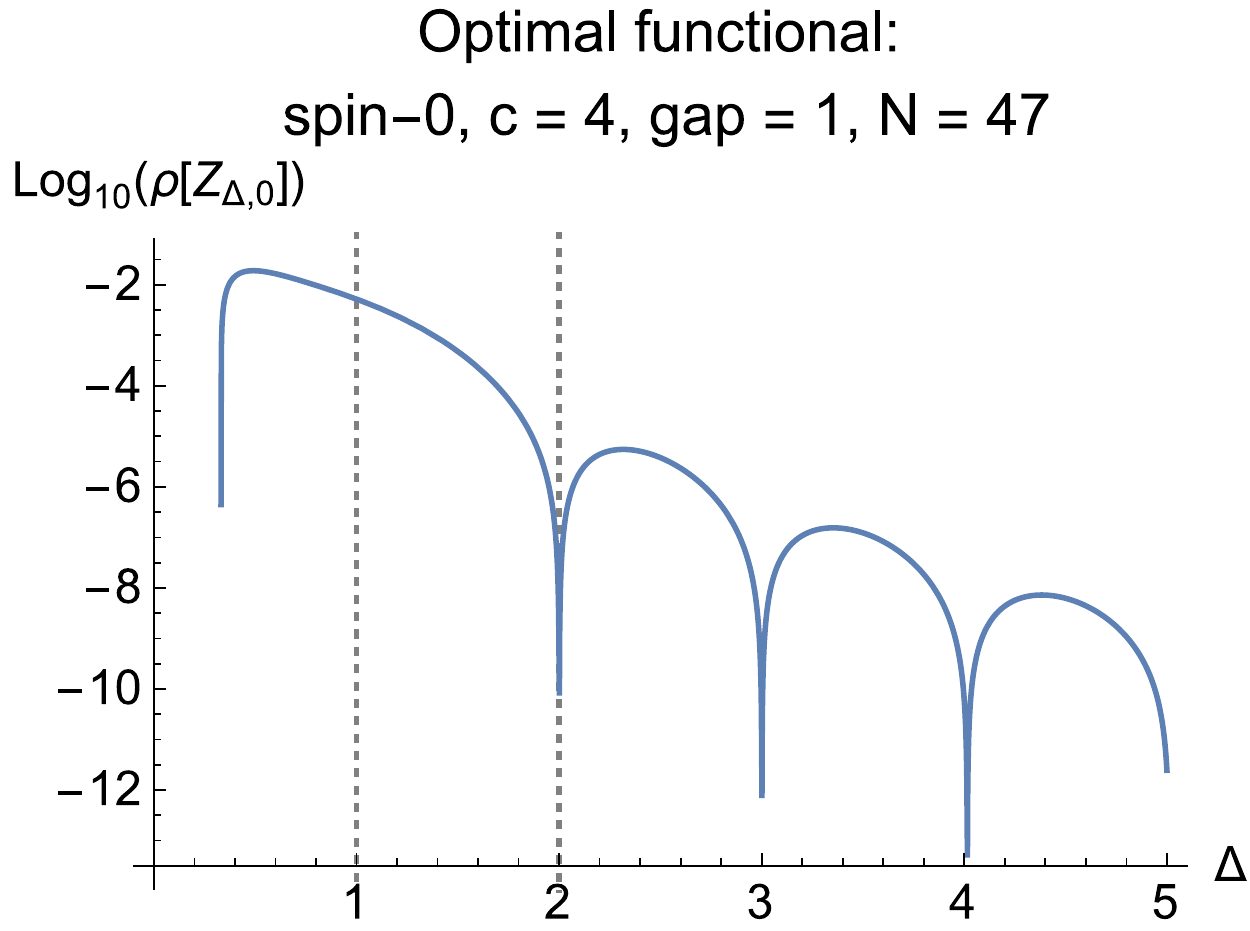}
}
\\
\subfloat{
\includegraphics[width=.33\textwidth]{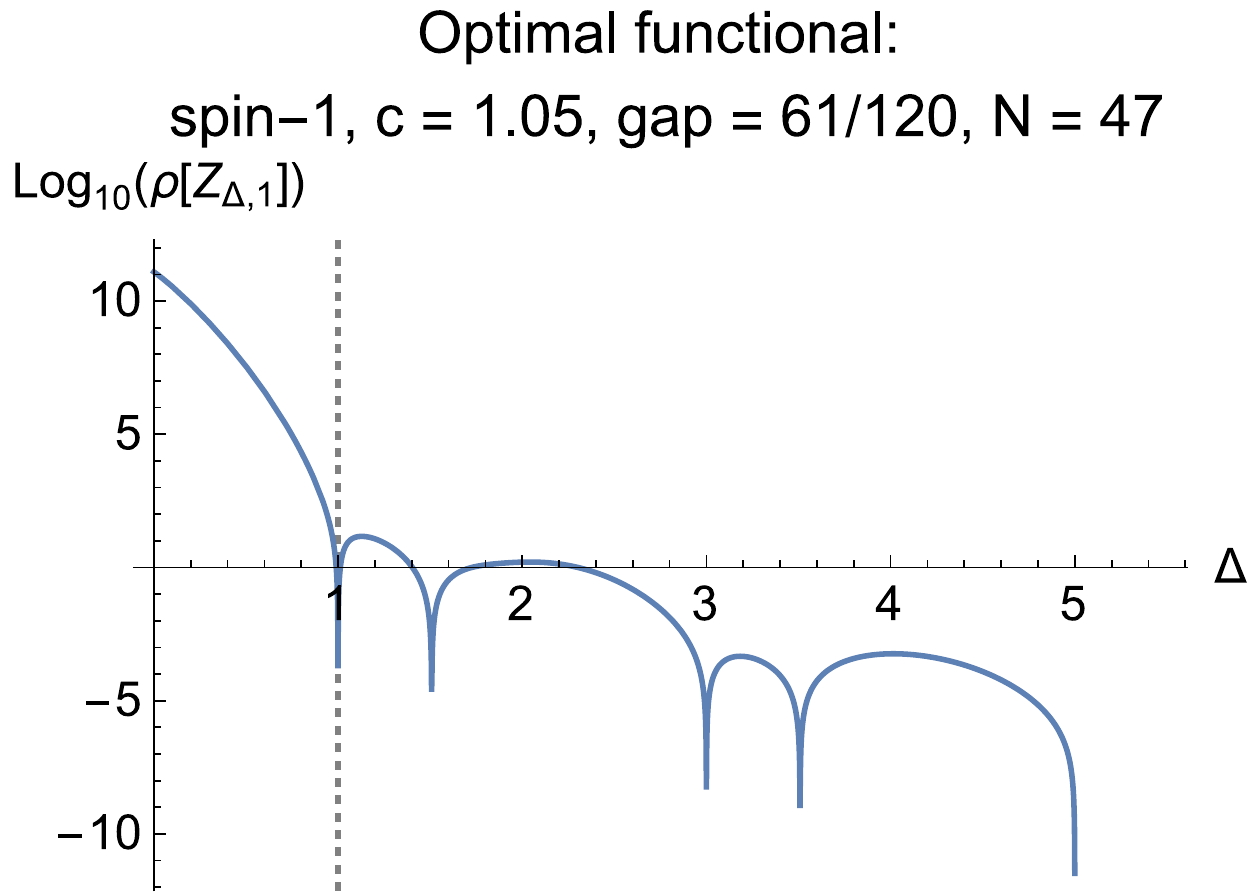}
}
\subfloat{
\includegraphics[width=.33\textwidth]{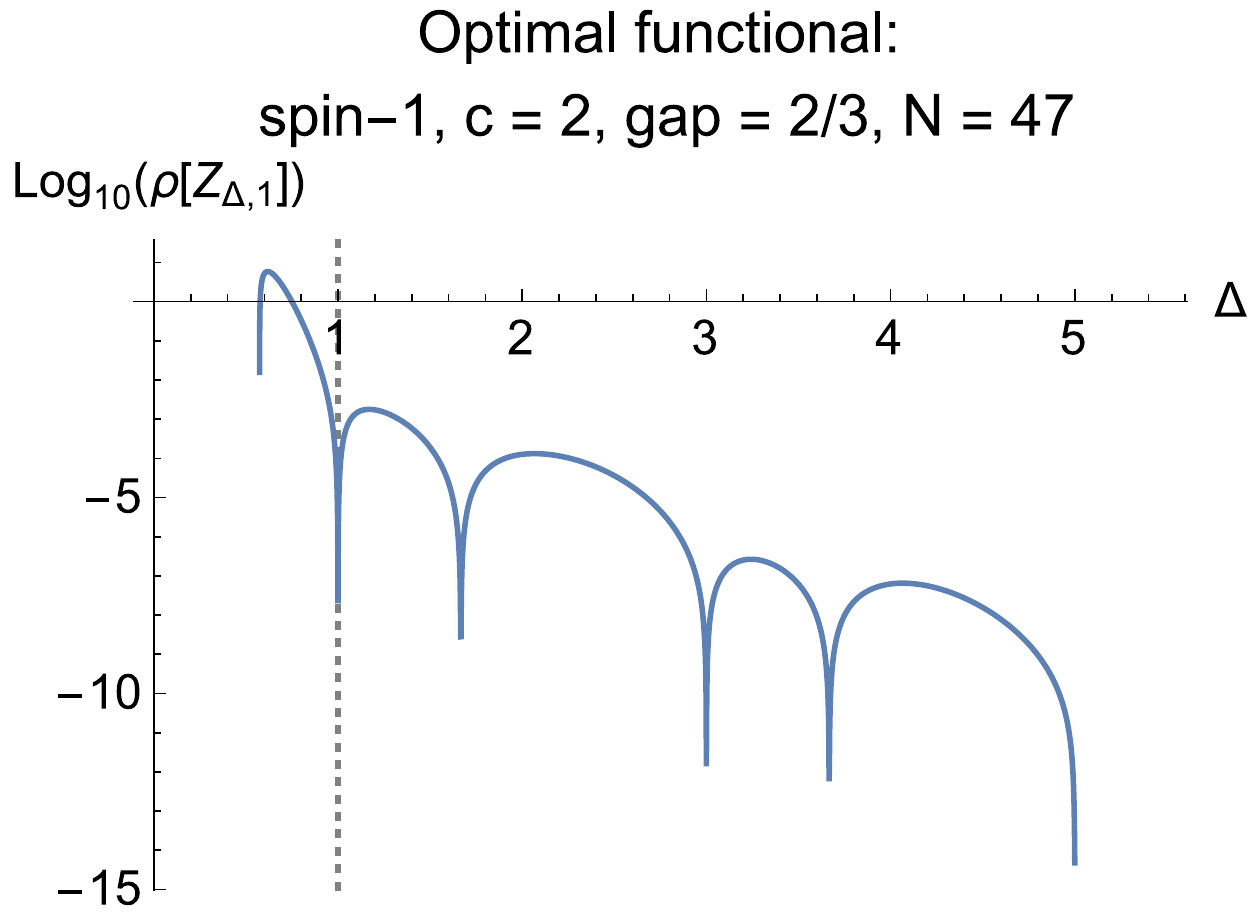}
}
\subfloat{
\includegraphics[width=.33\textwidth]{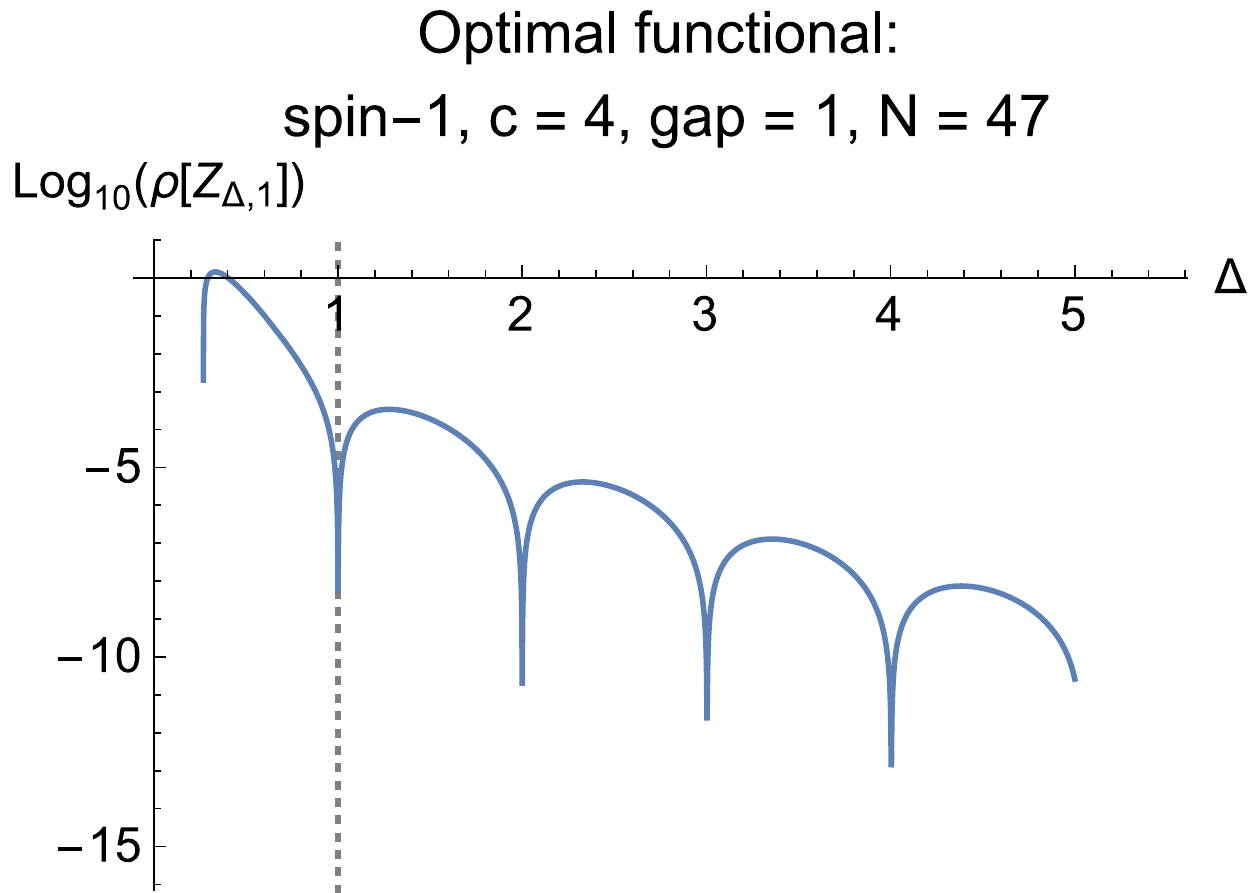}
}	
\caption{The action of the optimal functional on spin-0 and spin-1 reduced characters when the degeneracy of operators at the gap is maximized and the dimension gap bound is saturated for $c>1$. The dotted lines highlight the gap and the presence of marginal scalar primaries in the spin-0 case, and conserved spin-1 currents in the spin-1 case.}\label{fig:OptimalZeroesHigherCentralCharge}}
\end{figure}

Let us now consider the extremal spectra at higher values of the central charge when the gap is maximized. For $1\le c \le 4$, curiously, we find that conserved spin-1 currents and marginal scalar primaries generically occur in the extremal spectra when the gap is maximized --- Figure~\ref{fig:OptimalZeroesHigherCentralCharge} shows the evidence for this for a few values of the central charge. 

\begin{figure}[h!]
\centering
\subfloat{
\includegraphics[width=.48\textwidth]{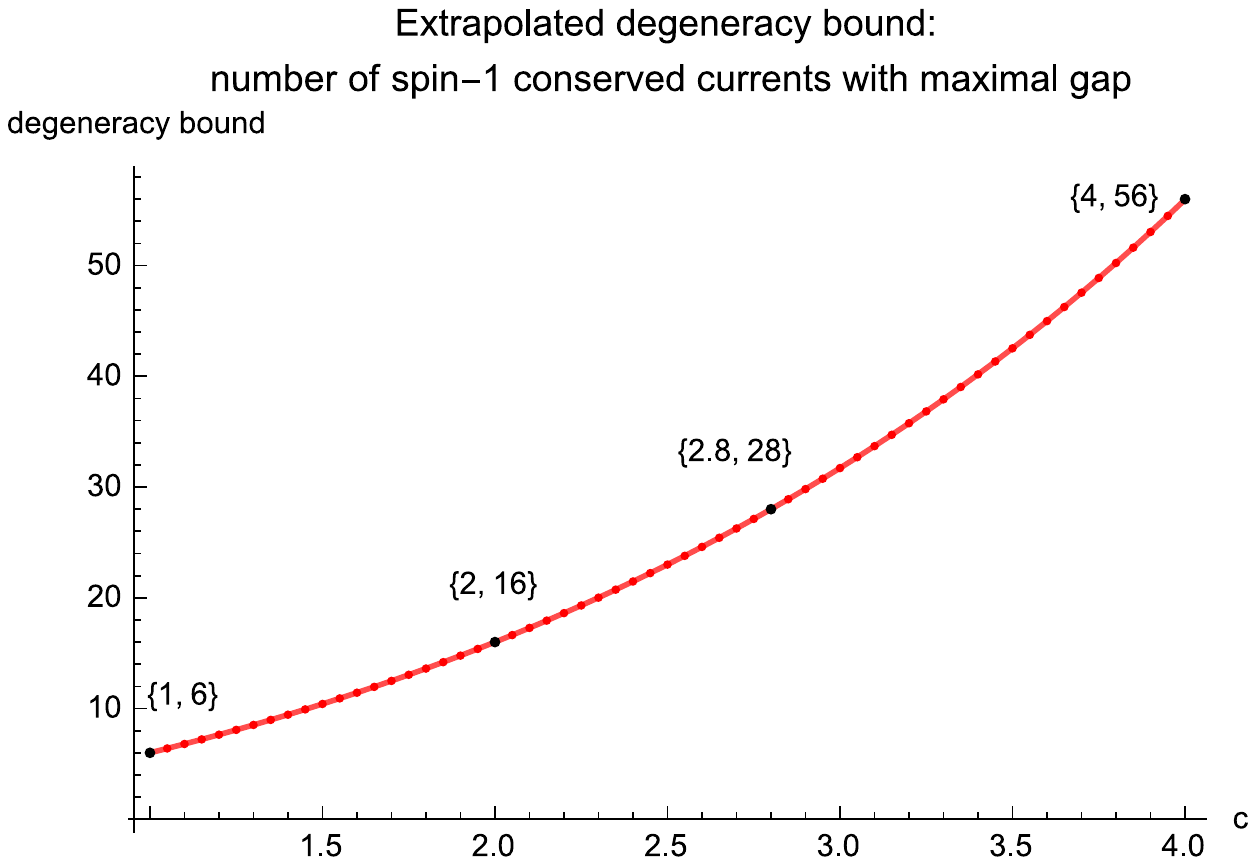}
}
\subfloat{
\includegraphics[width=.48\textwidth]{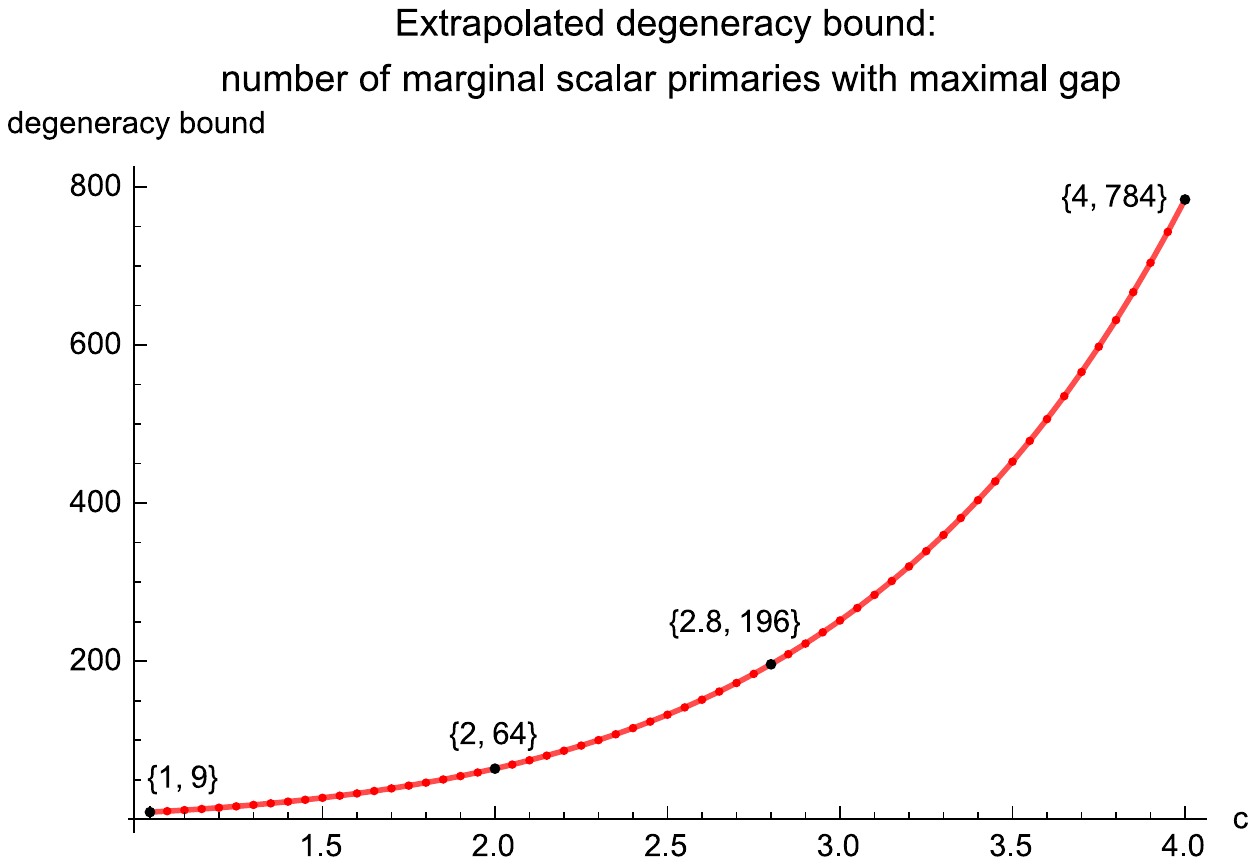}
}
\caption{{\bf Left:} The extrapolated upper bound on the total (holomorphic and antiholomorphic) number of conserved spin-1 currents as a function of the central charge when the gap is maximized. {\bf Right:} The extrapolated upper bound on the number of marginal scalar primaries as a function of the central charge. This is obtained by adding the naive bound obtained from (\ref{DegUpperBound}) with $(\Delta,s) = (2,0)$, which assumes a decomposition of the partition function into non-degenerate characters, to the bound on the total number of conserved spin-1 currents, cf., ${N_1\over 2}(q+\bar q) + N_0q\bar q = {N_1\over 2}[q(1-\bar q)+\bar q(1-q)] + (N_0+N_1) q \bar q$. In both plots, the black points denote the special theories discussed in section \ref{Sec:Kinks}. In particular, the extremal spectra at central charge $c=1,\,2,\,{14 \over 5},\,4$ are realized respectively by the $SU(2)$, $SU(3)$, $G_2$ and $SO(8)$ WZW models at level $1$.}\label{fig:ConservedSpin1Currents}
\end{figure}

Given the generic presence of conserved spin-1 currents and marginal scalar primaries in the extremal spectra when the upper bound on the dimension gap is saturated, one may then ask what the maximal degeneracies of these operators are subject to the maximal gap. Figure~\ref{fig:ConservedSpin1Currents} shows the growth of the upper bound of the degeneracy of spin-1 conserved currents subject to the maximal gap as a function of the central charge. 

Given the presence of spin-1 conserved currents, the upper bound obtained from (\ref{DegUpperBound}) with $(\Delta,s) = (2,0)$ does not exactly bound the number of marginal scalar primaries in the extremal spectrum: rather, it is a bound on the coefficient of $|\tau|^{1\over 2}\hat\chi_1\hat{\bar{\chi}}_1$ in the decomposition of the reduced partition function into \emph{non-degenerate} characters. To obtain the bound on the number of marginal scalar primary operators, one must sum up the bound obtained from (\ref{DegUpperBound}) together with the bound on the degeneracy of spin-1 conserved currents. The resulting upper bound on the degeneracy of marginal scalar primaries is plotted as a function of the central charge in Figure~\ref{fig:ConservedSpin1Currents}.

Of course, only when the upper bounds on the degeneracies at all weights converge to integers can the extremal spectrum with maximal gap be realized by a physical conformal field theory.
Based on the locations of the zeroes of the optimal functional and the values of the maximal degeneracies, we may then attempt to guess the CFTs that realize the extremal spectra with maximal gap. Let's consider in particular the case of $c=2,~\Delta_{\rm gap} = {2\over 3}$. The upper bound on the degeneracy of scalar primaries at the gap converges to 18, while the number of $\Delta=2$ scalar primaries is bounded above (in the case that the gap is $\Delta_{\rm gap} = {2\over 3}$) by 64.
From Figure~\ref{fig:OptimalZeroesHigherCentralCharge}, we see that the extremal spectrum also contains scalar primaries of dimension  $8\over 3$, 4, ${14\over 3},\ldots$.  The maximal degeneracies of these scalar primaries subject to $\Delta_{\rm gap} = {2\over 3}$ predicted by (\ref{DegUpperBound}) are
72, 64, 450, \ldots, respectively. The extremal spectrum that saturates these bounds is that of the $SU(3)$ WZW model at level 1. The partition function of this theory admits a decomposition into non-degenerate characters 
with precisely the operator dimensions of the $\Delta_{\rm gap}={2\over 3}$ extremal spectrum, with integer coefficients equal to the predicted maximal degeneracies: 
\begin{align}
	&(q\bar q)^{1\over 24} \left\{ \hat{Z}_{\rm ext}\left(2, {2\over 3}\right) - \hat{\chi}_0(\tau)\hat{\bar\chi}_0(\bar\tau) \right\} \nonumber\\
	=& |\tau|^{1\over 2}\left[18(q\bar q)^{1\over 3} +8(q+\bar q)+ 36(q^{4\over 3}\bar q^{1\over 3} + q^{1\over 3}\bar q^{4\over 3}) + 8(q^2+\bar q^2) + 48q\bar q
	+\ldots\right].
\end{align}

Let us now consider the theory that lives at the kink of the bounding curve in Figure~\ref{fig:DeltaGapExtrapolated-TAU}, with $c=4$ and $\Delta_{\rm gap} = 1$. The upper bound on the number of $\Delta = 1$ scalar primaries converges to 192, while the bound on $\Delta=2$ scalar primaries converges to 784.
Furthermore, the corresponding optimal functional has zeroes at every integer dimension when acting on spin-0 and spin-1 reduced characters. In fact, this theory is nothing but the CFT of 8 free fermions (with diagonal GSO projection), with partition function
\begin{equation}
Z_{\text{ext}}(4, 1) = {1\over 2}\left(\left|{\Theta_2(\tau)\over\eta(\tau)}\right|^8+\left|{\Theta_3(\tau)\over\eta(\tau)}\right|^8+\left|{\Theta_4(\tau)\over\eta(\tau)}\right|^8\right),
\end{equation}
where the $\{\Theta_a(\tau)\}$ are the Jacobi theta functions.
The expansion of this extremal partition function into non-degenerate characters yields precisely the operator spectrum and degeneracies predicted by (\ref{DegUpperBound}) and the zeroes of the optimal functional:
\ie
& (q\bar q)^{1\over 8} \left\{ \hat{Z}_{\text{ext}}(4,1)-\hat{\chi}_0(\tau)\hat{\bar\chi}_0(\bar\tau) \right\} 
\\
=& |\tau|^{1\over 2}\bigg[28(q+\bar q) + 192(q\bar q)^{1\over 2} + 105(q^2+\bar q^2) + 728 q\bar q + 1344(q^{1\over 2}\bar q^{3\over 2} + q^{3\over 2}\bar q^{1\over 2})+\ldots\bigg].
\fe
In particular, the 28 holomorphic conserved spin-1 currents correspond to the 28 fermion bilinears. 

\begin{figure}[h!]{
\centering
\subfloat{
\includegraphics[width=.45\textwidth]{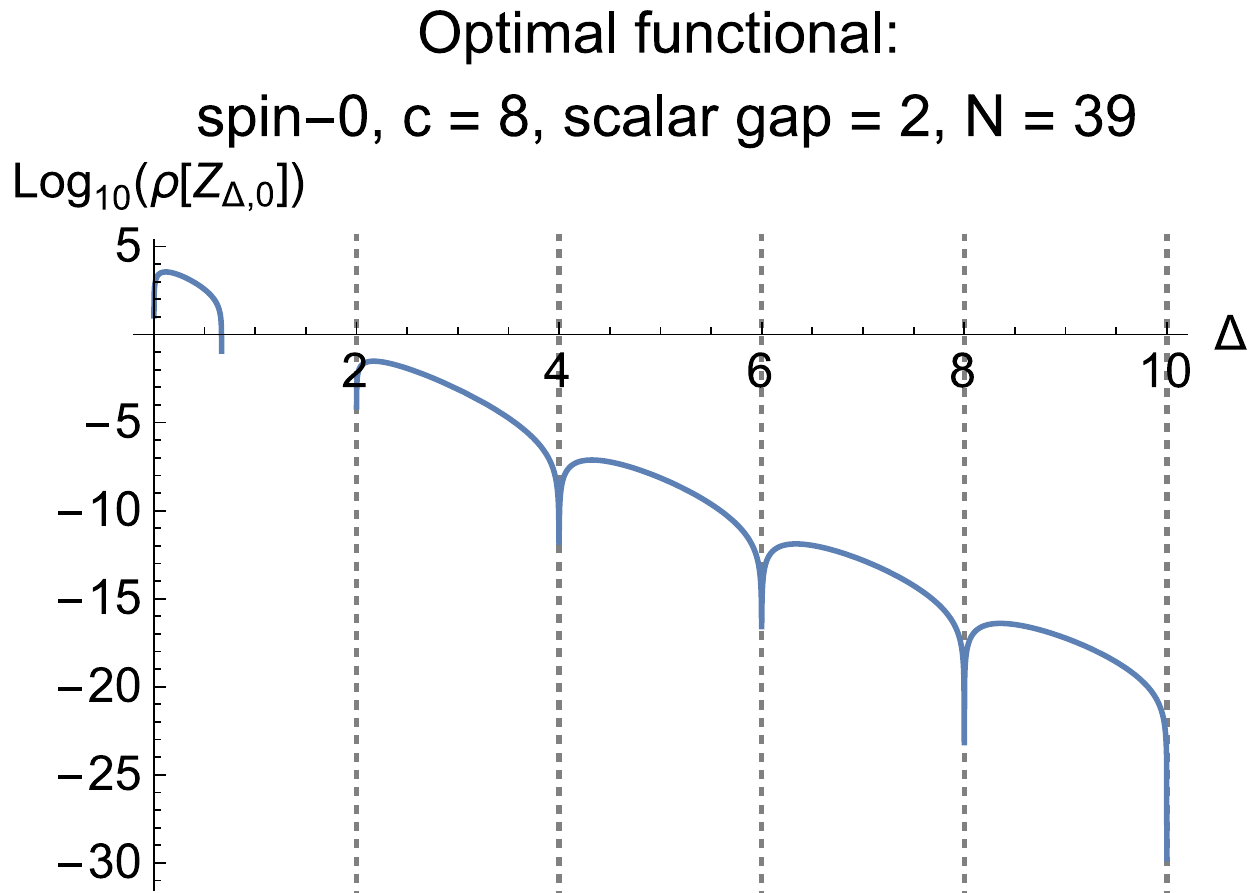}
}
\subfloat{
\includegraphics[width=.45\textwidth]{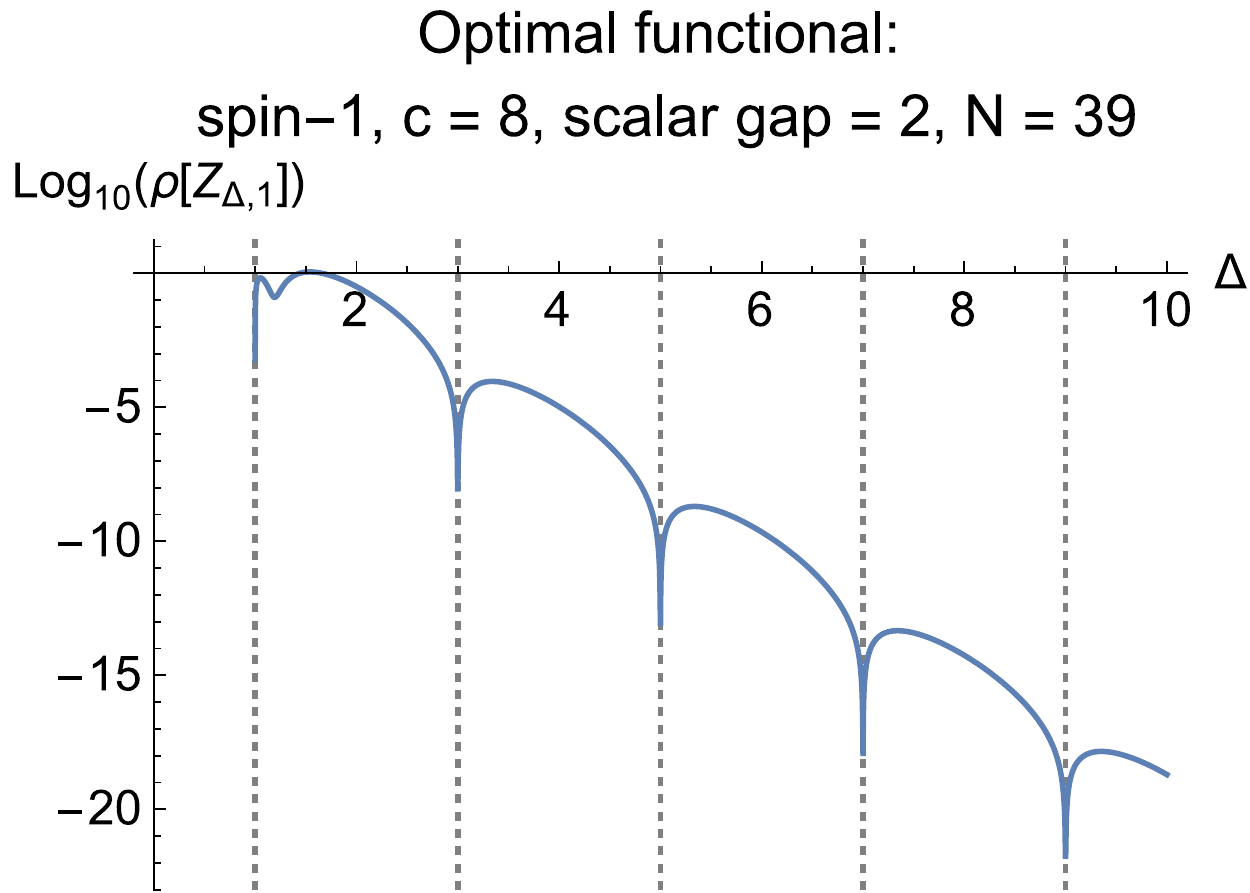}
}
\caption{The optimal functional acting on spin-0 and spin-1 reduced characters for $c=8$, $\Delta_{\rm gap}^{s=0}=2$.}\label{fig:OptimalZeroesc=8}}
\end{figure}

Another illustrative example is the case of the $c=8$ theory that saturates the upper bound on the gap for scalar primaries, $\Delta_{\text{gap}}^{s=0}=2$, populating the first kink in Figure~\ref{fig:ScalarDeltaGapExtrapolated}. Figure~\ref{fig:OptimalZeroesc=8} shows the action of the optimal functional that maximizes the degeneracy of the dimension-two scalar operators on the spin-0 and spin-1 reduced characters.
Notice that the functional has zeroes when acting on spin-0 characters of even integer dimensions and zeroes at odd integer dimensions when acting on spin-1 characters. Furthermore, maximizing the degeneracy of $\Delta=2$ scalar primaries subject to the scalar gap $\Delta_{\rm gap}^{s=0} = 2$ reveals an upper bound of 61504 marginal scalar primaries.
In fact, the extremal spectrum is nothing but that of 8 compact bosons on the $\Gamma_8$ Narain lattice, with holomorphically factorized partition function\footnote{We use the notation $Z_{\text{ext},s=0}(c,\Delta_{\text{gap}}^{s=0})$ to refer to the partition function of the CFT with the degeneracy of scalar primaries saturating the scalar gap $\Delta_{\text{gap}}^{s=0}$ maximized.}
\begin{equation}\label{ExtremalZc=8}
Z_{\text{ext},s=0}(8,2) =\left( j(\tau)\bar j(\bar \tau)\right)^{1\over 3}.
\end{equation}
The partition function (\ref{ExtremalZc=8}) admits a decomposition into non-degenerate characters with a spectrum of primary operators and non-negative integer coefficients predicted by the maximal degeneracies (\ref{DegUpperBound}) and the zeroes of the optimal functional
\begin{align}
&(q\bar q)^{7\over 24} \left\{ \hat{Z}_{\text{ext},s=0}(8,2)-\hat{\chi}_0(\tau)\hat{\bar\chi}_0(\bar\tau) \right\} \nonumber\\
=& |\tau|^{1\over 2}\bigg[248 (q+\bar q) + 3875(q^2+\bar q^2) + 61008 q\bar q + 30380 (q^3 + \bar q^3) + 957125(q^2\bar q+ q\bar q^2) + \ldots\bigg].
\end{align}

\section{Discussion and Open Questions}

By optimizing the linear functional acting on the modular crossing equation, we have uncovered a surprisingly rich set of constraints on the spectrum. However, the semi-definite programming approach becomes difficult at large values of the central charge: while we have concluded that the asymptotic slope of $\Delta_{\rm mod}(c)$ at large $c$ lies between ${1\over 12}$ and ${1\over 9}$, we still do not know its accurate value (which amounts to an upper bound on the mass of the lightest massive particle in a theory of quantum gravity in $AdS_3$ in Planck units \cite{Hellerman:2009bu}). We have identified the shape of the optimal linear functional numerically, and hopefully this will eventually lead to an analytic derivation of the optimal bound on the dimension gap at large $c$.

It is nonetheless clear from our results that, at large $c$, the basis (\ref{basis}) is inefficient for representing the optimal linear functional. Presumably, the latter is more appropriately expressed as an integral transform, rather than derivatives taken at $\tau=-\bar\tau=i$. Our preliminary attempt at a multi-point bootstrap approach \cite{Hogervorst:2013sma, Echeverri:2016ztu} has yielded results consistent with $\Delta_{\rm mod}$, but it did not improve the numerical efficiency due to the need for a polynomial approximation of functions of the conformal weight in implementing the semi-definite programming with SDPB.

In deriving most of our modular constraints, we have ignored the requirement that the degeneracies are integers. For instance, if the degeneracy at the maximal gap is not an integer, demanding that the degeneracy takes an integer value would slightly lower the upper bound on the gap. However, since the degeneracy bound grows exponentially with the central charge, the improvement of the bound by demanding integral degeneracy at the gap seems inconsequential.

One can also place bounds on the gap in the spectrum by considering the OPE of a pair of primaries (say at the gap), and using the crossing equation of the sphere 4-point function, by considering the decomposition of the 4-point function in Virasoro conformal blocks and imposing positivity of the coefficients. This is currently being investigated. Ultimately, one would like to combine crossing equation for the sphere 4-point function with the modular invariance of torus 1-point function. Perhaps the most efficient way to do this, instead of considering the crossing equations that involve many external operators, is to study the modular constraints from higher genus partition functions.

\section*{Acknowledgements} 

We are grateful to Chi-Ming Chang, Ethan Dyer, Tom Hartman, Simeon Hellerman, Christoph Keller, Elias Kiritsis, Petr Kravchuk, Jaehoon Lee, Eric Perlmutter, David Poland, Balt van Rees, Slava Rychkov, Ashoke Sen, T. Senthil, Shu-Heng Shao, Yifan Wang, Cenke Xu for important discussions. XY would like to thank the organizers of the workshops {\it Conformal Field Theories and Renormalization Group Flows in Dimensions $d>2$}, Galileo Galilei Institute for Theoretical Physics, Florence, Italy, {\it NCTS Summer Workshop on Strings and Quantum Field Theory}, National Tsing Hua University, Hsinchu, Taiwan, and {\it Strings 2016}, YMSC, Tsinghua University, Beijing, China, for their hospitality during the course of this work. This work is supported by a Simons Investigator Award from the Simons Foundation, and in part by DOE grant DE-FG02-91ER40654. YL would like to thank the hospitality of the Berkeley Center for Theoretical Physics during the course of this work. SC is supported in part by the Natural Sciences and Engineering Research Council of Canada via a PGS D fellowship and thanks the organizers of the summer school on quantum gravity, cosmology and particle physics at the Institut d'\'Etudes Scientifiques de Carg\`ese for hospitality during the course of this work.
The numerical computations in this work are performed using the SDPB package \cite{Simmons-Duffin:2015qma} on the Odyssey cluster supported by the FAS Division of Science, Research Computing Group at Harvard University.

\bibliographystyle{JHEP}
\bibliography{gapdraft.bib}

\end{document}